\documentclass[useAMS,graphics,usenatbib]{mn2e}
\usepackage{amssymb,color}
\usepackage{epsfig}
\usepackage{dcolumn}

\newcolumntype{d}[1]{D{.}{.}{#1}}

\newcommand{\Hz}{~\mbox{Hz}}
\newcommand{\erg}{~\mbox{erg}}

\newcommand{\Myr}{~\mbox{Myr}}

\newcommand{\eV}{~\mbox{eV}}
\newcommand{\Mpc}{~\mbox{Mpc}}
\newcommand{\cMpch}{~h^{-1}~\mbox{cMpc}}
\newcommand{\comMpch}{~h^{-1}~\mbox{comoving Mpc}}
\newcommand{\Mpch}{~h^{-1}~\mbox{Mpc}}

\newcommand{\ckpch}{~h^{-1}~\mbox{ckpc}}

\newcommand{\cmsqi}{~\mbox{cm}^{-2}}
\newcommand{\cmci}{~\mbox{cm}^{-3}}

\newcommand{\kpc}{~\mbox{kpc}}
\newcommand{\pc}{~\mbox{pc}}
\newcommand{\K}{~\mbox{K}}

\newcommand{\yr}{~\mbox{yr}}
\newcommand{\invyr}{~\mbox{yr}^{-1}}
\newcommand{\invs}{~\mbox{s}^{-1}}
\newcommand{\Msunh}{~h^{-1}~\mbox{M}_{\odot}}
\newcommand{\Msun}{~\mbox{M}_{\odot}}
\newcommand{\Zsun}{~Z_{\odot}}

\newcommand{\kms}{~\mbox{km s}^{-1}}

\title[Simulating Cosmic Reionization]{Spatially adaptive radiation-hydrodynamical simulations of galaxy formation 
  during cosmological reionization}

\author[Pawlik et al.]
 {Andreas H. Pawlik$^{1}$\thanks{E-mail:pawlik@mpa-garching.mpg.de}, Joop
   Schaye$^2$, Claudio Dalla Vecchia$^{3,4}$\\
   $^1$Max-Planck Institute for Astrophysics, Karl-Schwarzschild-Strasse 1, 85748 Garching, Germany\\
   $^2$Leiden Observatory, Leiden University, P.O. Box 9513, 2300 RA Leiden,The Netherlands\\
   $^3$Instituto de Astrof\'isica de Canarias, C/ V\'ia L\'actea s/n,38205 La Laguna, Tenerife, Spain \\
   $^4$Departamento de Astrof\'isica, Universidad de La Laguna, Av. del Astrof\'isico Franciso S\'anchez s/n, 38206 La Laguna, Tenerife, Spain}

\begin{document}

\date{}

\pagerange{\pageref{firstpage}--\pageref{lastpage}} \pubyear{xxxx}

\maketitle

\label{firstpage}

\begin{abstract}
We present a suite of cosmological radiation-hydrodynamical simulations of the
assembly of galaxies driving the reionization of the intergalactic medium
(IGM) at $z \gtrsim 6$.  The simulations account for the hydrodynamical
feedback from photoionization heating and the explosion of massive stars as
supernovae (SNe). Our reference simulation, which was carried out in a box of
size $25 \comMpch$ using $2\times 512^3$ particles, produces a reasonable
reionization history and matches the observed UV luminosity function of
galaxies. Simulations with different box sizes and resolutions are used to
investigate numerical convergence, and simulations in which either SNe or
photoionization heating or both are turned off, are used to investigate the
role of feedback from star formation. Ionizing radiation is treated using
accurate radiative transfer at the high spatially adaptive resolution at which
the hydrodynamics is carried out. SN feedback strongly reduces the star
formation rates (SFRs) over nearly the full mass range of simulated galaxies
and is required to yield SFRs in agreement with observations. Photoheating
helps to suppress star formation in low-mass galaxies, but its impact on the
cosmic SFR is small. Because the effect of photoheating is masked by the strong
SN feedback, it does not imprint a signature on the UV galaxy luminosity
function, although we note that our resolution is insufficient to model
star-forming minihaloes cooling through molecular hydrogen transitions. Photoheating 
does provide a strong positive feedback on reionization 
because it smooths density fluctuations in the IGM, which lowers
the IGM recombination rate substantially. Our simulations demonstrate a tight
non-linear coupling of galaxy formation and reionization, motivating the need
for the accurate and simultaneous inclusion of photoheating and SN
feedback in models of the early Universe.
\end{abstract}


\section{Introduction}
The first sources of ionizing radiation in our universe transformed the cold
and neutral cosmic gas that was present shortly after the Big Bang into the
hot and ionized plasma that we observe in the intergalactic medium (IGM)
today. Research into this transformation, which took place in the first
billion years, during the Epoch of Reionization, is crucial for understanding
the formation and evolution of galaxies, including our own galaxy, the Milky
Way (for reviews see, e.g., \citealp{Barkana2001}; \citealp{Morales2010};
\citealp{Ricotti2010}; \citealp{Bromm2011}; \citealp{Zaroubi2013}; \citealp{Natarajan2014}). A
number of new telescopes are currently being developed or are already underway
to unravel the astrophysics of these early times, including the {\it Low
  Frequency Array} (LOFAR; e.g., \citealp{Zaroubi2012}), the {\it Murchison
  Widefield Array} (MWA; e.g., \citealp{Lidz2008}), the {\it Square Kilometer
  Array} (SKA; e.g., \citealp{Mellema2013}), the {\it James Webb Space
  Telescope} (JWST; e.g., \citealp{Zackrisson2011}), and many others (e.g.,
\citealp{FOB2006}; \citealp{Fanreview2006}; \citealp{Pritchard2012}). These
investments at the observational frontier have spurred the construction of
models of the early Universe to guide the design of observational campaigns
and to help interpret the observations once the data arrives.
\par
Most of the current models are built on the observationally supported notion
that reionization is driven mainly by star-forming galaxies (e.g.,
\citealp{Choudhury2007}; \citealp{Volenteri2009}; \citealp{Loeb2009}; \citealp{Robertson2010};
\citealp{Raicevic2011}; \citealp{Fontanot2012}). A natural implication of these models is a strong
coupling between reionization and galaxy formation caused by ionizing
radiative feedback (e.g., \citealp{Ciardi2005}). Photoionization heating by
star-forming galaxies typically raises the temperature in the reionized gas to
a few times $10^4 \K$. The associated increase in gas pressure increases the Jeans
scale, i.e., the scale below which pressure prevents the collapse of gas into
gravitationally bound objects (e.g., \citealp{Gnedin2000}). This boils gas
out of low-mass dark matter (DM) haloes and impedes the accretion of gas from the
IGM onto them (e.g., \citealp{Rees1986}; \citealp{Shapiro1994}; 
\citealp{Barkana1999}; \citealp{Gnedin2000}; \citealp{Dijkstra2004}; \citealp{Finlator2011};
\citealp{Petkova2011}; \citealp{Noh2014}).  Photoheating
therefore leads to a reduction in the fuel from which stars form, which lowers
the star formation rates (SFRs) and the ionizing emissivities of low-mass
galaxies. This makes it more difficult for galaxies to reionize the IGM and,
therefore, photoheating provides a negative feedback on reionization.
\par
Photoheating also provides a positive feedback on reionization. The increase
in the gas pressure and Jeans mass smooths density fluctuations in the
IGM. This decreases the gas clumping factor, $C_{\rm IGM} \equiv \langle
n_{\rm H}^2 \rangle / \bar{n}_{\rm H}^2$, where the angular brackets indicate
the average in the IGM, and $\bar{n}_{\rm H}$ is the cosmic mean hydrogen
density (e.g., \citealp{Wise2005}; \citealp{Pawlikclump2009};
\citealp{Finlatorclump2012}; \citealp{Shull2012}; \citealp{So2014}).  Because
the rate at which the IGM recombines is proportional to the gas clumping
factor (e.g., \citealp{MHR1999}), photoheating decreases the IGM recombination
rate. Fewer ionizing photons are then needed to keep the gas ionized, which 
facilitates reionization (e.g., \citealp{Pawlikclump2009};
\citealp{Shull2012}; \citealp{Finlatorclump2012}; \citealp{Sobacchi2014}).
This positive feedback from photoheating will be especially strong if X-ray
sources preheat the gas before it is reionized (e.g., \citealp{Madau2004};
\citealp{Ricotti2005}; \citealp{Haiman2011}; \citealp{Jeon2014};
\citealp{Xu2014}; \citealp{Knevitt2014}).
\par
The feedback from photoheating may be accompanied by feedback from the
explosion of stars as supernovae (SNe; e.g., \citealp{Dekel1986};
\citealp{Wyithe2013}). SNe are thought to suppress star formation in low-mass
galaxies by expelling gas in winds, which makes it more difficult for galaxies
to reionize the IGM. However, the ejection of gas may open additional
low-density channels through which ionizing photons may escape the ISM more
easily, thus making it easier for galaxies to reionize the IGM (e.g.,
\citealp{Yajima2009}; \citealp{Wise2009}; \citealp{Paardekooper2011};
\citealp{Kimm2014}).  The suppression of star formation by SNe may be
amplified in a nonlinear manner by the feedback from photoheating
(\citealp{Pawlik2009}; \citealp{Finlator2011};
\citealp{Hopkins2013}). Locally, SNe may also increase the rates at which
stars form as interstellar gas is compressed to star-forming densities in
colliding SN shocks (e.g., \citealp{Geen2013}). The wide range of
possibilities by which star formation may feed back on the formation of
structure and the reionization of the IGM makes modelling reionization a
challenging task.
\par
Cosmological simulations are generally considered to be among the most
powerful techniques to investigate the impact of stellar feedback on
reionization and galaxy formation (e.g., \citealp{Trac2011};
\citealp{Finlator2012}), but are also computationally highly demanding. A
principal computational challenge is the required large dynamic range. The
characteristic size of individual ionized regions is of order 10 comoving Mpc
(cMpc; e.g., \citealp{Furlanetto2006}). The redshift at which these regions
percolate to reionize the Universe shows spatial fluctuations due to the
modulation of galaxy formation by structure formation modes on still larger
scales, of order $100 \cMpch$ (e.g., \citealp{Barkana2004};
\citealp{Iliev2014}). The concurrent need to resolve the Jeans mass in the IGM
as well as the population of atomically cooling galaxies with masses above
$\sim 10^8 \Msun$, the prime candidates for driving reionization, extends the
spectrum of relevant scales down to a few kpc, and less (e.g.,
\citealp{Gnedin2006}; \citealp{Bolton2009}).
\par
A second computational challenge is the large expense of the radiative
transfer (RT) of ionizing photons. A straightforward and accurate treatment of
the RT consists of tracing photons from each ionizing source along straight
rays, one source at a time (for an overview of RT methods see, e.g.,
\citealp{Iliev2006}; \citealp{Iliev2009}; \citealp{Trac2011}). The
computational cost of such approaches therefore increases linearly with the
number of ionizing sources. This renders simulations of cosmological volumes
that contain large numbers of sources computationally demanding. The issue
becomes particular severe if the ionizing radiation from recombining ions
needs to be treated accurately (e.g., \citealp{Maselli2003}; \citealp{Hasegawa2010};
\citealp{Raicevic2014}). The high cost of ray-based RT methods has triggered
the development of moment-based RT methods that are computationally less
expensive, but also less accurate (e.g., \citealp{Gnedin2001};
\citealp{Aubert2008}; \citealp{Finlator2009}; \citealp{Rosdahl2013}; \citealp{Norman2013}).
\par
To simplify the problem, cosmological hydrodynamical simulations of galaxy
formation often assume the gas to be in ionization equilibrium with a
spatially uniform ionizing background (e.g., \citealp{Miralda2000};
\citealp{Pawlikclump2009}; \citealp{Duffy2014}). Such simulations enable the
investigation of the response of the gas in galaxies and the IGM to heating by
ionizing photons at relatively low computational cost, but must assume the
redshift at which the ionizing background is instantaneously turned on. The
main drawbacks of such simulations are that they do not yield insights in the
timing and morphology of reionization, and that fluctuations in the intensity
due to local sources (e.g., \citealp{Gnedin2010}; \citealp{Yajima2012};
\citealp{Rahmati2013}) and the ability of the gas to throw shadows and
self-shield (e.g., \citealp{Abel2007}; \citealp{Rahmati12013};
\citealp{Altay2013}) are ignored. Sometimes, self-shielding is approximately
accounted for by setting the intensity of the ionizing background to zero at
densities above which the gas is expected to be neutral in the absence of
local sources (e.g., \citealp{Nagamine2010}; \citealp{Paardekooper2013};
\citealp{Vogelsberger2013}).
\par
Other simulations focus on the accurate treatment of the RT of ionizing
photons at the expense of a detailed treatment of the feedback from radiative
heating (e.g., \citealp{Nakamoto2001}; \citealp{Razoumov2002};
\citealp{Ciardi2003}; \citealp{Iliev2006}; \citealp{McQuinn2007};
\citealp{Finlator2009}; \citealp{Aubert2010}; \citealp{Baek2012}). In this
type of simulations, the computational complexity is reduced by computing the
RT and the evolution of structure separately, ignoring the impact of
photoionization heating on the dynamics and distribution of the gas. Often, to
simplify the problem further, the dynamics of the gas is ignored altogether
and assumed to trace that of the DM (e.g., \citealp{Iliev2006}). A
semi-analytical model then specifies the location of the galaxies and the
rates at which these galaxies form stars. Sometimes, the semi-analytical model
is calibrated to approximately account for the radiation-hydrodynamical response to
photoheating (e.g., \citealp{Ciardi2006}; \citealp{Iliev2007}).
\par
The most direct but also computationally most expensive way to investigate the
coupling of reionization and galaxy formation, is to carry out
three-dimensional cosmological radiation-hydrodynamical simulations. Several
of these simulations now exist, including the pioneering works by
\cite{Gnedin1997} and \cite{Gnedinreion2000}. However, even after more than a
decade of intense research, the field is still in its infancy.
Radiation-hydrodynamical simulations have only very recently begun to approach
the cosmological scales and the high resolution needed to capture the relevant
physics (\citealp{Norman2013}; \citealp{Gnedin2014}; \citealp{Kimm2014}).
\par
Most earlier radiation-hydrodynamical simulations of reionization were
restricted to relatively small volumes that are less representative of the
cosmological mean (e.g., \citealp{Gnedin2000}; \citealp{Ricotti2008};
\citealp{Petkova2011}; \citealp{Finlator2011}; \citealp{Hasegawa2013}). Only
some simulations afforded ray-based RT methods that ensure the accurate
propagation of ionization fronts in inhomogeneous density fields
(\citealp{Hasegawa2013}). The majority of simulations employed the
computationally less expensive but also numerically diffusive moment-based RT
methods (e.g., \citealp{Gnedinreion2000}; \citealp{Petkova2011};
\citealp{Finlator2011}; \citealp{Norman2013}; \citealp{So2014}). In some
cases, the computational expense needed to be reduced further by lowering the
spatial resolution at which the RT is carried out, thus requiring a
sub-resolution model for the propagation and consumption of photons at
hydrodynamically resolved scales (\citealp{Finlator2011}).
\par
In this work we present radiation-hydrodynamical simulations of reionization
in cosmological volumes at high spatially adaptive resolution, implementing
accurate RT using the TRAPHIC code (\citealp{Pawlik2008}). TRAPHIC solves the
time-dependent RT equation by tracing photon packets at the speed of light and
in a photon-conserving manner through the simulation box. TRAPHIC thus enables
an accurate treatment of shadows and self-shielding. The photon packets are
transported directly on the spatially adaptive, unstructured grid defined by
the SPH particles, thus exploiting the full dynamic range of the hydrodynamic
simulation. The large number of ionizing sources typical of these simulations
does not pose a problem because TRAPHIC employs a photon packet merging
technique that renders the computational cost of the RT independent of the
number of sources. TRAPHIC is implemented in the galaxy formation code GADGET
(last described in \citealp{Springel2005}), and this enables us to account for
the radiation-hydrodynamical feedback from star formation and reionization.
\par
We focus our investigations on the roles of radiative heating (which we also
refer to as photoheating) and SNe in shaping the SFR and reionization history
in the physically motivated and observationally supported models of the early
Universe that our simulations provide. It is currently not feasible to provide
an ab initio description of the internal structure of galaxies in cosmological
volumes, even in substantially less expensive simulations in which the
detailed transfer of ionizing photons is ignored (e.g., \citealp{Trac2011};
\citealp{Scannapieco2012}; \citealp{Schaye2014}). Simulations like ours
therefore rely on sub-resolution models of the interstellar medium (ISM). This
introduces free parameters that require careful calibration, e.g., using
zoomed simulations that achieve a higher resolution by focusing on a smaller
computational volume, or observations. The two main parameters relevant to
simulations of galaxy formation during reionization are the fraction $f_{\rm
  SN}$ of the energy released in SNe and other unresolved stellar evolution
processes that is available to heat the ISM, and the fraction $f_{\rm
  esc}^{\rm subres}$ of the ionizing photons that escapes the unresolved ISM
and are available to reionize the gas.
\par
This paper is organized as follows. In Section~\ref{Sec:Sim} we describe our
numerical techniques and the set of simulations we have carried out. In
Section~\ref{Sec:Results} we describe our results. We start in
Section~\ref{Sec:Reionization} with an overview of the simulations and a
comparison with observational constraints. This is continued in
Section~\ref{Sec:Feedback} and extended by an analysis of the impact of
stellar feedback on galaxy formation and reionization. In
Section~\ref{Sec:Discussion} we 
discuss the sensitivity of our results to variations in
physical parameters as well as some of the main limitations inherent to our
numerical approach. In Section~\ref{Sec:Summary} we conclude with a brief
summary. Lengths are expressed in physical units, unless noted otherwise.

\section{Simulations}
\label{Sec:Sim}
\subsection{Gravity and hydrodynamics}

\begin{table*}
\begin{center}
\caption{Simulation parameters:
  simulation name;
  comoving size of the simulation box, $L_{\rm box}$;
  number of DM particles, $N_{\rm DM}$;
  mass of DM particles, $m_{\rm DM}$;
  mass of gas particles, $m_{\rm gas}$;
  comoving gravitational softening scale, $\epsilon_{\rm soft}$;
  purpose. The number of SPH particles initially equals $N_{\rm DM}$ (it decreases
  during the simulation due to star formation). All simulations are carried out down to redshift 
  $z = 6$, except for simulation L12N512, which was stopped at $z = 6.8$ because of the 
  large computational expense.
  \label{tbl:params}}
\begin{tabular}{llllllll}

\hline
\hline  
simulation & $L_{\rm box}$ & $N_{\rm DM}^{1/3}$ & $m_{\rm DM}$  &$m_{\rm gas}$
&$\epsilon_{\rm soft}$ & purpose &\\
           & $[\cMpch]$    &              & $[\Msun]$  & $[\Msun]$& $[\ckpch]$    &&\\           
\hline

{L25N512}        & $25.0$ & $512$ & $1.00 \times 10^7$ &$2.04 \times 10^6$ & 1.95    & reference &\\

{L12N256}        & $12.5$ & $256$ & $1.00 \times 10^7$ &$2.04 \times 10^6$ & 1.95   &  small box &  \\

{L12N512}        & $12.5$ & $512$ & $1.25 \times 10^6$ &$2.55 \times 10^5$ & 0.98  &  high resolution &  \\

{L50N512}        & $50.0$ & $512$ & $8.02 \times 10^7$ &$1.63 \times 10^7$ & 3.91  & large box &  \\

{L25N256}        & $25.0$ & $256$ & $8.02 \times 10^7$ &$1.63 \times 10^7$ & 3.91    & low resolution &  \\

\hline
\label{tab:sims}
\end{tabular}
\end{center}
\end{table*}

We use a modified version of the N-body/TreePM Smoothed Particle Hydrodynamics
(SPH) code GADGET (\citealp{Springel2005}; our specific version is derived
from that discussed in \citealp{Schaye2010}) to perform a suite of
cosmological radiation-hydrodynamical simulations of galaxy formation down to
redshift $z = 6$ (see Table~\ref{tab:sims}).
\par
The simulations are initialized at redshift $z = 127$. Initial particle
positions and velocities are obtained by applying the Zeldovich approximation
(Zeldovich 1970) to particles arranged along a uniform grid of glass-like
structure (\citealp{White1996}). We use a transfer function for matter
perturbations generated with CAMB (\citealp{Lewis2002}) and apply it to
describe perturbations in both the dark matter and the gaseous components. We
adopt the $\Lambda$CDM cosmological model with parameters $\Omega_{\rm m} =
0.265$, $\Omega_{\rm b} = 0.0448$ and $\Omega_{\Lambda} = 0.735$, $n_{\rm s} =
0.963$, $\sigma_8 = 0.801$, and $h = 0.71$ (\citealp{Komatsu2011}). These
parameters are consistent with the most recent constraints from observations
of the Cosmic Microwave Background by the {\it Planck} satellite
(\citealp{Planck2014}; \citealp{Planck2015}).
\par
Our reference simulation utilizes $2 \times 512^3$ DM and gas particles in a
box of size $25 \Mpch$. This implies a DM particle mass of $m_{\rm DM} = 0.7
\times 10^7 \Msunh \approx 10^7 \Msun$ and a gas particle mass of $m_{\rm gas}
= 1.4\times 10^6 \Msunh \approx 2 \times 10^6\Msun$. Atomically cooling haloes
with mass (Eq.~3.12 in \citealp{Loeb2010})
\begin{equation}
M_{\rm vir} \ge 10^8 \Msun \left(\frac{T_{\rm vir}}{10^4 \K}\right)^{3/2} \left(\frac{\mu}{0.6}\right)^{-3/2}  \left(\frac{1+z}{10}\right)^{-3/2}, 
\end{equation}
where $\mu$ is the mean atomic weight, are therefore resolved by at least $M_{\rm vir}  (\Omega_{\rm m} - \Omega_{\rm
  b}) \Omega_{\rm m}^{-1}  m_{\rm DM}^{-1} \sim 10 [(1+z)/10]^{-3/2}$ DM particles.
\par
\par
We apply a comoving Plummer-equivalent gravitational softening length, $\epsilon_{\rm soft}$,
equal to 1/25 of the mean DM particle separation to both DM and
baryonic particles, i.e., 
$\epsilon_{\rm soft}  \approx 1.95 \ckpch \ (L/25.0 \cMpch)\ (N_{\rm
    DM}^{1/3}/512)^{-1}$, where $N_{\rm DM}$ is the number of DM
particles.  SPH
quantities are estimated by averaging inside a sphere containing $N_{\rm ngb}
= 48$ neighboring gas particles and adopting the entropy conserving
formulation of SPH (\citealp{Springel2002}). The SPH kernel, i.e., the
radius of this sphere, is prevented from falling below $10^{-2} \epsilon_{\rm
  soft}$. 
\par

\subsection{Chemistry and cooling}
\label{sec:chemistry}
While our simulations account for the thermal feedback from the explosion of
massive stars as core-collapse SNe (see Section~\ref{Sec:SN}), for simplicity,
we do not follow the ejection and transport of metals. We therefore follow the
non-equilibrium chemistry and radiative cooling of the gas, assuming that the
gas is of primordial composition. Molecule formation is assumed to be
suppressed by the Lyman-Werner background expected to pervade the universe
soon after the formation of the first stars (e.g., \citealp{Haiman1997};
\citealp{Ricotti2001}; \citealp{Wise2005}; \citealp{Greif2006};
\citealp{Ahn2009}), a process not resolved in our simulations. Our simulations
thus ignore that molecular hydrogen may form in self-shielded regions
(e.g., \citealp{Wolcott2011}). 
\par
We consider all relevant atomic radiative cooling processes, using the rate
coefficients in \cite{Pawlik2011}: cooling by collisional ionization,
collisional excitation of atomic lines, the emission of free-free and
recombination radiation, and Compton cooling by the CMB. The species fractions
and the gas temperature are advanced in time using the explicit subcycling
time integration scheme presented in \cite{Pawlik2011}.  Once stars form and
emit radiation, the chemical and thermal evolution of the gas is also affected
by photoionization and photoheating, and by the injection of energy by SNe, as
we describe below.
\par
Solving for the chemistry and cooling of the gas is by far the most expensive
operation in our simulations. To speed up the calculations and unless
otherwise noted, we set $X = 1$, i.e., we neglect helium. In
Section~\ref{Sec:Parameters}, we will describe a preliminary comparison with a
simulation in which we follow the radiative heating and cooling by helium
assuming $X = 0.75$.
\par
Our simulations do not have sufficient resolution and do not capture the
physics required to describe the multiple phases of the dense gas in galaxies
that are expected to develop above densities $n_{\rm H}\sim 10^{-2}-10^{-1}
\cmci$ (e.g., \citealp{Schaye2004}) as observed in the local universe. We
therefore employ a subgrid model to describe the thermodynamics of the dense
gas, imposing a pressure floor $P\propto \rho^{\gamma_{\rm{eff}}}$ on gas with
densities $n_{\rm{H}} > n_{\rm{H}}^{\rm c}$, where $n_{\rm{H}}^{\rm c} \equiv
0.1 \cmci $, normalized to $P/k = 10^3 \cmci \K$ at the critical density
$n_{\rm{H}}^{\rm c}$. We use $\gamma_{\rm{eff}} = 4/3$ for which both the
Jeans mass and the ratio of the Jeans length and the SPH kernel of gas at the
pressure floor are independent of the density, thus preventing spurious
fragmentation due to a lack of numerical resolution (\citealp{Bate1997};
\citealp{Schaye2008}). Note that, unlike in simulations that impose a polytropic
equation of state, in our simulations, gas with densities above the critical
density can have pressure above the pressure floor. Our implementation follows \cite{Schaye2008}
and \cite{DallaVecchia2012}, to which
we refer the reader for further details. In our simulations adopting a hydrogen mass fraction $X
= 1$, we use a critical density higher than $n_{\rm{H}}^{\rm c}$ by a factor
$1/0.75$ to facilitate comparisons with simulations in which $X = 0.75$.

\subsection{Star formation}
\label{Sec:sf}
We employ the star formation implementation of \cite{Schaye2008}, to which we
refer the reader for details. Briefly, gas with densities $n_{\rm H} > n_{\rm
  H}^\star$ and temperatures $T < T^\star$ is allowed to form stars using a
pressure-dependent rate that reproduces the Kennicutt-Schmidt relation
observed in the local universe (\citealp{Kennicutt1998}) in idealized
simulations of isolated disk galaxies,
\begin{equation}
\dot \Sigma_\star = 1.515 \times 10^{-4} \Msun \yr \kpc^{-2} \left(\frac{\Sigma_{\rm
  gas}}{1 \Msun \pc^{-2}}\right)^{1.4}, 
\end{equation}
where $\dot \Sigma_\star$ is the SFR surface density and $\Sigma_{\rm gas} $
the gas surface density. The last equation has already been renormalized by a
factor of $1/1.65$ to account for the fact that it assumes the
\cite{Salpeter1955} IMF whereas we are using a Chabrier IMF. This conversion
factor between SFRs has been computed using the \cite{Bruzual2003} population
synthesis code for model galaxies of age $> 10^7 \yr$ forming stars at a
constant rate and is insensitive to metallicity.
\par
We set $n_{\rm H}^\star = n_{\rm H}^{\rm c}$ and use $T^\star \equiv \max(10^5
\K, 10^{0.5} T_{\rm floor})$, where $T_{\rm floor}$ is the density-dependent
temperature implied by the pressure floor below which particles are prevented
from cooling. Moreover, to avoid numerical artefacts at high redshifts when
the cosmological mean gas density is high, we require a minimum overdensity of
57.7 for gas to be allowed to form stars.
\par
The star formation law is interpreted stochastically, and the probability that
a star-forming gas particle is turned into a star particle in a time interval
$\Delta t$ is given by $\min(\Delta t/ \tau_\star, 1)$, where $\tau_\star
\equiv \Sigma_{\rm gas}/ \dot \Sigma_\star$ is the gas consumption time. Gas
particles are converted to star particles assuming a conversion efficiency of
100\%, i.e., the masses of the star particles are identical to those of the
gas particles from which they are formed. At higher resolution,
star formation and the associated feedback may be more extended in time and
in space. This may affect the
impact of star formation on the gas and therefore alter, e.g., the resolved
escape fraction of ionizing photons.

\subsection{Ionizing luminosities}
We interpret the star particles as simple stellar
populations, i.e., coeval stellar clusters that are characterized by an
initial mass function (IMF), metallicity, and age. 
\par
We compute the time-dependent hydrogen ionizing luminosities of these
star formation bursts using the population synthesis models of
\cite{Schaerer2003}. Since \cite{Schaerer2003} does not
tabulate ionizing luminosities for the Chabrier IMF, which is the IMF
used by our star formation recipe, we adopt the
ionizing luminosities of the \cite{Schaerer2003} models with Salpeter initial
mass function in the range $1-100 \Msun$ and metallicity $Z =
0.02 = \Zsun$. This yields ionizing luminosities close to those of
stellar populations drawn from a Chabrier IMF in the range $0.1-100
\Msun$ with subsolar
metallicity $Z = 10^{-3} = 0.05\Zsun$ and is consistent with the IMF used in our
star formation recipe and appropriate for stellar populations
at $z \gtrsim 6. $ (e.g., \citealp{Maio2010}; \citealp{Wise2014}; 
\citealp{Jeon2014b}).  
\par
We multiply the luminosities of each star particle by the sub-resolution
escape fraction $f_{\rm esc}^{\rm subres}$ to model the removal of photons due
to absorption in the unresolved ISM. Some previous works that carried out RT
simulations of reionization have calibrated the ISM escape fraction to yield
reionization histories in agreement with current observational constraints
(e.g., \citealp{Aubert2010}; \citealp{Finlatorclump2012};
\citealp{Ciardi2012}). However, as we will discuss below, the interpretation
of these constraints is subject to substantial systematic uncertainties and
limited by our incomplete understanding of the physical processes at play. We
therefore prefer to set $f_{\rm esc}^{\rm subres} = 1$, and use our
simulations to explore the dependence of the reionization history on box size
and resolution. In Section~\ref{Sec:Parameters}, we will briefly discuss the
impact of variations away from $f_{\rm esc}^{\rm subres} = 1$ on our results.
\par

\subsection{Ionizing radiative transfer}
We use the RT code TRAPHIC \citep{Pawlik2008,Pawlik2011} to transport ionizing
photons. TRAPHIC solves the time-dependent RT equation in SPH simulations by
tracing photon packets emitted by source particles through the simulation box
in a photon-conserving manner. The photon packets are transported directly on
the spatially adaptive set of SPH particles and hence the RT exploits the full
dynamic range of the hydrodynamical simulations. A directed transport of the
photon packets radially away from the sources is accomplished despite the
irregular distribution of SPH particles by guiding the photon packets inside
cones. A photon packet merging technique renders the computational cost of the
RT independent of the number of radiation sources.  In the following, we
provide a brief overview of TRAPHIC in order to motivate the meaning of the
numerical parameters of the RT specified below. The reader is referred to the
descriptions in \citet{Pawlik2008}, \citet{Pawlik2011} and \citet{Pawlik2013}
for further discussion.

\subsubsection{Method}

The transport of radiation starts with the emission of photon packets
by star particles in $N_{\rm EC}$ tessellating emission cones. The photons in each
photon packet are shared among the $\tilde{N}_{\rm
  ngb} \lesssim N_{\rm ngb}$ neighboring SPH particles residing in the cones.  In
cones containing zero neighbors, an additional, so-called virtual particle is
inserted to which the photon packet is assigned. Star particles emit
photons using emission time steps $\Delta t_{\rm em}$, in between which
the orientation of the cones is randomly changed to increase the
sampling of the volume with photons. The spectrum of the emitted
radiation is discretized using $N_{\nu}$ frequency bins, and each photon
packet carries photons from one of these bins. Therefore, the
number of photon packets emitted per emission time step is
$N_{\nu}\times N_{\rm EC}$.
\par
The newly emitted photons are assigned a propagation direction
parallel to the central axis of the associated emission cone, and,
together with any other photon in the simulation box,
are then propagated to the downstream neighbors of the SPH
particle at which they reside. A particle is a downstream neighbor if
it is among the $\tilde{N}_{\rm ngb}$ neighboring gas particles and
resides in the regular transmission cone centred on the
propagation direction and subtending a solid angle of $4 \pi/N_{\rm
  TC}$. The parameter $N_{\rm TC}$ defines the angular
resolution of the RT. If there is no downstream neighbor inside a
transmission cone, a virtual particle is created to which the photons
are then propagated. The transmission cones confine the propagation of
photons to the solid angle in which they were originally emitted. The
transport of photons occurs at the user-specified speed
$\tilde{c}$ and is discretized using RT time steps $\Delta t_{\rm r}$.
\par
A given SPH particle may receive several photon packets within the
same RT time step $\Delta t_{\rm r}$. These photon packets are collected
according to their propagation directions using a set of $N_{\rm RC}$
tessellating reception cones. Photon packets whose propagation
directions fall in the same reception cone are merged and replaced by
a single new photon packet. Each reception cone subtends a solid angle
$4\pi / N_{\rm RC}$. Hence, the parameter $N_{\rm RC}$ determines
the angular resolution of the merging. In the absence of virtual
particles, the merging strictly limits the maximum number of photon packets in
the simulation box to $N_{\rm RC} \times N_{\nu} \times N_{\rm SPH}$,
where $N_{\rm SPH}$ is the number of SPH particles, and renders the
computation time independent of the number of sources. Note that the
angular resolution at which photon packets are merged may be chosen 
independently of the angular resolution at which photon packets are transferred.
\par
Photons are absorbed as they propagate through the gas from SPH particles to
their neighbors depending on the optical depth between the two neighboring
particles, respecting photon conservation (\citealp{Abel1999a};
\citealp{Mellema2006}).  The absorption of photons within each frequency bin
is treated in the grey approximation using photoionization cross
sections\footnote{While TRAPHIC can treat ionization of helium, in this work
  we set X = 1, as discussed in Section~\ref{sec:chemistry}.}
\begin{equation}
\langle \sigma_{\rm HI} \rangle_{\nu} \equiv \int_{\nu_{\rm l}}^{\nu_{\rm h}} {d\nu \frac{4 \pi J_{\nu}(\nu)}{h_{\rm P}\nu} 
						  \sigma_{\rm HI}(\nu)} 
						  \times
						   \left[ \int_{\nu_{\rm l}}^{\nu_{\rm h}} 
						  {d\nu \frac{4 \pi J_{\nu}(\nu)}{h_{\rm P} \nu} }  \right]^{-1},
\end{equation}
where $J_{\nu}(\nu)$ is the spectrum, and $\nu_{\rm l}$ and $\nu_{\rm h}$ are
the low and high energy limits of frequency bin $\nu$. The number of absorbed
photons determines the photoionization rate $\Gamma_{\gamma \rm HI, \nu}$ of
HI in the given frequency bin $\nu$ defined by \citep[e.g.,][]{Osterbrock2006}
\begin{equation}
\Gamma_{\gamma \rm HI, \nu} = \langle \sigma_{\rm HI} \rangle_{\nu} \int_{\nu_{\rm l}}^{\nu_{\rm h}} {d\nu \frac{4 \pi J_{\nu} (\nu)}{h_{\rm P} \nu} }.
\end{equation}
The photoionization rate implies a photoheating rate given by 
$\mathcal{E}_{\gamma \rm HI, \nu} = \langle \varepsilon_{\rm HI} \rangle_{\nu} \Gamma_{\gamma \rm HI, \nu}$, where
\begin{eqnarray}
\langle \varepsilon _{\rm HI} \rangle_{\nu} &=& \left[  \int_{\nu_{\rm l}}^{\nu_{\rm h}} {d\nu \frac{4 \pi J_{\nu} (\nu)}{h_{\rm P} \nu} 
						   \sigma_{\rm HI} (\nu) (h_{\rm P}\nu - h_{\rm P}\nu_{\rm HI})} \right] \nonumber \\
						  &\times& 
						   \left[ \int_{\nu_{\rm l}}^{\nu_{\rm h}} {d\nu \frac{4 \pi J_{\nu}(\nu)}{h_{\rm P} \nu} }  
						   \sigma_{\rm HI} (\nu) \right]^{-1}						  
\end{eqnarray}
is the grey excess energy of frequency bin $\nu$, and $h\nu_{\rm HI} = 13.6
\eV$ is the photoionization threshold energy of HI. The photoionization and
photoheating rates are then passed to a chemistry solver to update the HI
fraction and the temperature of the gas.

\subsubsection{Numerical parameters}
In the simulations presented in this work, we set $\tilde{N}_{\rm ngb} = 32$,
and choose an angular resolution of the transport of $N_{\rm TC} = 8$ and of
the merging of $N_{\rm RC} = 8$.  Sources emit photons into $N_{\rm EC} = 8$
directions using emission time steps $\Delta t_{\rm em} = \min (10^{-1}\Myr,
\Delta t_{\rm r})$. Photons are transported at a speed $\tilde{c} = c$, where
$c$ is the speed of light, using time steps of size $\Delta t_{r} = \min
(10^{-1}\Myr, \Delta t_{\rm hydro})$, where $\Delta t_{\rm hydro}$ is the
smallest GADGET particle time step. We use a single frequency bin with
bounding energies located at $[13.6, \infty] \eV$. We use the fits to the
frequency-dependent photoionization cross sections by \cite{Verner1996} and
adopt a black body spectral shape of temperature $T_{\rm bb} = 5\times 10^4
\K$, consistent with the parameters of the adopted population synthesis models, 
to compute the grey photoionization and photoheating rate associated with
that bin. This gives $\langle \sigma_{\rm HI} \rangle =
2.93\times10^{-18}\cmsqi$ and $\langle \varepsilon_{\rm HI} \rangle = 3.65
\eV$. The RT assumes periodic boundary conditions, i.e., photons leaving the
box through any given face are reinserted at the periodically opposing
face. The dependence of the RT on the parameters has been discussed in
\cite{Pawlik2008} and \cite{Pawlik2011}, to which we refer the
reader. Additional tests can be found in the appendices of \cite{Pawlik2013}
and \cite{Rahmati12013}. For simplicity, we treat recombination radiation in
the on-the-spot approximation, i.e., we compute recombination rates using Case
B (e.g., \citealp{Raicevic2014}; \citealp{Tanaka2014}).
\par

\subsection{Energy injection by core-collapse supernovae}
\label{Sec:SN}
The feedback from the explosion of stars as core-collapse SNe is modeled as an
injection of thermal energy into the ISM surrounding the star
particles\footnote{We refer to this thermal feedback as SN feedback for
  simplicity of presentation. However, our implementation is agnostic to the
  physical processes that inject the energy, and therefore may be interpreted
  as collectively describing all promptly acting feedback processes associated
  with the formation of stars that are not explicitly treated otherwise. Such
  processes include stellar winds as well as radiation pressure (for a
  discussion see, e.g., \citealp{Schaye2014}).}. Our numerical implementation,
which has been designed to control spurious radiative energy losses due to the
limited numerical resolution, is described in
\cite{DallaVecchia2012}. Core-collapse SNe inject thermal energy after a delay
of $30 \Myr$ after the birth of the star particle, approximately corresponding
to the maximum lifetime of stars that end their lives as core-collapse SNe.
For each core-collapse SN that occurs, $f_{\rm SN} \times 10^{51} \erg$ of
thermal energy is injected. The energy is distributed stochastically to an
average subset of $1.34$ of the 48 neighboring SPH particles, instantaneously
increasing their gas temperature by $\Delta T = 10^{7.5} \K$ and assuming a
mean atomic weight $\mu = 0.6$ appropriate for fully ionized gas. We set
$f_{\rm SN} = 1$, which results in a good agreement of the star formation
history in our reference simulation and observational constraints at $z > 6$
(see Figure~\ref{fig4}).
\par
\par
\subsection{Identification of galaxies}
We post-process our simulations to extract haloes using the friends-of-friends
(FOF) halo finder, with linking length equal to 1/5th of the mean DM
inter-particle distance, built into the substructure finder Subfind
(\citealp{Springel2001}; \citealp{Dolag2009}). For each FOF halo, we use
Subfind to identify the particle for which the gravitational potential is
minimum, and let its location mark the halo center. We obtain the virial
radius as the radius of the sphere centered on the halo center within which
the average matter density equals 200 times the redshift-dependent critical
density of the universe. The total mass inside this sphere is the virial mass
of the halo. The total SFR of the halo is the sum of the SFRs of the gas
particles it contains. We will also compute the baryon fraction of the halo,
which is the ratio of the total mass in gas and stars inside the virial radius
and the virial mass.
\section{Results}
\label{Sec:Results}
In Section~\ref{Sec:Reionization} we provide an overview of the simulations
listed in Table~\ref{tab:sims}. Thereafter, in Section~\ref{Sec:Feedback}, we
discuss how stellar feedback from SNe and photoheating impact the formation
and evolution of galaxies and the reionization of the gas. We will make use of
additional simulations identical to those in Table~\ref{tab:sims}, except that
we have turned off SNe and/or photoheating to help isolate the impact of the
feedback these processes provide. Most observational works on UV galaxies
assume a Salpeter IMF to infer SFRs, while our simulations assume a Chabrier
IMF. Where necessary, we have divided observationally inferred SFRs by a
factor of 1.7 to account for the difference between the two IMFs (see
Section~\ref{Sec:sf}). Note that the high-resolution (high-res) simulation
L12N512 has been stopped at $z = 6.8$ due to its large computational expense.
\par
\par

\begin{figure*}
  \begin{center}
    \includegraphics[width=0.45\textwidth,clip=true, trim=10 0 10 0,
      keepaspectratio=true]{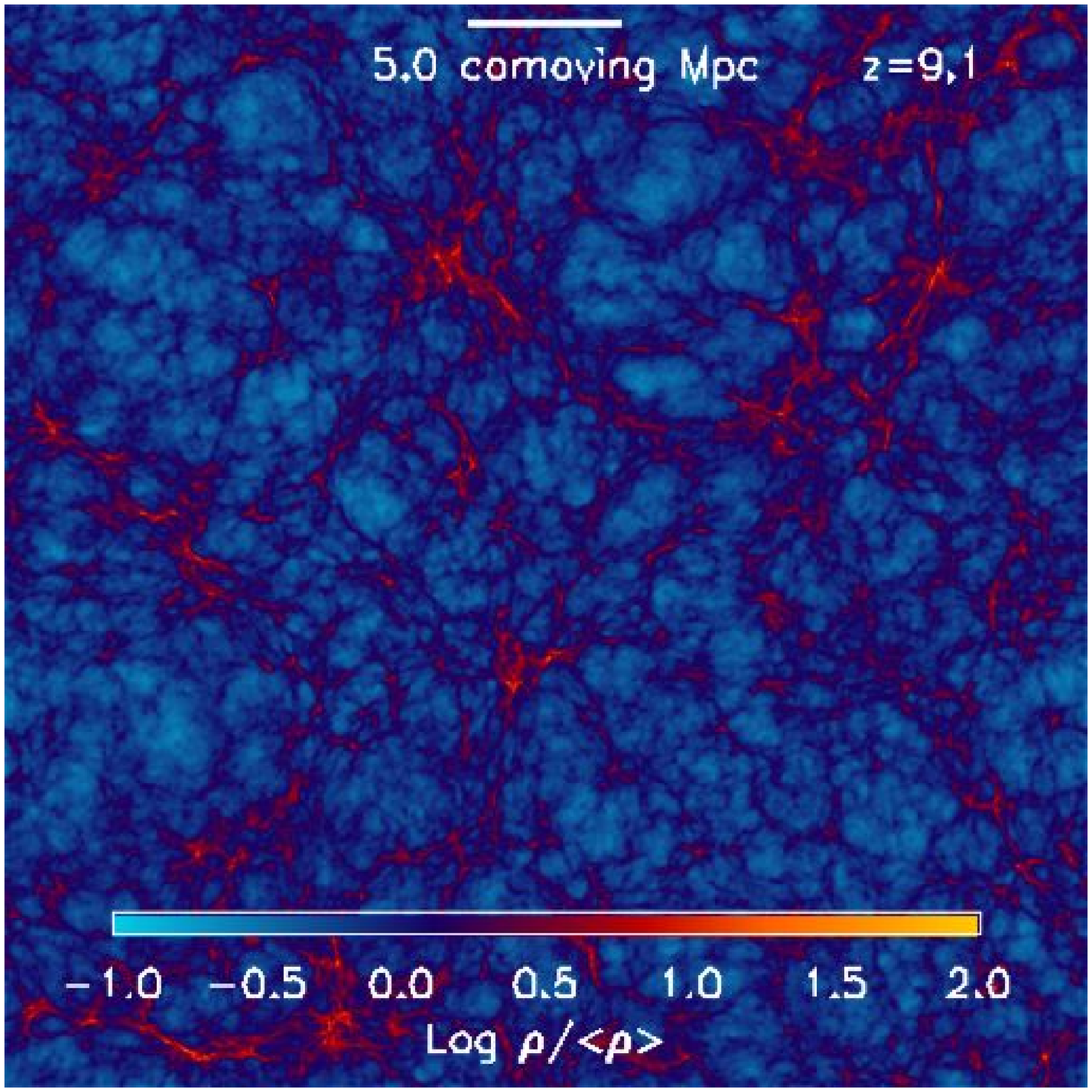}
    \includegraphics[width=0.45\textwidth,clip=true, trim=10 0 10 0,
      keepaspectratio=true]{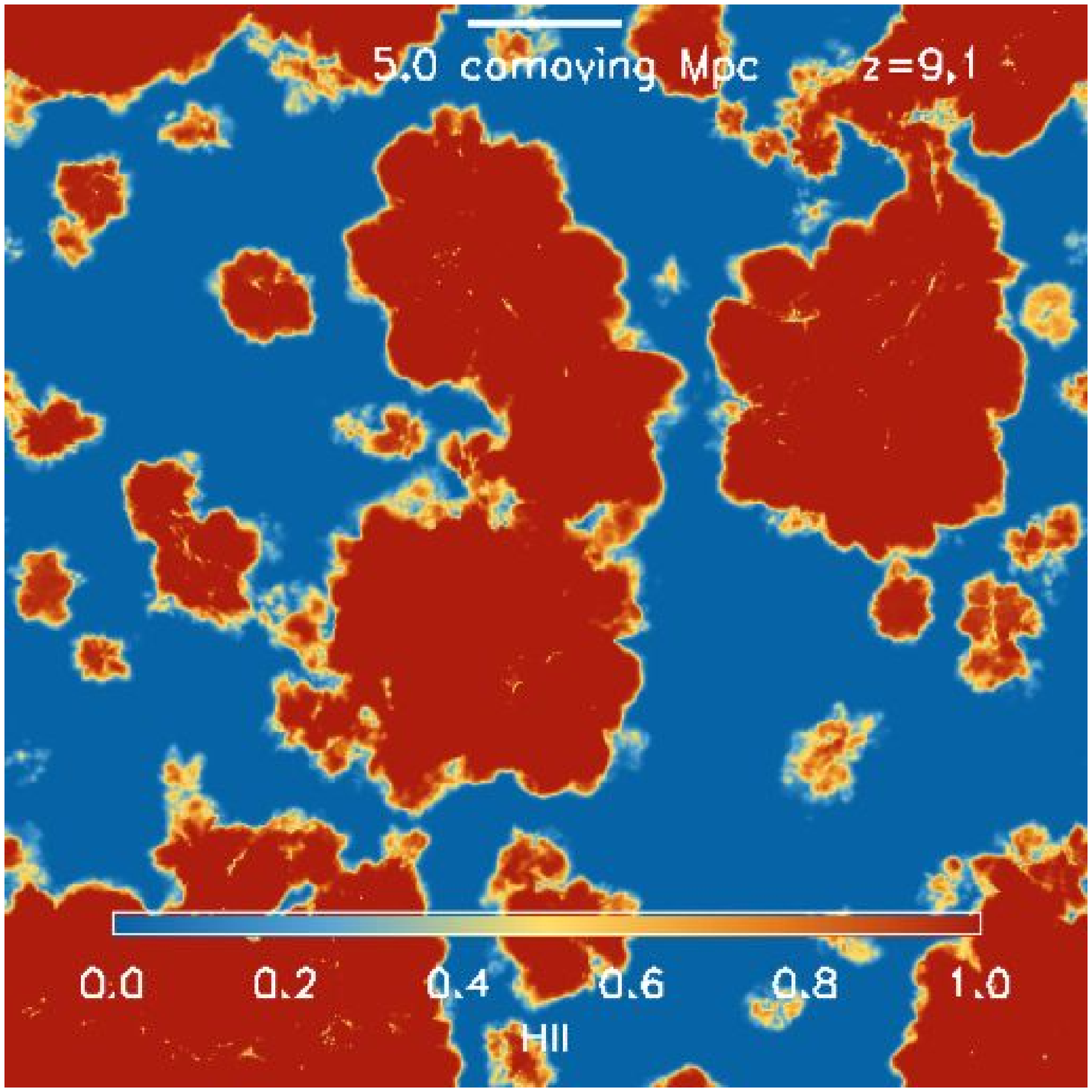}\\
 \includegraphics[width=0.45\textwidth,clip=true, trim=10 0 10 0,
      keepaspectratio=true]{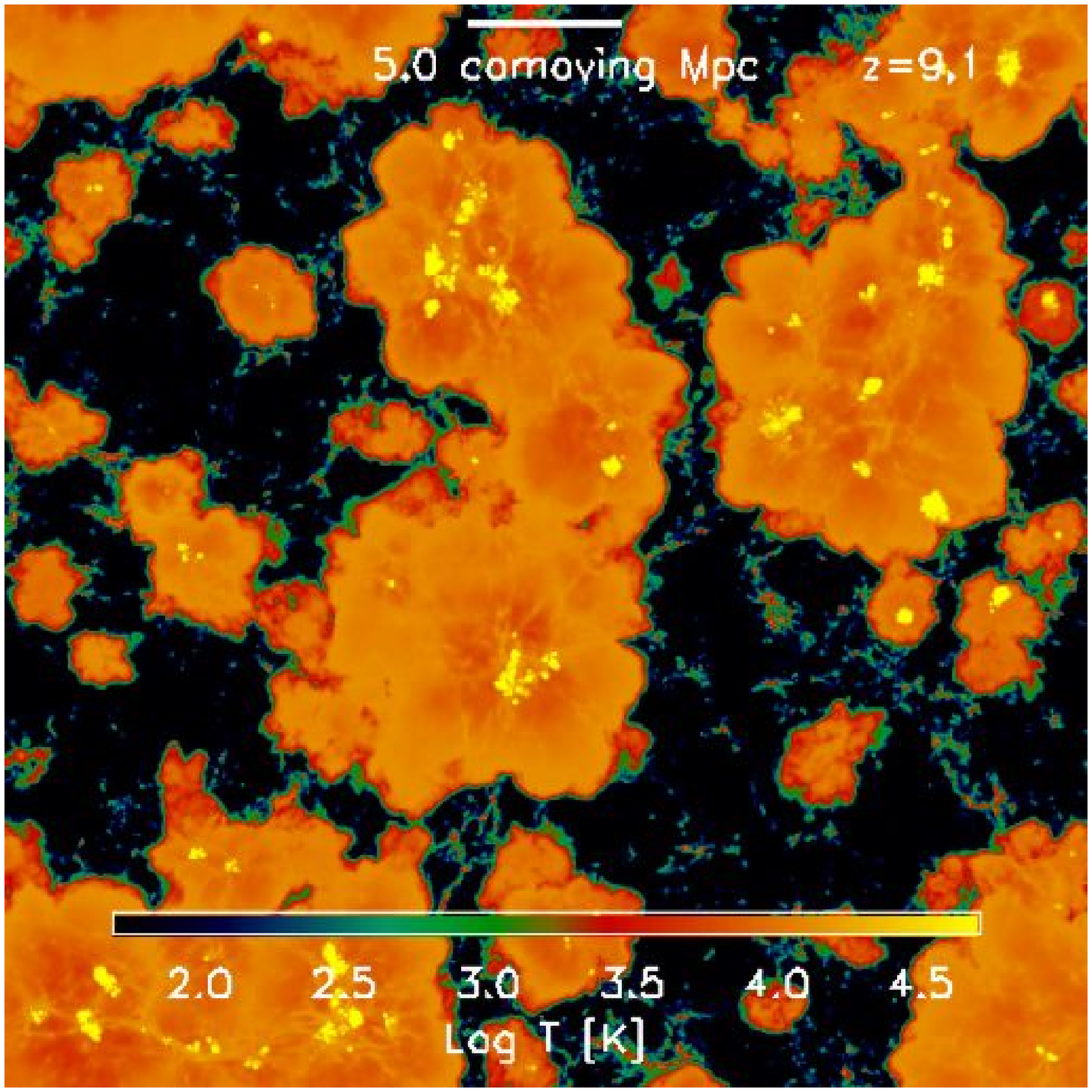}
    \includegraphics[width=0.45\textwidth,clip=true, trim=10  0 10 0,
      keepaspectratio=true]{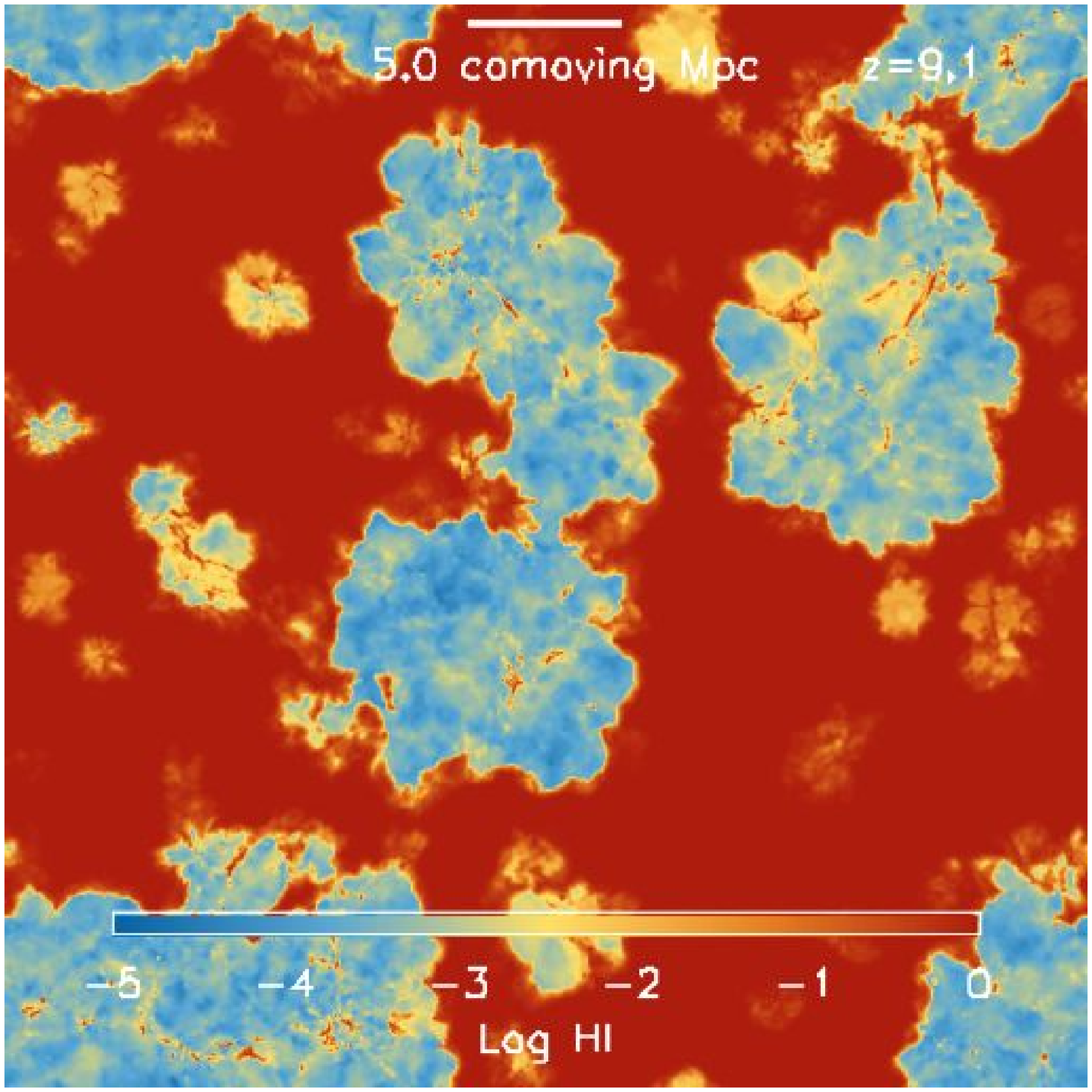}

    \end{center}
  \caption{Images of the gas overdensity, the ionized fraction, the
    temperature, and the neutral fraction (left to right, top to
    bottom) in the reference simulation L25N512 at $z \approx 9$.
    Each image is an average in a central slice of linear extent $25
    \cMpch$ and thickness $2.5 \cMpch$ cut perpendicularly to one of the
    axes of the simulation box. Reionization is already well underway (the 
    volume-averaged ionized fraction is $x^{\rm v}_{\rm HII} \approx 0.4$)
    and has heated the gas in ionized regions to temperatures $\sim 2
    \times 10^4 \K$. The gas near galaxies may reach temperatures up
    to $\sim 10^7 \K$ as it is heated by SNe and structure formation
    shocks. Our spatially adaptive RT simulations reveal a network of
    self-shielded neutral clumps and filaments inside the ionized
    regions.} \label{fig1}
\end{figure*}

\begin{figure*}
  \begin{center}
    \includegraphics[width=0.49\textwidth,clip=true, trim=0 40 20 0,
      keepaspectratio=true]{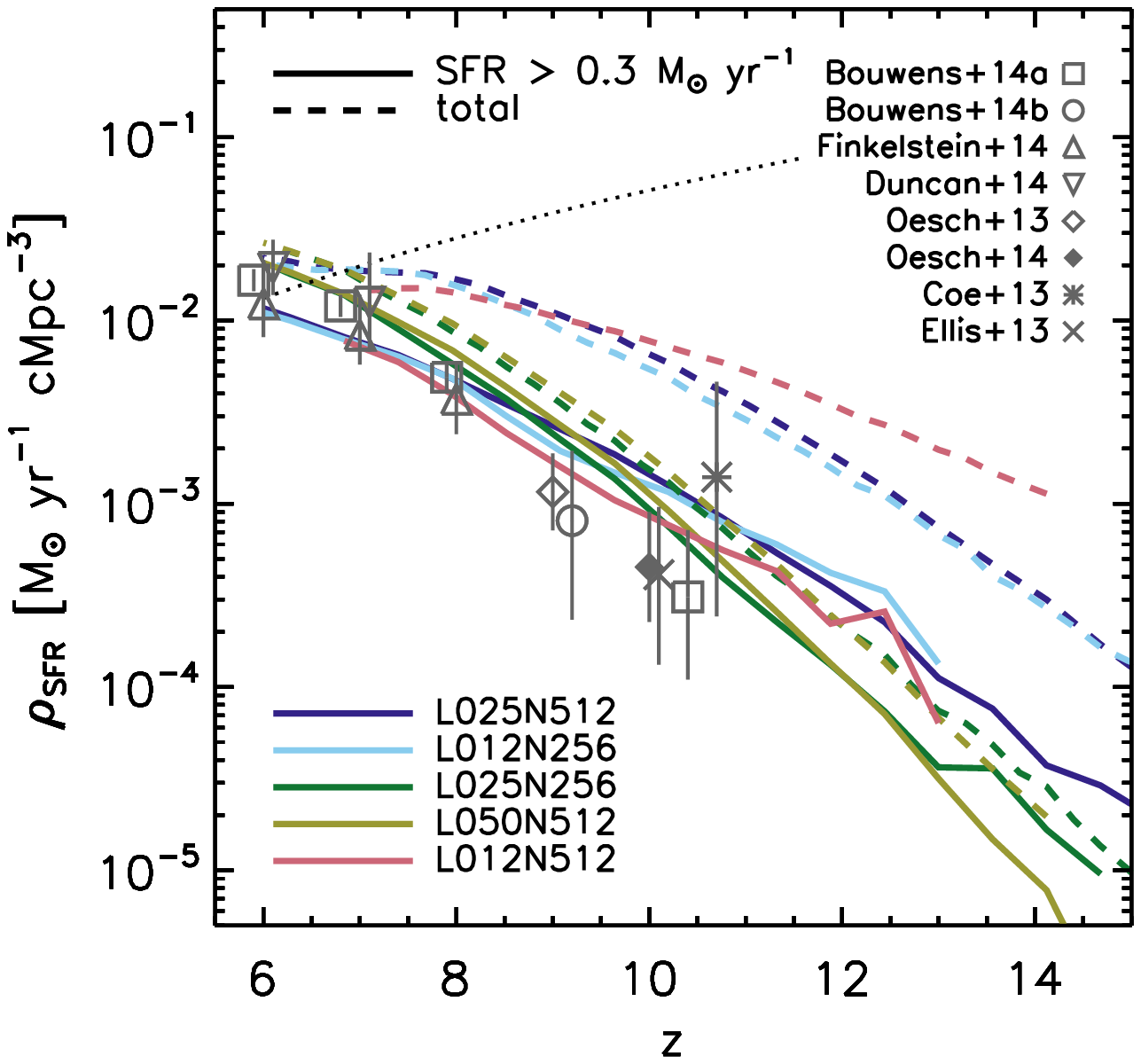}
    \includegraphics[width=0.49\textwidth,clip=true, trim=0 40 20 0,
      keepaspectratio=true]{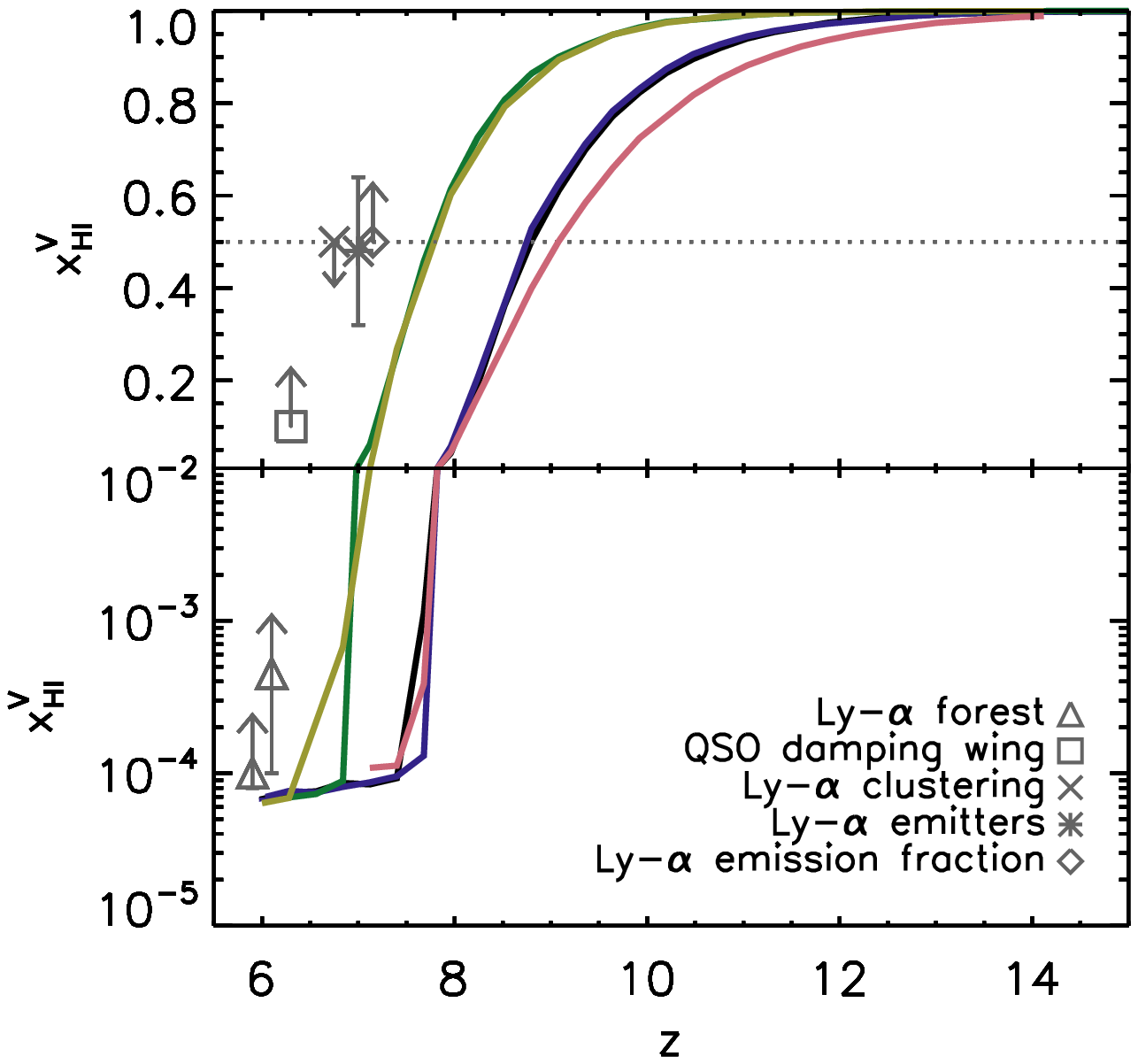} \\
    \includegraphics[width=0.49\textwidth,clip=true, trim=0 40 20 0,
      keepaspectratio=true]{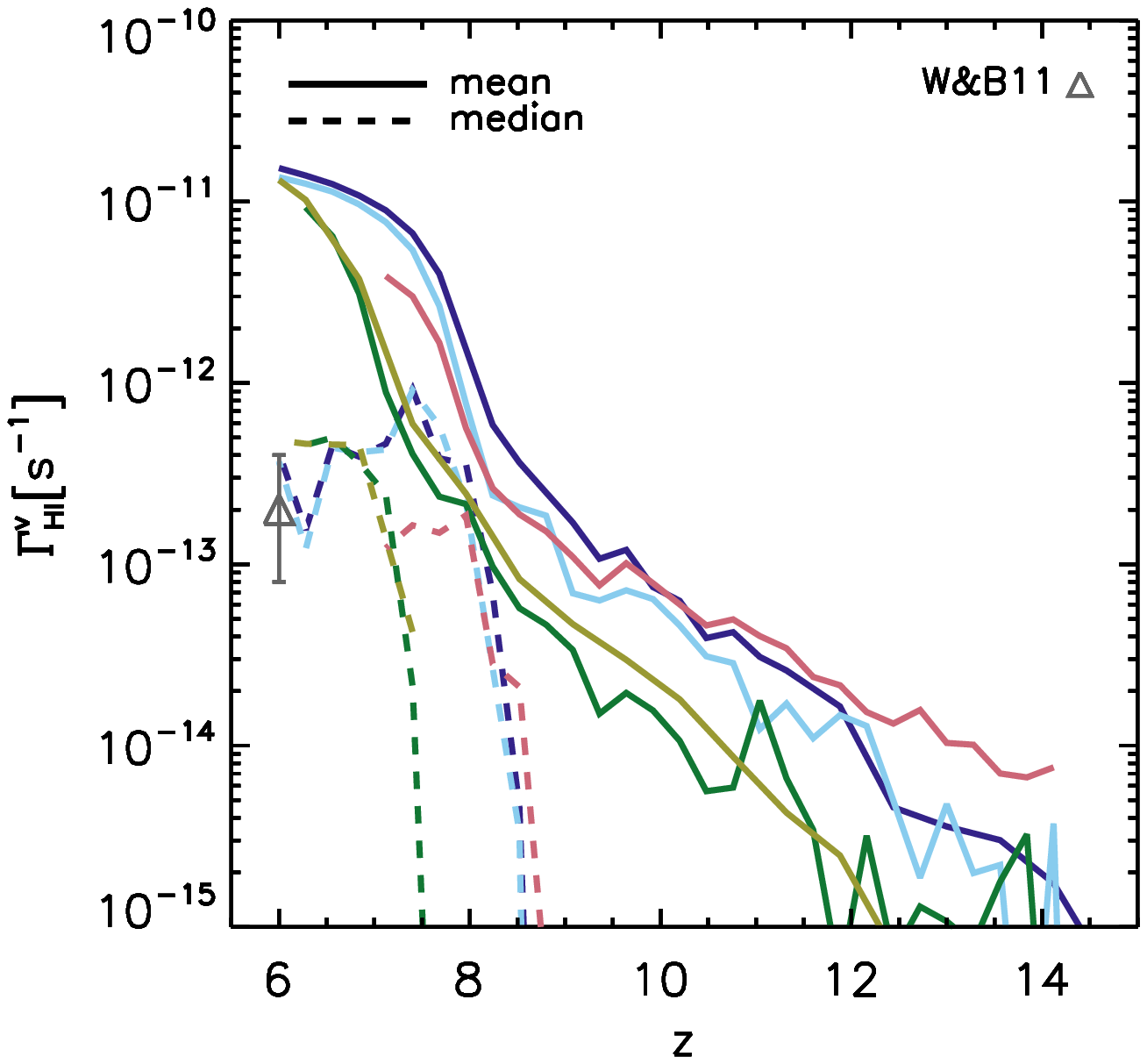}
    \includegraphics[width=0.49\textwidth,clip=true, trim=0 40 20 0,
      keepaspectratio=true]{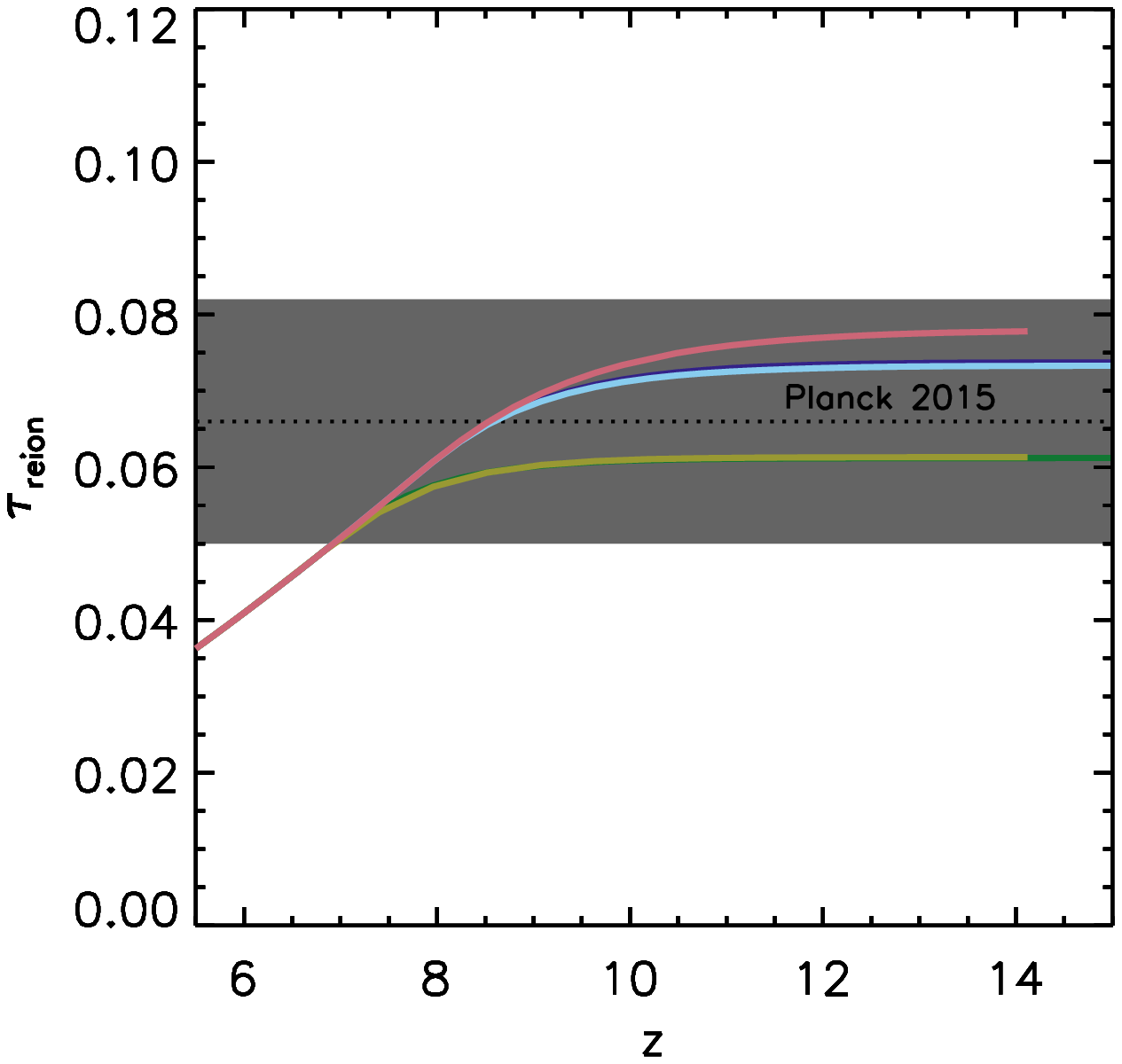}
  \end{center}
  \caption{Effect of box size and resolution on the star formation and
    reionization histories. Curves of different colors correspond to different
    simulations as indicated in the legend. {\it Top left:} comoving SFR
    density. Dashed and solid curves show the total SFR density and the SFR
    density including only the contribution from galaxies with SFRs greater
    than $0.3\Msun\invyr$, the current detection limit (assuming a Chabrier
    IMF, corresponding to a limiting UV magnitude $M_{\rm AB} \gtrsim -17$, see
    Figure~\ref{fig4}). The dotted curve shows the critical SFR
    (Eq.~\ref{Eq:sfrcrit}), assuming $C_{\rm IGM}/f_{\rm esc} = 25$.
    Observational data is taken from \protect\cite{Bouwens2014},
    \protect\cite{Finkelstein2014}, \protect\cite{Bouwens2014b},
    \protect\citet[their SED models]{Duncan2014}, \protect\cite{Oesch2013},
    \protect\cite{Oesch2014}, \protect\cite{Coe2013}, and
    \protect\citet[revised down by a factor of two following
      \citealp{Oesch2013}]{Ellis2013}, as indicated in the legend, and divided
    by 1.7 if necessary to convert from the Salpeter IMF to the Chabrier IMF
    used here. When necessary, we have corrected the published SFR densities
    upwards to account for contribution from galaxies down to the current
    detection threshold, following \protect\cite{Bouwens2014}. {\it Top right:}
    volume-weighted mean neutral hydrogen fraction. The horizontal dotted line
    marks a neutral fraction $x_{\rm HI}^{\rm v} = 0.5$ to guide the eye. The
    (model-dependent) constraints on the neutral fraction are taken from
    \protect\citet[QSO damping wings]{Schroeder2013},
    \protect\citet[Ly-$\alpha$ clustering]{Ouchi2010},
    \protect\citet[Ly-$\alpha$ emitters]{Ota2008}, \protect\citet[Ly-$\alpha$
      emission fraction]{Dijkstra2011} and \protect\citet[Ly-$\alpha$
      forest]{Fan2006}, following the discussion of Figure~5 in
    \protect\cite{Robertson2013}. {\it Bottom left:} mean (solid) and median
    (dashed) volume-weighted hydrogen photoionization rate. The triangle with
    error bars marks the observational constraint from
    \protect\cite{Wyithe2011}. {\it Bottom right:} electron scattering optical
    depth towards reionization. The horizontal dotted line shows the latest
    estimate from observations by the Planck satellite, and the
    grey band indicates the associated 1-sigma error interval.} 
\label{fig2}
\end{figure*}

\subsection{Overview of simulations}
\label{Sec:Reionization}
Figure~\ref{fig1} shows images of the gas densities, temperatures and neutral
and ionized hydrogen fractions in our reference simulation L25N512 at $z
\approx 9$. By this redshift, reionization has already proceeded
significantly, and individual reionized regions are just about to percolate
the simulation box.
\par
Most of the gas inside the ionized regions is photoheated to about $\sim
2\times 10^4 \K$, the characteristic temperature of stellar HII regions.  Near
galaxies, gas may be heated to much higher temperatures in gas-dynamical
shocks that accompany the accretion of gas in massive haloes and the explosion
of stars as SNe.  Some of the gas in the ionized regions is sufficiently dense
to shield from the ionizing radiation and remains neutral. The high resolution
enabled by our spatially adaptive RT technique is key to tracking this gas
phase. For comparison, a RT simulation on a uniform grid with a number of grid
cells equal to the number of SPH particles employed here would have limited
our ability to resolve the structure of the ionized gas to scales above
$\gtrsim 70 \kpc$ comoving. This is larger than the typical scale of
self-shielded (Lyman-limit) systems that are thought to be among the main
consumers of ionizing photons (e.g., \citealp{Schaye2001};
\citealp{Furlanetto2005}; \citealp{Gnedin2006}). Matching our spatially
adaptive resolution using a uniform grid would require $512 \times 25 = 12800$
grid points per dimension, which is beyond current computational capabilities.
\par
Figure~\ref{fig2} compares a set of basic observables extracted from
our simulations with current observational constraints. The 
SFR density of currently observationally accessible galaxies with SFRs 
$\ge 0.3 \Msun \invyr$ (assuming a Chabrier IMF, see Figure~\ref{fig4}), 
is insensitive to changes in box size and resolution and in good agreement with
observations. The comparison with the total SFR density suggests that
current observations uncover only a fraction of stars forming at $z
\gtrsim 6$. 
\par
There is a significant dependence of the total SFR on
resolution. Because density fluctuations are increased at higher
resolution, the reference simulation L25N512 yields initially a higher
total SFR than the low-resolution (low-res) simulation L25N256, and a lower total SFR
than the high-res simulation L12N512.  On the other hand, near the final
simulation redshift at $z = 6$, the total SFR is slightly smaller in the
reference simulation than in the low-res simulation, and slightly
larger than in the high-res simulation. This inversion in
the dependence on resolution is caused by the stronger suppression of star
formation by stellar feedback at higher resolution.
\par
The top right panel of Figure~\ref{fig2} shows that the reionization history,
i.e., the evolution of the neutral hydrogen fraction, is insensitive to the
size of the simulation box $L \gtrsim 12.5 \Mpch$. This rapid convergence may
partly result from our use of identical amplitudes and phases that
characterize the initial Gaussian random density fluctuations from which our
simulations start. Indeed, a larger variance in the reionization histories on
these scales is seen by comparing correspondingly sized sub-volumes of
larger-scale simulations (\citealp{Iliev2006}; \citealp{Iliev2014};
\citealp{Gnedin2014}). However, this variance is caused primarily by the large
range of environments sampled by the subvolumes, each of which may have a
different mean density (\citealp{Iliev2014}).  The top right panel of
Figure~\ref{fig2} shows further that there is a significant dependence of the
reionization history on resolution.  At higher resolution, reionization ends
at higher redshifts. This mostly reflects the increased SFRs at increased
resolution that our simulations show during the early and middle phases of
reionization.
\par
In our simulations, reionization completes at higher redshifts than suggested
by constraints on the neutral fraction from observations of Lyman-$\alpha$
emitting galaxies and high-redshift quasars (e.g., Figure~5 in
\citealp{Robertson2013}). However, interpretations of these observations are
dominated by large systematic uncertainties (for a discussion see, e.g.,
\citealp{Robertson2013}; \citealp{Jensen2013}; \citealp{Dijkstra2014};
\citealp{Mesinger2014}; \citealp{Taylor2014}). At $z = 6$, the simulated mean
neutral fractions are in reasonable agreement with observational constraints
from the Lyman-$\alpha$ forest (\citealp{Fan2006}), which are thought to be
relatively robust (e.g., \citealp{Fan2006}; but see, e.g.,
\citealp{McGreer2011}; \citealp{Gnedin2014};
\citealp{Becker2014}). Nevertheless, it is important to keep in mind that our
simulations may underestimate the duration and overestimate the redshift of
reionization because the limited size of the simulated volume may bias the
reionization histories (e.g., \citealp{Iliev2014}; \citealp{Gnedin2014}), and
because of our choice of a large sub-resolution escape fraction $f_{\rm
  esc}^{\rm subres} = 1.0$.
\par
A useful check for consistency between the total SFR densities and
the reionization histories is provided by comparison with the critical
SFR density $\rho_{\rm SFR}^{\rm c}$ needed to keep the IGM ionized
(\citealp{MHR1999}; updated as in \citealp{Pawlikclump2009}),
\begin{eqnarray}
\rho_{\rm SFR}^{\rm c} &\approx& 0.013 \Msun \invyr \Mpc^{-3} \nonumber
\\ &\times& \left(\frac{C_{\rm IGM}}{5}\right) \left( \frac{f_{\rm
    esc}}{0.2}\right)^{-1} \left(\frac{1+z}{7}\right)^3, \label{Eq:sfrcrit}
\end{eqnarray}
where $f_{\rm esc}$ is the fraction of ionizing photons that, on average,
escape galaxies to ionize the IGM (e.g., \citealp{Razoumov2006};
\citealp{Gnedin2008}; \citealp{Wise2009}; \citealp{Paardekooper2013};
\citealp{Yajima2011}), and we have assumed a Chabrier IMF, which implies an
ionizing emissivity per unit stellar mass larger, and therefore a critical SFR
density smaller, by a factor of about $1.7$ compared to that implied by a
Salpeter IMF (see Section~\ref{Sec:sf}).  An IGM clumping factor $C_{\rm IGM}
= 5$ is a conservative estimate from our simulations (see Figure~\ref{fig8}),
and consistent with earlier works (\citealp{Pawlikclump2009}; see the summary
of results in \citealp{Finlatorclump2012}). 
\par
In this case, the simulated
star-forming galaxies can keep the gas at $z \lesssim 7$ ionized provided $f_{\rm esc} \gtrsim
0.3$. Whether such a relatively large escape fraction is physically plausible
is a matter of active research (see, e.g., the recent overview in
\citealp{Benson2013}). Computing the escape fraction from our simulations
would require additional RT computations beyond the current work. Observations
currently probe SFRs $\gtrsim 0.3 \Msun \invyr$ (assuming a Chabrier IMF;
e.g., \citealp{Bouwens2014}, \citealp{Finkelstein2014}; \citealp{Duncan2014};
Figure~\ref{fig4}).  Under the above assumptions, at $z = 6$, the observed
population of galaxies alone is capable of keeping the Universe ionized,
lending support to scenarios in which reionization is driven by star-forming
galaxies (e.g., \citealp{Pawlikclump2009}; \citealp{Finkelstein2012};
\citealp{Cai2014}).
\par
The simulations predict a strong increase in the volume-averaged mean and
median hydrogen photoionization rates as the simulation volumes are
reionized (bottom left panel of Figure~\ref{fig2}). 
This signals a rapid build-up of an ionizing background, which sets
the equilibrium neutral hydrogen abundance in the reionized IGM (e.g.,
\citealp{Gnedinreion2000}; \citealp{Aubert2010}). The median photoionization
rate is in good agreement with constraints at $z = 6$ from observations of,
e.g., the Lyman-$\alpha$ forest (\citealp{Bolton2007}) and quasar near-zones
(e.g., \citealp{Wyithe2011}). However, the mean photoionization rate exceeds the median by more
than an order of magnitude. The excess in the mean photoionization rate with
respect to the observational constraints is 
similar to that seen in previous simulations of reionization (e.g.,
\citealp{Aubert2010}; \citealp{Finlator2011}).
\par
The optical depth towards reionization, $\tau_{\rm reion} = \int^{z_{\rm
    start}}_{z = 0} n_{\rm e} c dt$, where $n_{\rm e}$ is the number density
of free electrons and $z_{\rm start} \gtrsim 30$ is the redshift at which
reionization begins, is an important integral constraint on the reionization
history (e.g., \citealp{Alvarez2006}; \citealp{Shull2008}; \citealp{Ahn2012}).
For the computation of the optical depth, we assume a hydrogen mass fraction
$X = 0.75$, and that the fraction of ionized hydrogen is equal to that in our
simulations, which assume $X = 1$.  Furthermore, we make the standard
assumptions that hydrogen is fully ionized at $z < 6$, that helium is neutral
(singly ionized) whenever hydrogen is neutral (ionized) at $z \ge 3$, and that
helium is doubly-ionized at $z < 3$ (e.g., \citealp{Iliev2005}).
\par
The bottom right panel of Figure~\ref{fig2} shows that our simulations yield
optical depths consistent with the most recent constraints derived from
observations of the cosmic microwave background, $\tau_{\rm reion} = 0.066 \pm
0.016$ (\citealp{Planck2015}). Our simulations
lack the resolution and the physics to follow the formation of stars in
minihaloes, i.e., haloes with virial temperatures below $10^4 \K$. Ionization by
these stars is expected to increase the optical depth by $\lesssim 0.03$
(e.g., \citealp{Shull2008}; \citealp{Ahn2012}; \citealp{Wise2014}).  If these
stars evolve into accreting black holes, the optical depth may be further
increased due to the pre-ionization of the IGM by X-rays (e.g.,
\citealp{Ricotti2004}). If the additional contribution to ionization from
minihaloes and X-ray sources is indeed that high, and if we were
to add this contribution to our estimates of the optical depth,
then this would yield values in slight excess of the
Planck observational estimates. However, consistency with observations 
may still be achieved by lowering the ISM escape fraction of ionizing photons, which
is a free parameter in our simulations. In this case, 
reionization in our simulations would occur later and yield
a lower optical depth (see Figure~\ref{fig9}).
\begin{figure}
  \begin{center}
    \includegraphics[width=0.49\textwidth,clip=true, trim=0 10 10 20,
      keepaspectratio=true]{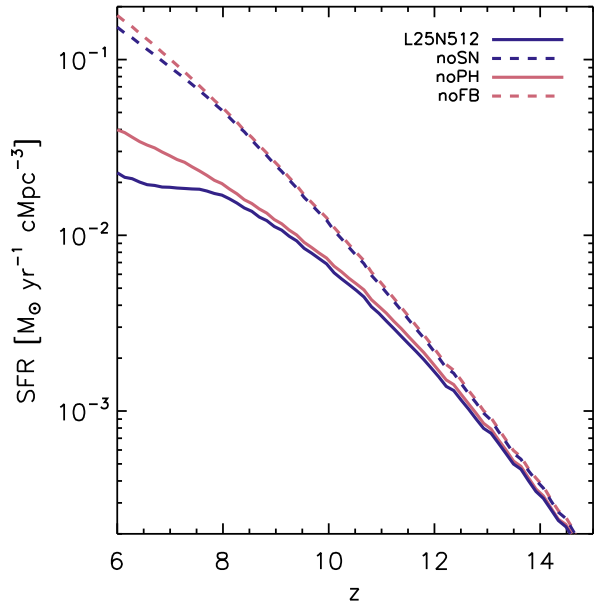}
  \end{center}
  \caption{Effect of stellar feedback on the SFR history in the reference simulation 
    L25N512. Shown are SFR densities in the simulations without feedback (noFB; red dashed), 
  with only SN feedback (noPH; red solid), with only photoheating (noSN; blue dashed), and with both SN feedback and photoheating
  (L25N512; blue solid). SNe strongly reduce the cosmic SFR. On
  the other hand, photoheating has a comparatively small impact on the cosmic
  SFR, and this impact is nearly negligible in the absence of SNe. }
  \label{fig3}
\end{figure}

\par
Current cosmological simulations of reionization often have difficulties in
simultaneously matching observational constraints on the neutral fraction, the
photoionization rate and the optical depth (e.g., \citealp{Aubert2010};
\citealp{Finlator2011}). Figure~\ref{fig2} shows that our simulations are no
exception. Many works have argued that these
difficulties may signal a significant role by physical processes that are
currently not understood or unknown, and that are currently only poorly
modelled or lacking altogether in cosmological simulations. Such processes
include, among others, an evolving IGM escape fraction (e.g.,
\citealp{Kuhlen2012}; \citealp{Alvarez2012}; \citealp{Mitra2013};
\citealp{Ferrara2013}) and the absorption of photons by Lyman limit systems
(e.g., \citealp{Bolton2013}; \citealp{Mesinger2014}; which our reference
simulation begins to resolve). Some works
have successfully matched the observational constraints by a careful
calibration of parameters (e.g., \citealp{Ciardi2012}). In Section~\ref{Sec:Parameters} we confirm
that calibrating the ISM escape fraction of ionizing photons helps to improve the match
with observational constraints. However, it is also important to acknowledge
that current observational constraints on the neutral fraction at $z \gtrsim
6.0$ are highly model-dependent and too weak to single out reionization
scenarios (e.g., \citealp{Dijkstra2014}; \citealp{Gnedin2014}). Investigating
these issues is beyond the aims of the current work. Here we will focus on how
feedback from photoheating and SNe impacts the formation of galaxies and the
reionization of the IGM in physically motivated scenarios of reionization.

\par
\subsection{Stellar feedback}

\par
\begin{figure}
  \begin{center}
    \includegraphics[width=0.49\textwidth,clip=true, trim=15 30 20 00,
      keepaspectratio=true]{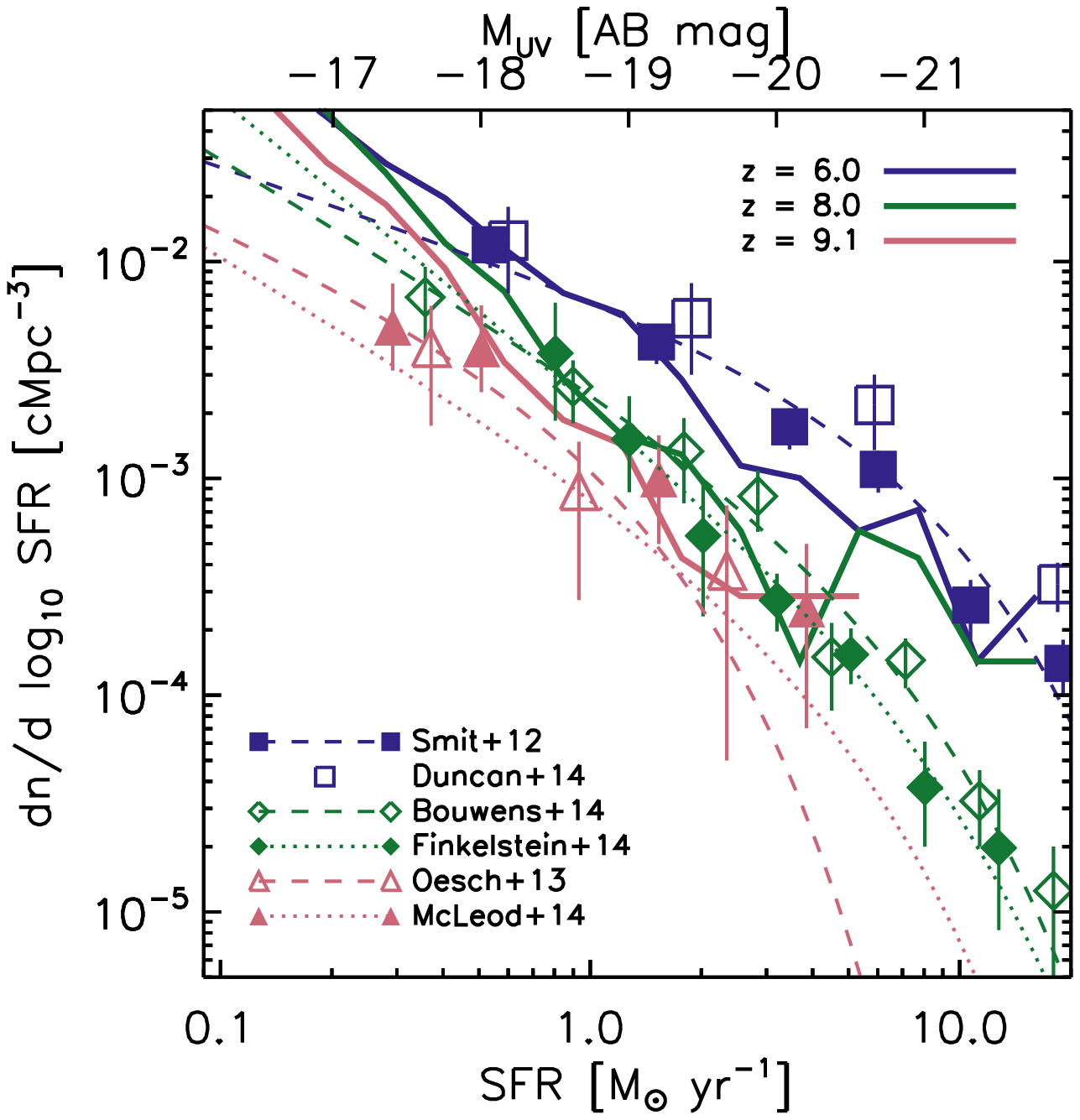}
  \end{center}
  \caption{Evolution of the UV luminosity function in the reference
    simulation L25N512 between $z = 6-9$ (solid curves). Symbols show observational constraints
    from \protect\cite{Smit2012} at $z = 6$ 
    (filled squares, corrected for dust) and \protect\citet[open
    squares, corrected for dust]{Duncan2014}, from \protect\citet[open diamonds]{Bouwens2014} and
    \protect\citet[filled diamonds]{Finkelstein2014} at $z = 8$ and from
    \protect\citet[open triangles]{Oesch2013} and 
    \protect\citet[filled triangles, their estimate assuming density evolution]{McLeod2014}
    at $z \approx 9$. The comparison at $z \approx 7$ is included in 
    Figure~\ref{fig5} for clarity. We have omitted any upper limits reported in these works. 
    Dashed and dotted curves are the corresponding
    Schechter fits, as indicated in the legend. The simulated UV luminosity function is in good agreement
    with the observational constraints across the entire range of
    redshifts.}
  \label{fig4}
\end{figure}

\label{Sec:Feedback}
In this section we investigate how stellar ionizing radiation and SNe impact
the formation and evolution of galaxies and the reionization of the IGM. We
will compare the simulations listed in Table~\ref{tab:sims}, which include
both photoheating and SNe, with simulations in which photoheating and SNe are
disabled but which are otherwise identical. We use the suffix {\it noFB} to
distinguish the latter simulations from the former. We further compare with
simulations that include SNe but not photoheating, distinguished by the suffix
{\it noPH}, and with simulations which include photoheating but not SNe,
distinguished by the suffix {\it noSN}. We focus on our reference simulation
L25N512 to illustrate the physics at play, and we assess the impact of box
size and resolution on our conclusions.
\par
\subsubsection{Impact on the cosmic SFR history}
\label{Sec:FeedbackSF}
Figure~\ref{fig3} quantifies the impact of stellar feedback on the evolution
of the total cosmic SFR density in comparisons of our reference simulation
L25N512 with simulations that differ only in the inclusion of the stellar
feedback processes. The comparison with simulation L25N512-noFB, in which both
photoionization heating and SNe were turned off, reveals the strong
suppression of star formation by stellar feedback. The comparison with
simulations L25N512-noPH, in which only photoheating is turned off, and
L25N512-noSN, in which only SNe are turned off, shows that the suppression of
the SFR due to SNe is much stronger than the suppression of the SFR due to
photoheating.
\par
SNe have a significant impact already at early times, $z \lesssim 10$, long
before the IGM in the simulation is substantially ionized.  On the other hand,
the impact of photoheating becomes noticeable only at $z \lesssim 8$, when the
average ionized fraction approaches unity. The belated impact of photoheating
on the SFR suggests a stronger role of suppression of star formation due to
the illumination of galaxies by the ionizing background than due to the
exposure to local ionizing radiation from stars inside the galaxies. However,
the impact of photoheating on the cosmic SFR remains comparatively small also
after reionization.
\par
Figure~\ref{fig3} also shows that photoheating suppresses star formation more
strongly in the simulation with SNe than in the one without SNe (compare the
difference between the two solid curves with the difference between the two
dashed curves). On the other hand, SNe suppress star formation more strongly
in the simulation with photoheating than in that without (compare the
difference between the two blue curves with the difference between the two red
curves). Photoheating and SNe thus amplify each other in suppressing the
formation of stars. These results are in good agreement with our earlier work
in \cite{Pawlik2009}, in which we first reported this effect in simulations in
which the gas was in photoionization equilibrium with a uniform ionizing
background instantaneously turned on at $z = 9$, and in which SN feedback was
implemented by kicking gas particles in winds (see also
\citealp{Finlator2011}; \citealp{Hambrick2011}; \citealp{Hopkins2013}).
\par
\par
\begin{figure*}
  \begin{center}
    \includegraphics[width=0.45\textwidth,clip=true, trim=10 20 20 0,
      keepaspectratio=true]{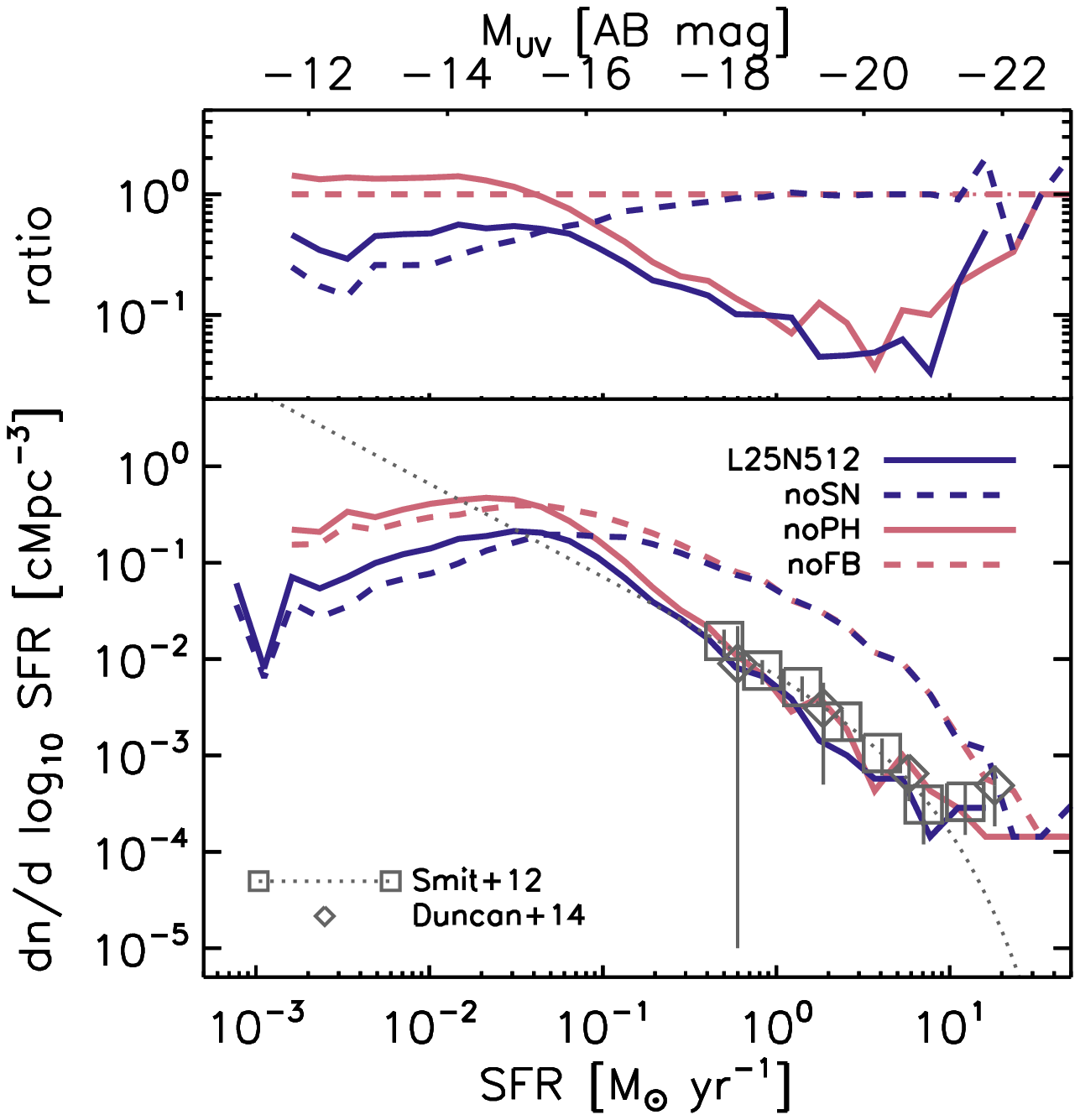}
    \includegraphics[width=0.45\textwidth,clip=true, trim=10 20 20 0,
      keepaspectratio=true]{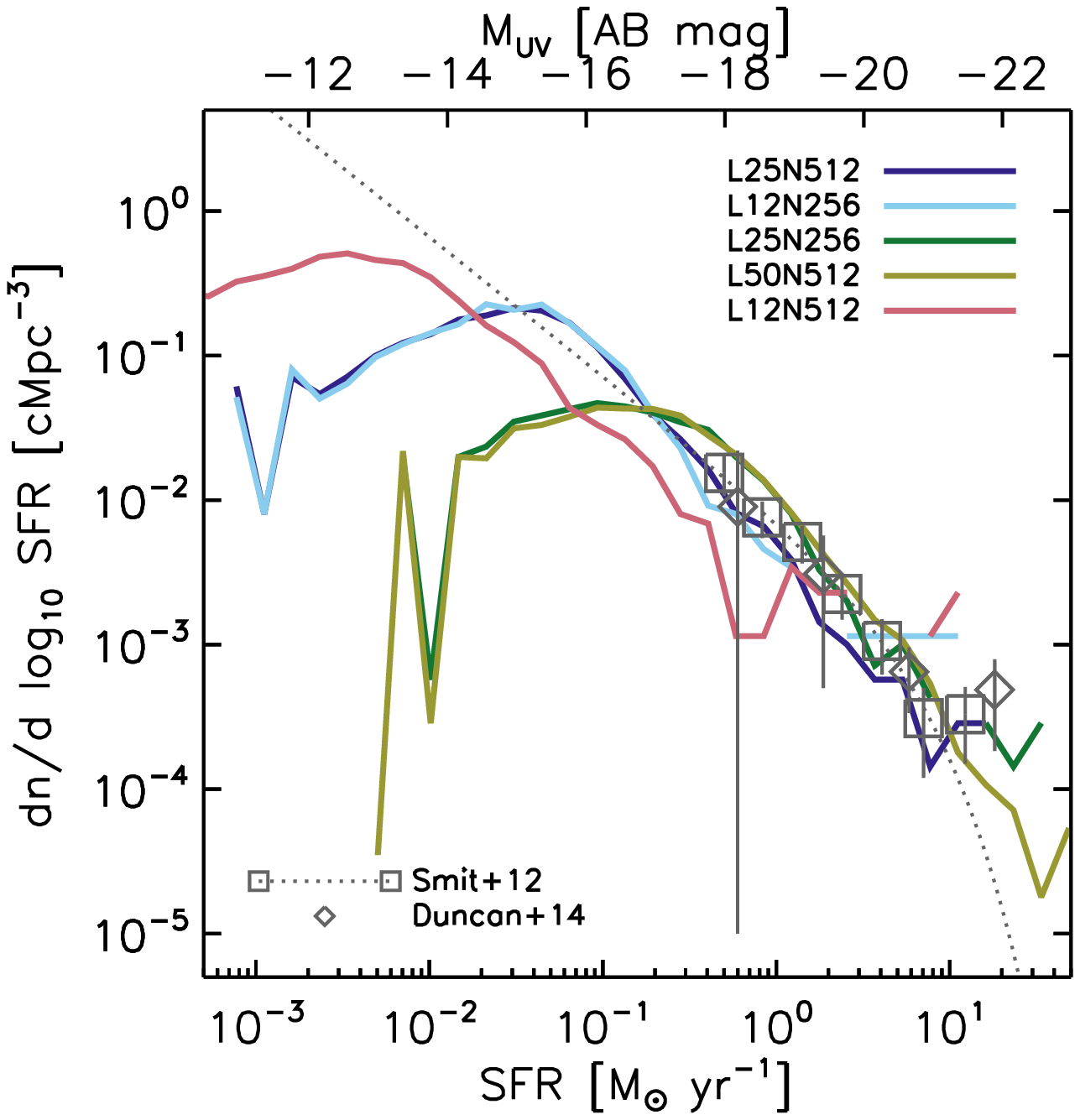}

  \end{center}
  \caption{Comparison of simulated and observed SFR functions at $z \approx
    7$. Squares show the observed dust-corrected SFR function from
    \protect\cite{Smit2012}, which is based on the dust-uncorrected SFR
    functions from \protect\cite{Bouwens2011}. We have divided the observed
    data, which assumes a Salpeter IMF, by 1.7 to enable comparisons with our
    simulations, which assume a Chabrier IMF. The grey dotted curve shows the
    corresponding Schechter fit. Diamonds show the observed dust-corrected SFR
    function from \protect\citet[their SED models]{Duncan2014}.  {\it Left:}
    effect of stellar feedback in the reference simulation L25N512. The top
    panel shows the ratio of the SFR functions in the bottom panel and the SFR
    function in the absence of stellar feedback extracted from simulation
    L25N512-noFB. Feedback from SNe is critical to matching the observed SFR
    function. The inefficiency of SNe in suppressing star formation at low
    SFRs is primarily a numerical artefact of the limited
    resolution. Photoheating affects the SFR function at SFRs $\lesssim 0.2
    \Msun \invyr$, but this is largely masked by the impact of SNe, which is
    strong at SFRs $\gtrsim 0.05 \Msun \invyr$. Photoheating thus does not
    imprint a noticeable characteristic scale in the UV luminosity
    function. {\it Right:} dependence on box size and resolution. At higher
    resolution, the explosion of massive stars as SNe suppresses star
    formation more strongly, although the effect is relatively small and
    therefore all curves are close to each other and to observational
    estimates. The turnover in the SFR functions at low SFRs is primarily a
    result of the limited resolution and the lack of low-temperature gas
    physics.} \label{fig5}
\end{figure*}
\par

\subsubsection{Impact on the UV luminosity function}
Another important observable of galaxy formation is the UV luminosity
function, i.e., the volume density of galaxies per unit logarithmic UV
luminosity. It quantifies the contributions to the cosmic SFR density from
galaxies with different SFRs, and therefore helps us to understand the
processes that shape the cosmic SFR density discussed
above. Figure~\ref{fig4} shows the UV luminosity function at $z \approx 6-9$
in our reference simulation L25N512. We converted SFRs and UV luminosities
using $ {\rm SFR} / (\Msun \invyr) = 0.6 \times 1.25 \times 10^{-28} L_{\rm
  UV}/ (\erg \invs \Hz^{-1})$ (e.g., \citealp{Bouwens2014};
\citealp{Kennicutt1998}), which yields the SFR function (e.g.,
\citealp{Smit2012}). The factor 0.6 converts the original relation, which is
based on a Salpeter IMF, to a relation assuming the Chabrier IMF used here
(see Section~\ref{Sec:sf}). Luminosities are related to AB magnitudes using $
-2.5 \log_{10} [L_{\rm UV}/ (4\pi 10^2 \pc^2)] - 48.6$.  Our reference
simulation L25N512 yields a SFR function in good agreement with the
observations above the limiting SFR of $\gtrsim 0.3 \Msun \invyr$ (assuming a
Chabrier IMF) currently probed by observations. Small difference exist at $z
\gtrsim 9$, where observational estimates are characterized by large
uncertainties, and at $z \lesssim 7$, especially at SFRs 
$\gtrsim 2\times 10 \Msun \invyr$, where dust corrections may be significant.
\par
In the left-hand panel of Figure~\ref{fig5} we investigate how feedback
impacts the UV luminosity function, where we focus for simplicity on a single
characteristic redshift $z \approx 7$. The
comparison of the reference simulation with simulation L25N512-noFB, in which
both photoheating and SNe are turned off, shows that suppression of star
formation by stellar feedback is a critical element of simulations that
attempt to match the observed SFR function.  In our reference simulation,
reionization heating and SNe suppress star formation at nearly complementary
scales. The SFR function in simulation L25N512-noSN, in which only SNe are
turned off and which serves to demonstrate the isolated impact of
photoheating, approaches that in simulation L25N512-noFB at SFRs $\gtrsim 0.2
\Msun \invyr$.  On the other hand, the SFR function in simulation
L25N512-noPH, in which only photoheating is turned off and which serves to
demonstrate the isolated impact of SNe, approaches that in simulation
L25N512-noFB at SFRs $\lesssim 0.05 \Msun \invyr$.
\par
The inefficiency of photoheating at high SFRs is expected if SFRs correlate
positively with halo mass, as is the case in our simulations
(Figure~\ref{fig6}), since reionization affects mostly the gas in low-mass
haloes (e.g., \citealp{Barkana1999}; \citealp{Dijkstra2004}; \citealp{Okamoto2008}). However, the
decreased efficiency of SN feedback at low SFRs is likely a numerical artefact
of the limited resolution. Indeed, we show in Appendix~\ref{App1} that in
galaxies with low SFRs, star formation is more strongly suppressed by SNe at
higher resolution.  Note that our simulations also likely underestimate the
local impact of photoheating by stars, especially in the lowest-mass
star-forming galaxies (\citealp{Rahmati2013}), thus likely raising the
relative importance of photoionization by the cosmological background.
\par
\subsubsection*{Dependence on resolution and box size}
The right-hand panel of
Figure~\ref{fig5} shows how our results depend on resolution and box size.
While our reference simulation yields SFR functions in good agreement with the
observations, simulations at lower and higher resolution significantly
overpredict and underpredict the observed SFR functions in the range of SFRs
in which SN feedback is efficient. This can be understood by noting that the
rate at which the SN heated gas cools and loses energy depends on resolution
(Eq.~17 in \citealp{DallaVecchia2012}). At higher resolution, the SN heated
gas can maintain a high ratio of the local radiative cooling time and the
sound-crossing time across a resolution element, a requirement for efficient
feedback, out to higher densities, increasing the ability of the SN heated gas
to prevent the formation of new stars. Because at higher resolution, the gas
fraction in low-mass haloes is more strongly reduced by stellar feedback, also
the hierarchically assembling larger-mass haloes grow systematically more
baryon-deficient as resolution is increased (see also, e.g.,
\citealp{Finlator2011}).
\par
The differences between the simulated and observed SFR functions in the
high-res and low-res simulations may be reduced by increasing or lowering the
fraction $f_{\rm SN}$ of the energy each SN injects. This would be reasonable
since our simulations lack the resolution and the physics that is needed to
accurately model the radiative losses in the ISM. This strategy would enable
one to investigate numerical convergence in simulations that reproduce
observational constraints on star formation independent of resolution (e.g.,
\citealp{Gnedin2014}; \citealp{Schaye2014}). The current set of simulations,
on the other hand, highlights the dependence on resolution of the SN energy
fraction required to match observational constraints on star formation. This
dependence demonstrates clearly that current cosmological simulations have
limited power in predicting the properties of star-forming galaxies from first
principles. Note however, that the difference in the simulated and observed
SFR functions is relatively small, of the order of the difference introduced
by uncertainties in the conversion between UV luminosities and SFRs (see
discussions in, e.g., \citealp{Munoz2011}; \citealp{Boylan2014};
\citealp{Bouwens2014}).  Therefore, the agreement between simulated and
observed SFR functions is still very good.
\par
\par
Finally, we investigate the dependence on box size. The reference simulation
overpredicts the abundance of the most intense star-forming galaxies in
comparison with observations. This is likely due to the finite size of the
box, which biases estimates of the abundance of the most massive and therefore
rarest galaxies. The comparison with the small and large box simulations
L12N256 and L50N512 supports box size as a limiting factor in accurate
determinations of the SFR function at the highest SFRs. Additional processes
that may impact models of the SFR function at high SFRs include feedback from
massive black holes (e.g., \citealp{diMatteo2005}; \citealp{Booth2009}), which
we have ignored here, and obscuration by dust (e.g., \citealp{Cen2014};
\citealp{Cai2014}).  While we compare to observations of dust-corrected SFRs,
there are large systematic uncertainties in the amount of dust as well as its
properties, especially at the high redshifts of interest here (e.g.,
\citealp{Smit2012}; \citealp{Duncan2014}). It is also important to keep in
mind that observational estimates of the SFR function are subject to cosmic
variance (e.g., \citealp{Finkelstein2012}).
\par
\subsubsection*{The UV luminosity function as a probe of reionization}
The reduction of the SFRs of low-mass galaxies due to photoheating may affect
the slope of the SFR function at low SFRs. Several (semi-)~analytical works
have proposed to search for this type of signature of reionization in the SFR
function (e.g., \citealp{Barkana2000}; \citealp{Mashian2013}).  Because the
mass scale below which reionization suppresses star formation depends on the
timing of reionization, locating this signature may constrain the reionization
history (e.g., \citealp{Barkana2006}; \citealp{Munoz2011}). Unfortunately, the
imprint of photoheating in the SFR function may be quite weak and therefore
difficult to detect, because of intrinsic scatter that causes galaxies of the
same mass to show different SFRs (e.g., \citealp{Barkana2000}), because of
hierarchical assembly that leaves also relatively massive haloes deficient in
baryons as they assemble from low-mass photoevaporated progenitors (e.g.,
\citealp{Barkana2000}; \citealp{Finlator2011}), and because of the averaging
in volumes in which reionization may proceed in a patchy manner (e.g.,
\citealp{Barkana2006}).
\par
In our simulation L25N512-noSN, which does not include SN feedback and
therefore lets us isolate the impact of radiative heating, photoheating indeed
causes a pronounced change in the slope of the SFR function at SFRs $\lesssim
0.2 \Msun \invyr$. However, in simulation L25N512 that includes both
photoheating and SN feedback, the change in the slope of the UV luminosity
function due to photoheating is strongly masked by the suppression of star
formation by SNe at SFRs $\gtrsim 0.05 \Msun \invyr$. In the presence of SNe,
the change in the slope of the SFR function occurs at lower SFRs, near the
scale at which the lack of low-temperature physics prevents gas cooling and
star formation in our simulations. This result is robust to changes in
resolution explored here. We thus conclude that if SN feedback is as efficient
as suggested by our simulations, detecting the signature of reionization
heating in the SFR function will be challenging. However, it is important to keep in
mind that our simulations ignore that stars may also form in minihaloes cooling
through molecular hydrogen transitions, which would be more susceptible to 
feedback from photoheating.
\par
The steep rise of the SFR function in the presence of radiative heating and
SNe at low SFRs down to the SFRs that our simulations do not accurately
resolve is consistent with that seen in previous works (e.g.,
\citealp{Finlator2011}; \citealp{Jaacks2012}; \citealp{Cai2014}), and with
extrapolations of the observed SFR function (e.g., \citealp{Smit2012};
\citealp{Bouwens2014}; \citealp{Finkelstein2014};
\citealp{Duncan2014}). Because the mass scale below which photoheating affects
the SFRs of galaxies is close to the mass scale below which efficient gas
cooling requires the presence of molecular hydrogen or metals, quantifying the
precise impact of reionization on the UV luminosity function will require 
more detailed, higher-resolution simulations of galaxy formation that include the
relevant gas physics. While this is challenging because it extends the
relevant dynamic range to still smaller scales, impressive first steps towards
such simulations have already been made (e.g.,
\citealp{Hasegawa2013}; \citealp{Gnedin2014b}).
\begin{figure*}
  \begin{center}
    \includegraphics[width=0.44\textwidth,clip=true, trim=0 20 30 0,
      keepaspectratio=true]{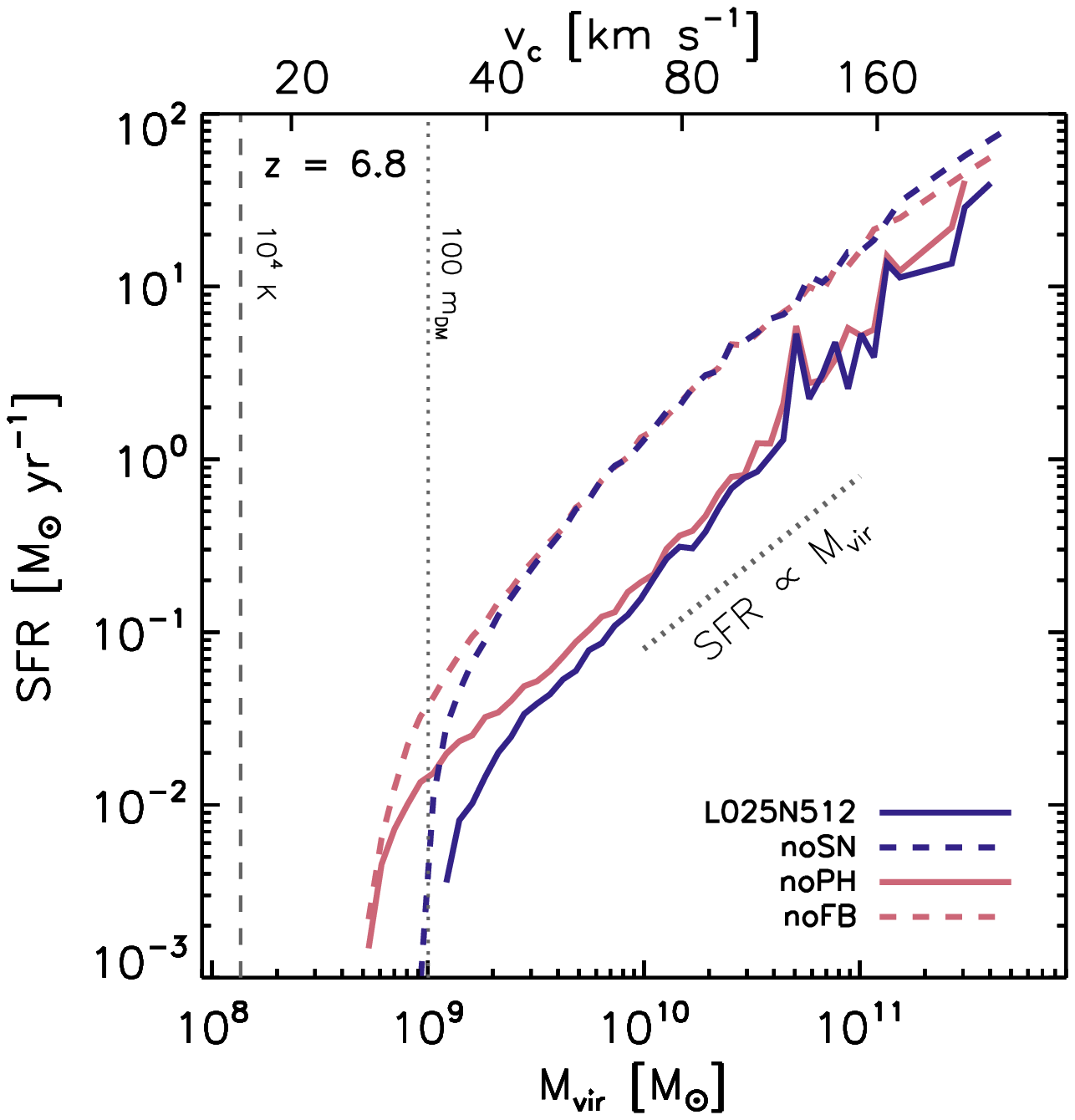}
    \includegraphics[width=0.44\textwidth,clip=true, trim=0 20 30 0,
      keepaspectratio=true]{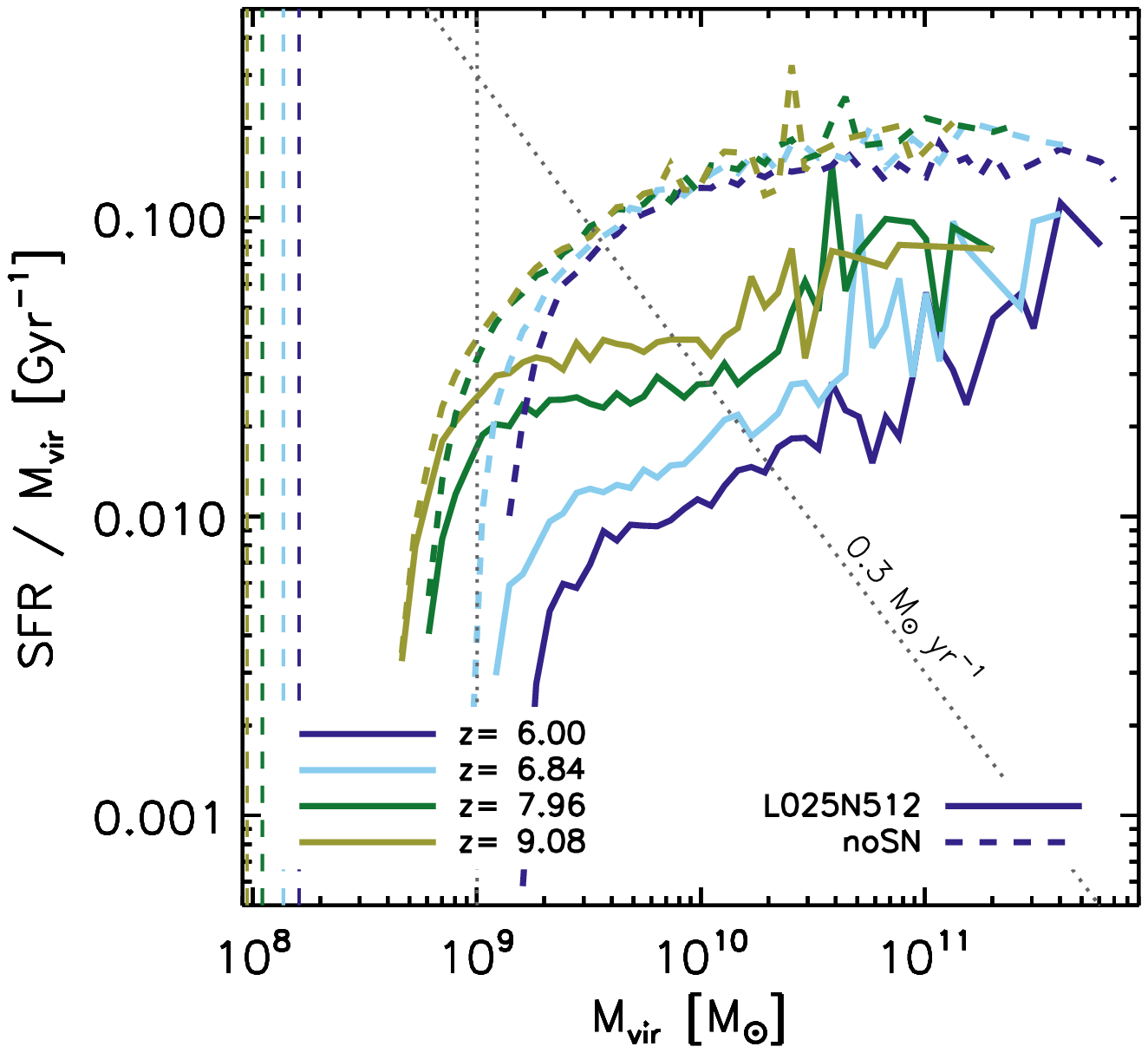} \\
     \includegraphics[width=0.44\textwidth,clip=true, trim=0 20 30 0,
       keepaspectratio=true]{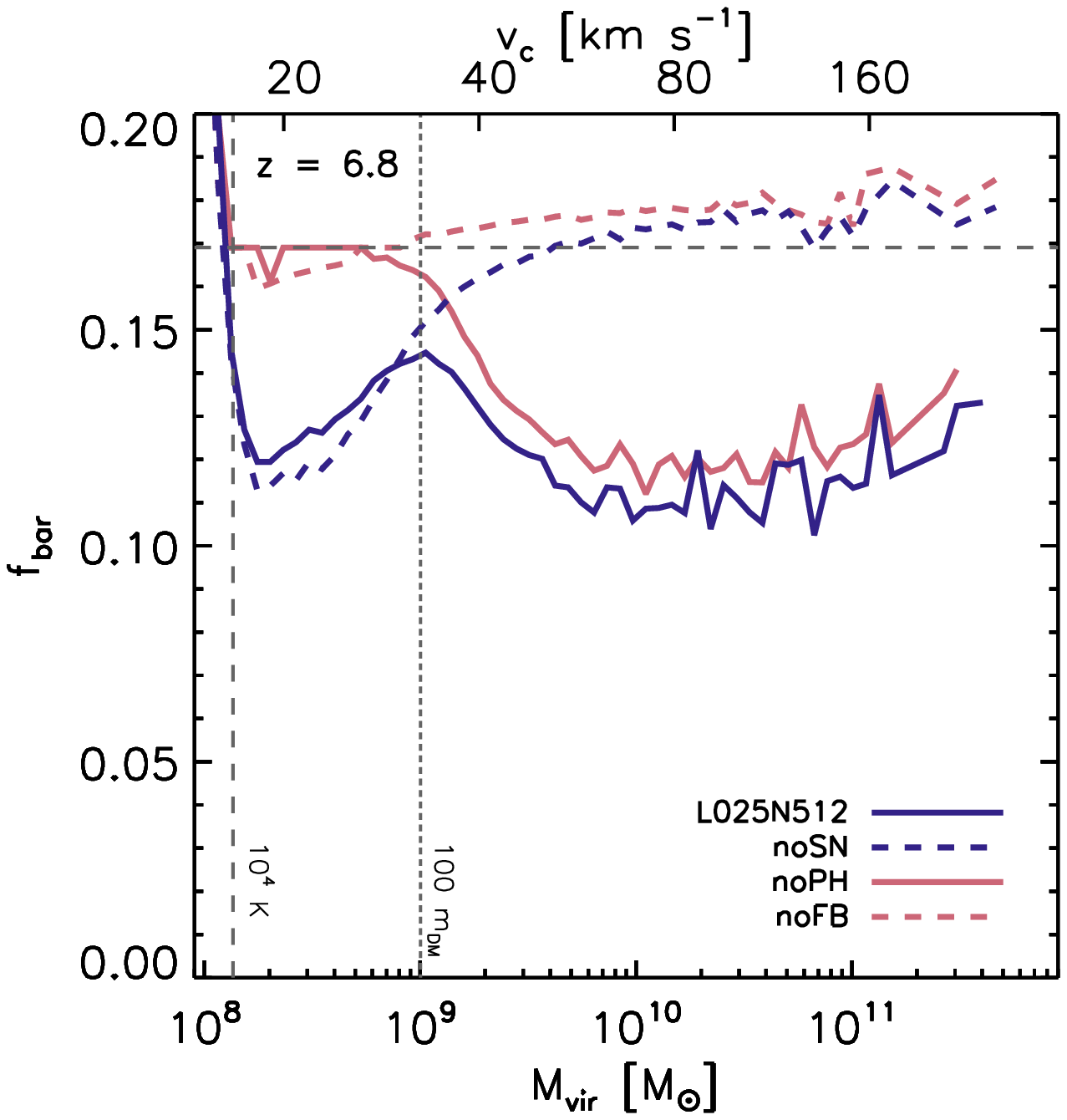}
    \includegraphics[width=0.44\textwidth,clip=true, trim=0 20 30 0,
      keepaspectratio=true]{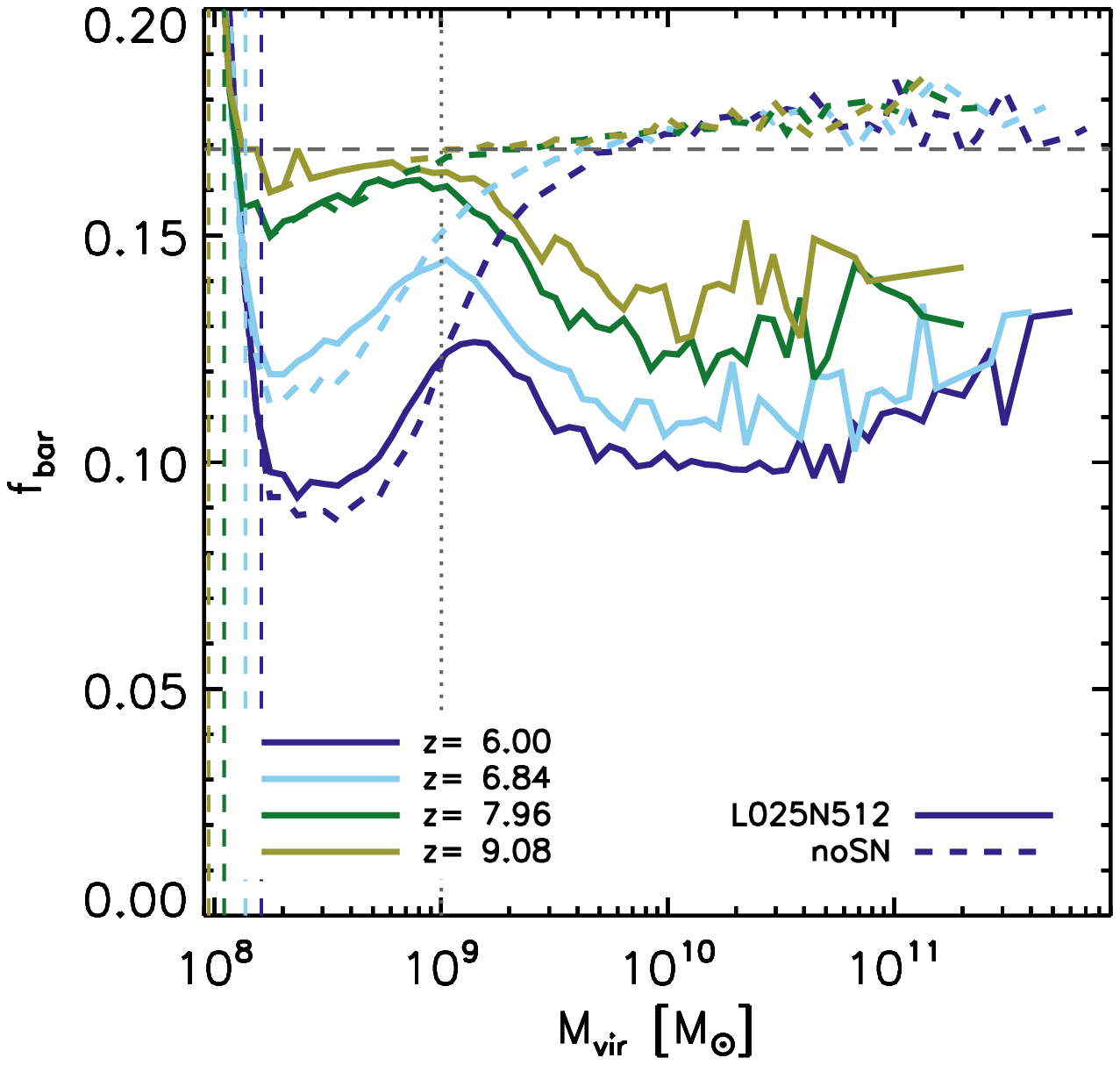}
 \end{center}
  \caption{Median SFRs (top) and median baryonic mass fractions $f_{\rm bar}$
    (bottom) of galaxies of virial masses $M_{\rm vir}$ and associated
    circular velocities $v_{\rm c} \equiv (G M_{\rm vir} / r_{\rm
      vir})^{1/2} = 17 (M/ 10^8 \Msun)^{1/3} [(1+z)/10]^{1/2} \kms$
    (e.g., Eq.~3.11 in \citealp{Loeb2010}). In each row, the left-hand panel shows
    the impact of stellar feedback in the reference simulation L25N512
    at a typical redshift $z \approx 7$ and the right-hand panel shows the
    dependence on redshift in that simulation. In the 
    right-hand panels, to isolate the impact of photoheating, we
    include results from simulation L25N512-noSN, in which SN
    feedback was turned off (dashed curves). In the top right panel, 
    we have divided the SFRs by the virial masses of the
    haloes hosting the galaxies to improve the clarity of the
    presentation. The dashed vertical lines mark the masses of haloes
    with virial temperatures $T_{\rm vir} = 10^4 \K$. The dotted
    vertical lines mark the mass of 100 DM particles. The dotted diagonal line in the
    top right panel marks a SFR of $0.3 \Msun \invyr$,
    which is the limiting SFR accessible in current observations
    (assuming a Chabrier IMF; see Figure~\ref{fig4}). The dashed horizontal lines in the
    bottom row panels mark the cosmic baryon fraction $\Omega_{\rm b}
    / (\Omega_{\rm b} + \Omega_{\rm DM})$. The strong suppression of
    star formation by photoheating and by the explosion of
    stars as SNe is accompanied by a strong reduction in the baryon
    fraction. The dependence on resolution is discussed in Figure~\ref{figa1}.}
  \label{fig6}
\end{figure*}

\subsubsection{Impact on galaxy properties}
\label{sec:galprop}
Inspecting the properties of individual galaxies helps gain insight into the
physical origin of the shape of the SFR function discussed above.
Figure~\ref{fig6} shows the median SFRs (top row) and the median baryonic mass
fractions $f_{\rm bar} \equiv (M_\star + M_{\rm gas}) / M_{\rm vir}$ (bottom
row) of galaxies in our simulations.  The left-hand panel in each row shows
the impact of feedback in the reference simulation at $z \approx 7$, and the
right-hand panel shows the evolution with redshift. In the top right panel, we
have divided the SFRs by the virial mass of the haloes hosting the galaxies to
improve the clarity of the presentation. Our reference simulation predicts
that the faintest galaxies accessible by current observations reside in haloes
with masses $M_{\rm vir} \sim 10^{10} \Msun$ (intersection of solid and
diagonal dotted curves in the top middle panel), consistent with observational
estimates based on matching the shapes of the DM halo mass function and the
observed UV luminosity function (e.g., \citealp{Trenti2010}).
\par
The top left panel of Figure~\ref{fig6} shows that in our reference
simulation, the median SFR at $z \approx 7$ drops sharply below $M_{\rm vir}
\sim 2 \times 10^9 \Msun$. The comparison with simulation L25N512-noSN in
which SNe were turned off, shows that this scale is set primarily
by feedback from reionization. In the absence of photoheating, star formation
continues in haloes with lower masses. However, even without feedback, galaxies
in haloes with masses $\lesssim 5 \times 10^8 \Msun$ do not form stars since
gas cooling and condensation is prevented by the finite resolution and the
lack of low-temperature coolants, such as molecular hydrogen and metals. SN
feedback suppresses star formation strongly across nearly the entire range of
galaxy masses.  It only becomes inefficient at the lowest masses at which our
finite resolution impacts the conversion of gas into stars, and at the highest
masses, at which our simulations are also subject to finite box size effects.
\par
\par
\par
\begin{figure}
  \begin{center}
    \includegraphics[width=0.49\textwidth,clip=true, trim=10 40 30 30,
      keepaspectratio=true]{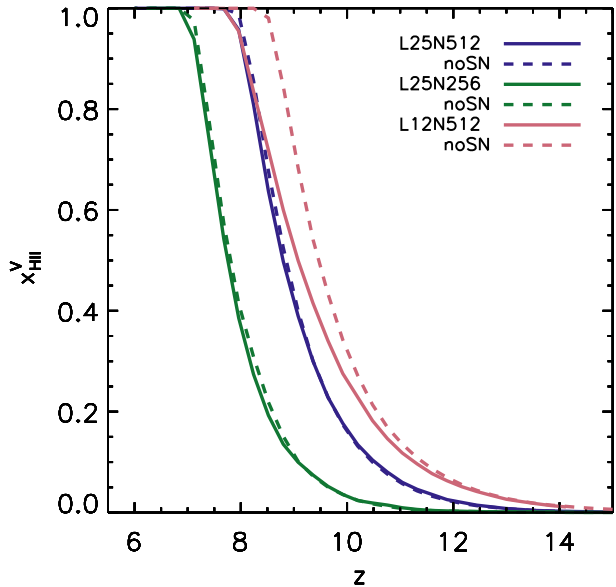}
   \end{center}
  \caption{Effect of SNe on the evolution of the volume-weighted mean
    ionized fraction $x_{\rm HII}^{\rm V}$ and its dependence on resolution. The
    solid curves show the reionization histories in simulations
    including feedback from ionizing radiation and SNe. The dashed
    curves show the reionization histories in the corresponding
    simulations in which SN feedback was turned off. The inclusion of
    SNe leads to at most a small delay in reionization,
    despite the strong suppression by SNe of the cosmic SFR
    (Figure~\ref{fig3}).}
  \label{fig7}
\end{figure}

Stellar feedback reduces star formation primarily because it reduces the
amount of dense gas, both by expelling gas and by limiting the rates at which
gas is accreted.  In our reference simulation, reionization strongly reduces
the gas fractions at halo masses $\lesssim 2 \times 10^9\Msun$. Reionization
also reduces, though less strongly, the gas fractions in up to $\sim 10$ times
more massive galaxies, which assemble in mergers of lower-mass
baryon-deficient galaxies and accrete gas from the reionized IGM (e.g.,
\citealp{Barkana2000}; \citealp{Finlator2011}; \citealp{Munoz2011}). SN
feedback most strongly impacts the gas fractions at masses $\gtrsim 10^9
\Msun$ in our reference simulation, again demonstrating that reionization and
SN feedback act mostly complementary at the reference resolution. We show in
Appendix~\ref{App1} that in our high-res simulation, both reionization and SN
feedback reduce the baryon fractions more strongly, and SN feedback extends
its impact to haloes with masses as low as $\gtrsim 2 \times 10^8 \Msun$.
\par
The scale at which photoheating becomes effective in suppressing star
formation and reducing the baryon fractions in our reference simulation
evolves significantly only after $z \approx 8$, which coincides with the
completion of reionization. A similar effect is seen in the evolution of the
baryon fractions. This late impact of photoheating suggests a larger role of
illumination by the external ionizing background than by localized ionizing
sources internal to the galaxies, in agreement with the discussion of the
belated impact of radiative heating on the cosmic SFR density above. However,
we caution that in our simulations the impact of local ionizing sources may be
underestimated due to the finite resolution. At least before reionization,
radiative feedback from local sources is expected to dominate over feedback
from external illumination. This is indeed seen in simulations of the first
stars and the start of reionization, which focus on smaller volumes and
therefore can afford a higher resolution and also accurately capture the
relevant low-temperature physics, primarily the cooling and chemistry of
molecular hydrogen (e.g., \citealp{Ricotti2005a}; \citealp{Wise2008};
\citealp{Pawlik2013}; \citealp{Jeon2015}).
\par

\begin{figure*}
  \begin{center}
    \includegraphics[width=0.49\textwidth,clip=true, trim=10 40 30 30,
      keepaspectratio=true]{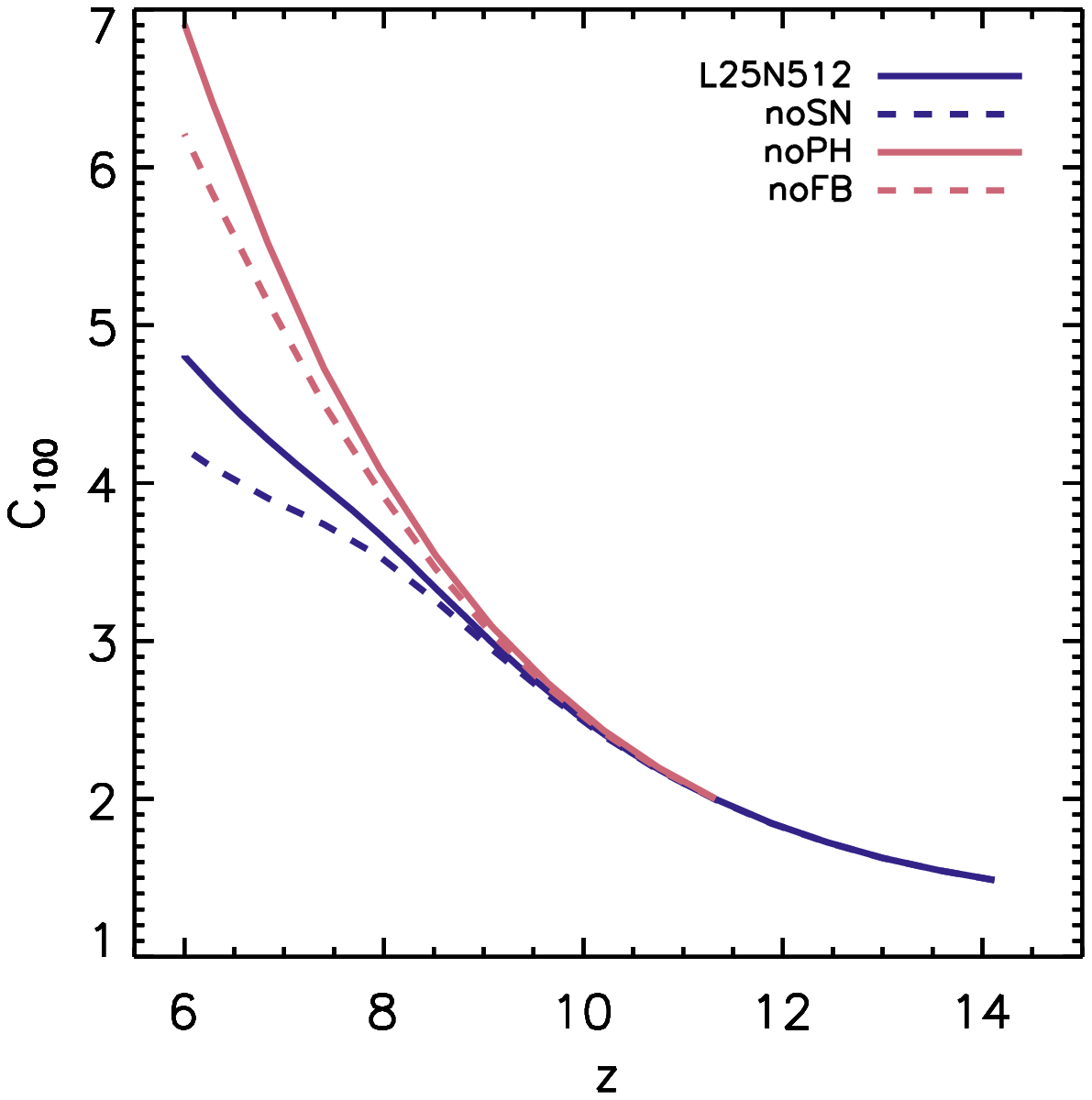}
    \includegraphics[width=0.49\textwidth,clip=true, trim=10 40 30 30,
      keepaspectratio=true]{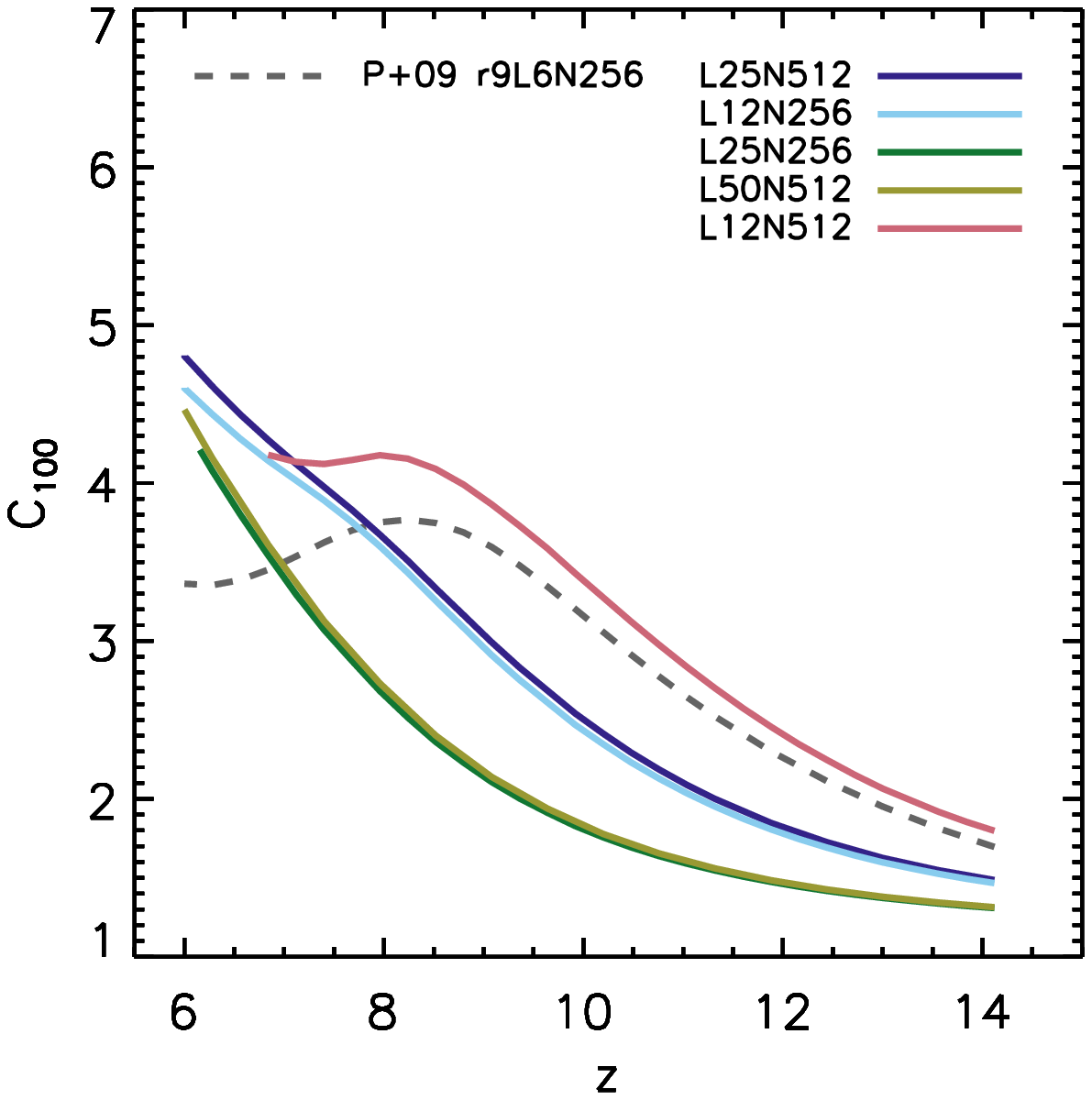}
  \end{center}
  \caption{Evolution of the IGM clumping factor $C_{100}$, which parametrizes
    the average recombination rate in gas with overdensity $\le 100$. {\it
      Left:} effect of stellar feedback in the reference simulation
    L25N512. Photoheating strongly reduces the clumping factor because it
    increases the Jeans mass in the reionized gas, providing a
    positive feedback on reionization. SN feedback increases the clumping
    factor by $\sim 15\%$ as it moves gas from galaxies to the IGM.  {\it
      Right:} dependence on box size and resolution.  The dashed curve shows
    the clumping factor from simulation r9L6N256 in
    \protect\cite{Pawlikclump2009}, in which the gas was heated by a uniform
    ionizing background turned on at $z \le 9$. The earlier result is in good
    agreement with the clumping factor in simulation L12N512, which employs a
    similar resolution and in which the IGM, on average, is reionized at a
    similar redshift. The small increase with respect to the earlier work is
    mostly due to the inclusion of SNe.}
  \label{fig8}
\end{figure*}

\subsubsection{Impact on reionization}
\label{Sec:reionimpact}

Both SNe and radiative heating may have a strong impact on
reionization.  We have seen above that SN feedback reduces the SFRs
and therefore the ionizing luminosities of galaxies, which makes it
more difficult for the galaxies to reionize the Universe. However, SNe
may also open low-density channels in the ISM, e.g. by expelling gas in winds, 
through which ionizing photons may escape more easily (e.g.,
\citealp{Dove2000}; \citealp{Fujita2003}; \citealp{Wise2009}; \citealp{Paardekooper2011}). This
makes it easier for galaxies to reionize the Universe. The net impact
of SNe on reionization is the result of the interplay of these two
processes, and is therefore difficult to predict. 
\par
We have also seen above that radiative heating may help SNe to suppress star
formation, which impedes reionization. On the other hand, radiative heating
raises the Jeans mass in the IGM, and this reduces the IGM clumping factor and
therefore the rate at which the IGM recombines.  If recombinations consume a
significant number of the ionizing photons that escape into the IGM, then
radiative heating will help keeping the ionized gas ionized and thereby
facilitate reionization (e.g., \citealp{Pawlikclump2009};
\citealp{Finlatorclump2012}; \citealp{So2014}; \citealp{Sobacchi2014}).
\par
Figure~\ref{fig7} shows that SN feedback has only a small impact on
the timing of reionization.  In the reference and low-res
simulations, the reionization histories with and without SN feedback
are nearly identical, and in the high-res simulation, SN feedback delays
reionization by $\Delta z \lesssim 0.5$. That the net impact of SNe on reionization is small
suggests that the strong reduction in the SFRs due to SNe - in the
high-res simulation by a factor $\sim 10$ by the end of reionization (Figure~\ref{fig3}) -
is partially compensated by an increase, also due to SNe, in the 
escape fraction of ionizing radiation. A plausible physical mechanism by which this
might be achieved, is the strong reduction of the baryon fractions by
SNe seen in our simulations (Figure~\ref{fig6}). However, the 
lack of a substantial impact on the reionization history 
could also be explained if reionization is driven mostly by the lowest-mass
galaxies in which SN feedback is less effective in suppressing star formation 
due to the limited resolution. RT computations of the
escape of ionizing photons into the IGM would help to identify which of these 
mechanisms is dominant, but this is beyond the scope of the current work.
\par
The left-hand panel of Figure~\ref{fig8} shows that the reduction in the IGM
recombination rate due to radiative heating is strong. We have followed
\cite{Pawlikclump2009} and have equated the IGM clumping factor, $C_{\rm IGM}$, to
$C_{100}$, which parametrizes the average recombination rate of gas with overdensities $\le
100$.  This enables us to separate recombinations in the IGM from
recombinations inside galaxies (see also \citealp{Miralda2000}). The latter
are typically parametrized by the ionizing escape fraction.  Alternative
definitions of the clumping factor are sometimes employed, e.g., by applying
additional selection criteria to identify the ionized IGM (e.g.,
\citealp{Kohler2007}; \citealp{Shull2012};
\citealp{Finlatorclump2012}; \citealp{Kaurov2014}). However, for typical reionization scenarios,
these definitions, which are all designed to achieve the same objective, i.e.,
locating the ionized gas in the IGM, generally yield very similar results
(see, e.g., the discussion in \citealp{Finlatorclump2012}).
\par
Our simulations likely underestimate the clumping factor before reionization,
since the Jeans mass in the unheated IGM is unresolved (e.g.,
\citealp{Emberson2013}). Thus, the positive radiative feedback from
photoheating on reionization is also underestimated. Note however, as shown in
\cite{Pawlikclump2009}, that the clumping factor at $z \approx 6$ is
insensitive to the redshift at which reionization occurs for reionization at
$z \gtrsim 8$. As already found in \cite{Pawlikclump2009}, SNe move gas from
galaxies to the IGM, and this leads to a slight increase in the clumping
factor (see also \citealp{Finlatorclump2012}).
\par
The right-hand panel of Figure~\ref{fig8} shows that the clumping factor
derived from the high-res simulation L12N512 is consistent with that in
simulation r9L6N256 of \cite{Pawlikclump2009}, which had similar resolution
but assumed that the IGM is heated by a uniform UV background turned on
instantaneously at $z = 9$. This agreement likely results because in our RT
simulations, reionization occurs sufficiently rapidly such that differences in
the time at which individual regions inside the simulated volume are reionized
are small and similar to the time it takes for the IGM to respond dynamically
to the increase in the Jeans mass in the simulation in which the gas is
instantaneously exposed to a uniform ionizing background. The slight increase
in the clumping factor by $\sim 15\%$ with respect to the earlier work is
mostly due to the inclusion of SN feedback, although, as shown in
\cite{Pawlikclump2009}, the differences in cosmological parameters also
contribute. For reionization occurring at redshifts $z \gtrsim 8 $, the IGM
clumping factor $C_{100}$ at $z \approx 6$ is numerically converged and
insensitive to a further increase in resolution and box size
(\citealp{Pawlikclump2009}).
\par

\section{Discussion}
\label{Sec:Discussion}
In this section we briefly discuss how changes in physical 
parameters affect the outcome of our simulations and also
mention some of the main physical processes our simulations have 
ignored.
\par

\subsection{Variations of physical parameters}
\label{Sec:Parameters}

\begin{figure*}
  \begin{center}
    \includegraphics[width=0.33\textwidth,clip=true, trim=0 30 20 00,
      keepaspectratio=true]{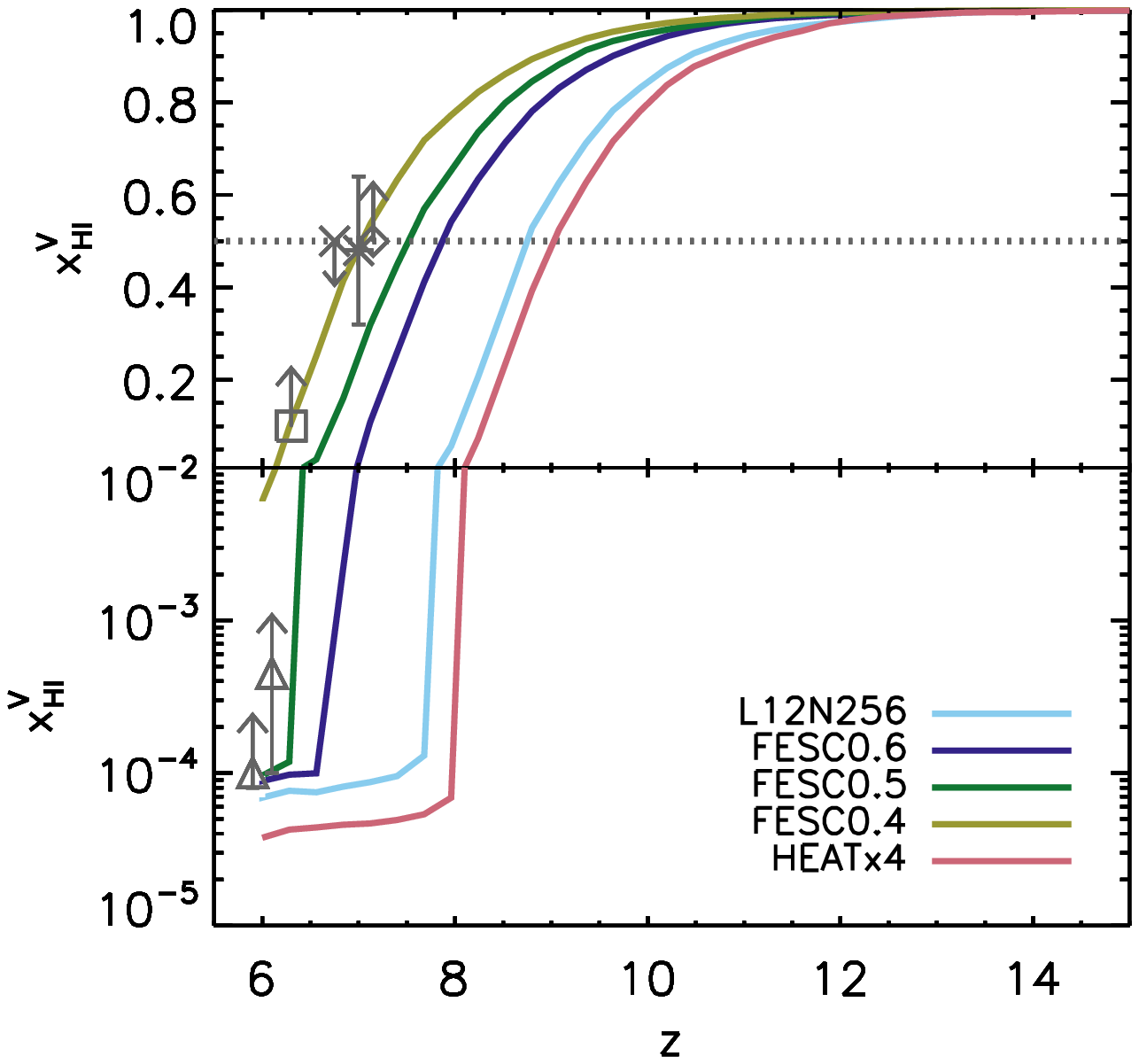}
    \includegraphics[width=0.33\textwidth,clip=true, trim=0 30 20 00,
      keepaspectratio=true]{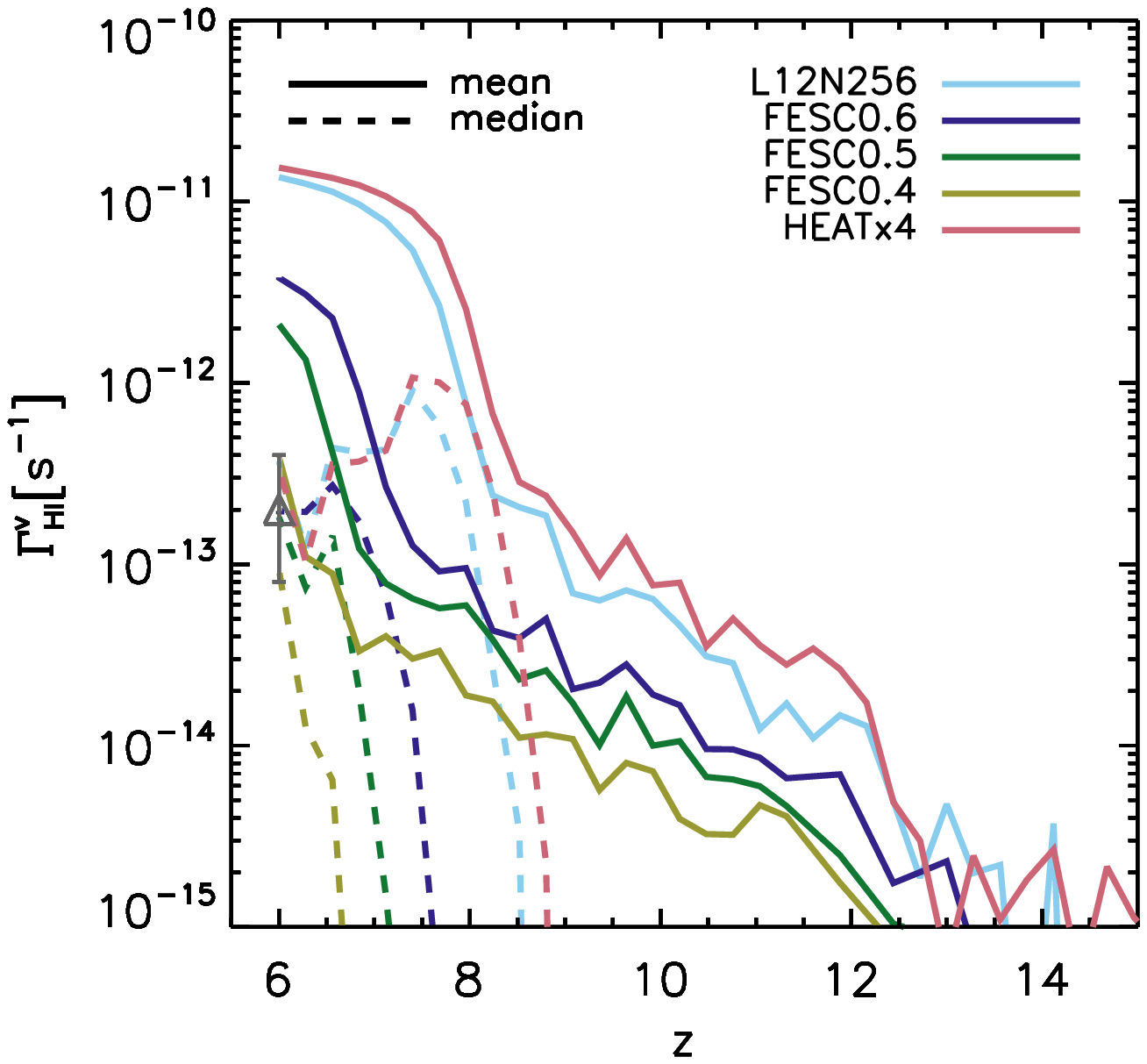}
    \includegraphics[width=0.33\textwidth,clip=true, trim=0 30 20 00,
      keepaspectratio=true]{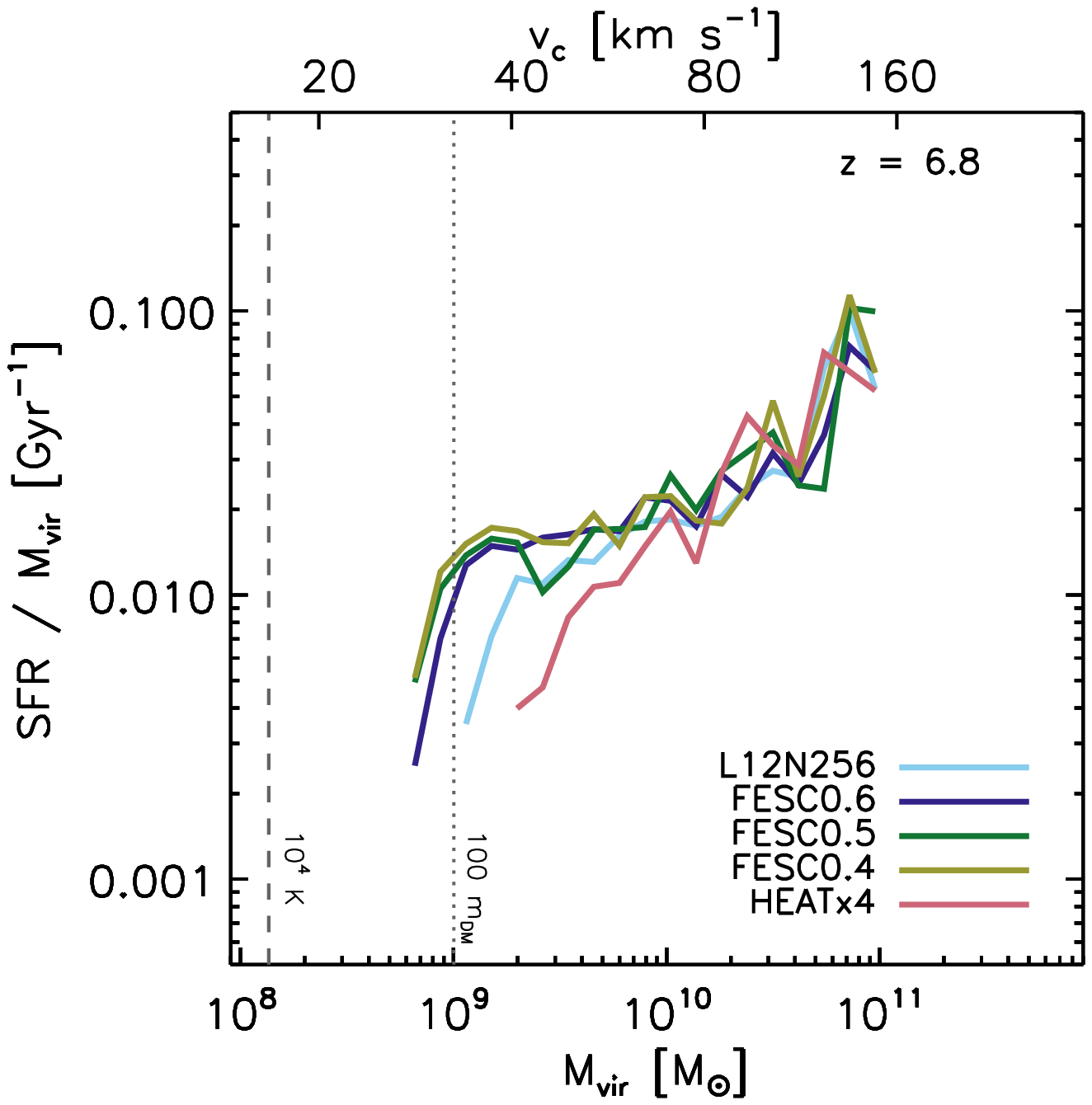}
  \end{center}
  \caption{Impact of variations in the physical parameters on the
    volume-weighted mean neutral fraction (left), the hydrogen photoionization
    rate (middle), and the median SFRs of galaxies at $z \approx 7$ (right,
    normalized by the virial masses) in simulation L12N256.  The vertical
    dashed line marks the mass corresponding to a virial temperature $10^4
    \K$, and the vertical dotted line marks the mass of 100 DM particles.
    Data points are as in Figure~\ref{fig2}. Reducing the sub-resolution
    escape fraction to $0.6, 0.5$, and $0.4$ (blue, green, and light
    green curves) implies that reionization occurs later, which reduces the
    impact of photoheating on the SFRs. Increasing the energy injected in each
    photoionization by a factor $4$ (red curves) leads to larger gas
    temperatures, which reduces recombination rates and increases the scale
    below which photoheating suppresses SF.}
  \label{fig9}
\end{figure*}

Figure~\ref{fig9} shows how variations in the physical parameters impact
the reionization history and the ability of galaxies to form stars. Towards
this end we have repeated the small-box simulation L12N256, which has the same
resolution as the reference simulation L25N512 but allows for a
computationally more efficient exploration of the parameter space, with
different values for some of the parameters.
\par
One of the main parameters of reionization simulations is the escape fraction
$f_{\rm esc}^{\rm subres}$ of the unresolved ISM.  Decreasing this fraction
from $1.0$ to $0.6, 0.5$ and $0.4$ causes reionization to occur at lower
redshifts, as expected, at which it extends over longer times, in good
agreement with similar parameter studies by \cite{Petkova2011} and
\cite{Hasegawa2013}. Since photoheating is delayed, low-mass haloes can
continue to form stars more efficiently down to lower redshifts. However, the
associated increase in the ionizing emissivity is smaller than the reduction
due to the decrease in the sub-resolution escape fraction, and so the impact
on reionization due to the change in the minimum mass of haloes is small (e.g.,
\citealp{Petkova2011}). In the current simulations, adopting an escape
fraction $f_{\rm esc}^{\rm subres} = 0.4$ yields excellent agreement with
observations of the evolution of the ionized fraction and the hydrogen
photoionization rate at $z = 6$ (but observational constraints are weak; see
the discussion in Section~\ref{Sec:Reionization}).
\par
The amount of energy injected in each photoionization depends on the spectrum
of the radiation sources and requires multi-frequency RT simulations for an
accurate computation including spectral hardening (e.g., \citealp{Abel1999b};
\citealp{Maselli2009}; \citealp{Pawlik2011}). Here we treat this energy as a
parameter (e.g., \citealp{Petkova2011}). Increasing it implies a slight
acceleration in reionization, and a slight decrease in the mean neutral
fraction after reionization. This is caused by the increase in the gas
temperatures implied by the higher photoheating rates, which in turn decreases
the rate at which hydrogen recombines (e.g., \citealp{Stiavelli2004};
\citealp{Pawlikclump2009}; \citealp{Finlatorclump2012}). Finally, the
increased gas temperature increases the negative impact of photoheating on the
efficiency of low-mass galaxies to form stars, raising the mass scale below
which star formation is strongly suppressed (see also, e.g.,
\citealp{Petkova2011}).

\par
Finally, we have carried out a preliminary comparison with results from a new
set of simulations of reionization similar to those presented here, which
feature increased physical realism and span a wider range of box sizes and
resolutions. These simulations will be discussed elsewhere in more detail and
include, among others, helium chemistry and cooling/heating, and feedback and
metal enrichment from AGB stars, core-collapse and Type Ia SNe, and
metallicity-dependent population synthesis. The treatment of helium was made
feasible by replacing the explicit chemistry solver described in
Sec.~\ref{sec:chemistry} with the implicit solver described in
\cite{Pawlik2013}, which is faster. Moreover, the simulations are designed, by
calibrating the sub-resolution escape fraction and the SN energy fraction, to match
the observed UV luminosity functions and to exhibit similar reionization
histories independent of resolution. A simulation employing the same
resolution as our reference simulation here and adopting a sub-resolution
escape fraction of $f^{\rm subres}_{\rm esc} = 0.5$, yields reionization and
SFR histories similar to those in the current simulation L12N256 adopting an
escape fraction $f^{\rm subres}_{\rm esc} = 0.6$.
\par

\subsection{Limitations}
Our simulations ignored a number of potentially relevant physical
processes. Most importantly, perhaps, our simulations ignored the chemistry of
and radiative cooling by molecular hydrogen, effectively assuming a soft UV
background that prevents the build-up of hydrogen molecules. This
approximation fails at the earliest stages of reionization, where it
artificially prevents the formation of stars inside low-mass minihaloes (e.g.,
\citealp{Wise2008b}; \citealp{Greif2008}; \citealp{Pawlik2013};
\citealp{Muratov2013}). This early population of stars may preionize the IGM
and provide a significant feedback on subsequent star formation and
reionization (e.g., \citealp{Ricotti2004b}; \citealp{Ahn2012}). Our
simulations have also ignored the enrichment with metals that accompanies the
explosion of stars as SNe. Metal-enrichment affects the rates at which gas
cools and forms stars (e.g., \citealp{Jappsen2009}; \citealp{Wiersma2009}),
which may be especially important in low-mass minihaloes in which radiative
cooling by atomic hydrogen is suppressed. Finally, we have only followed the
radiative feedback from the relatively soft ionizing radiation emitted by
metal-enriched stars. Other sources of radiation, such as zero-metallicity
stars (Pop~III stars) or X-ray emitting black holes, may be an important
source of ionization and feedback during reionization (e.g.,
\citealp{Madau2004}; \citealp{Ricotti2005}; \citealp{Alvarez2009};
\citealp{Haiman2011}; \citealp{Jeon2014}).

\section{Summary}
\label{Sec:Summary}
\par
We have carried out a suite of cosmological radiation-hydrodynamical
simulations of galaxy formation during reionization. The reference simulation
was run in a box of size $25 \cMpch$ and contained $512^3$ dark matter and
$512^3$ baryonic particles, thus resolving atomically cooling haloes with at
least $\gtrsim 10$ dark matter particles. Simulations using both
larger and smaller boxes and higher and lower resolution allowed us to
investigate the numerical convergence of our results. Simulations in which either SNe or
photoionization heating or both are turned off, enabled us to isolate and
investigate the impact of feedback from star formation. Ionizing photons were
transported using accurate and spatially adaptive RT, tracking the growth of
ionized regions and the build-up of an ionizing background at the native high
resolution at which the hydrodynamics was carried out. 
\par
Current cosmological simulations lack both the resolution and the physics to
provide an ab initio description of the structure of the interstellar gas and
the rate at which the gas cools. This necessitates the use of physically
motivated but resolution-dependent parameters to control the energy that each SN
injects and the fraction of ionizing photons that escape into the IGM. SFRs and
reionization histories are sensitive to these parameters and this impedes the
use of cosmological simulations in predicting these quantities from first
principles.  On the other hand, one may exploit this sensitivity and choose
parameters for which simulated SFRs and reionization histories are consistent
with current observational constraints, and investigate the implications of
such observationally supported models of galaxy formation. Here we have
focused on the role of feedback from SNe and photoheating, two processes that
critically shape galaxy formation and reionization.
\par
Our reference simulation yields SFR densities and a UV luminosity function in
excellent agreement with observational constraints and completes reionization
by $z \approx 8$. Increasing the resolution leads to a strong increase in the
cosmic SFR at high redshifts as it facilitates the condensation of gas into
low-mass galaxies. It leads to a mild decrease in the cosmic SFR at late
times, when star formation is strongly regulated by stellar feedback.  As a
consequence, near the end of reionization, our high-res simulation
yields a slightly smaller normalization of the UV luminosity function and our
low-res simulation yields a slightly larger normalization of the UV luminosity
function than our reference simulation and observations. Because of the
initial increase in the cosmic SFR, increasing the resolution also increases
the redshift at which the IGM is reionized. Increasing the size of the
simulation box above $12.5 \cMpch$ has only a minor impact on the SFR and
reionization histories.
\par
Photoheating reduces the baryon fractions and suppresses star formation
primarily in haloes with masses below $\lesssim 2\times 10^9 \Msun$. SNe, on
the other hand, reduce the baryon fractions and suppress star formation
primarily in haloes more massive than $\gtrsim 10^9 \Msun$. Therefore, the
currently observable cosmic SFR is more strongly suppressed by SNe than by
photoheating, and SN feedback alone is sufficient to match observational
constraints on the UV luminosity function. The inefficiency of SNe in the
lowest mass galaxies is primarily a consequence of the lack of low-temperature
gas physics and the limited resolution. Nevertheless, the feedback from SNe is
sufficiently strong to mask the impact of photoheating on the abundance of
low-mass atomically cooling star-forming galaxies. We thus do not find a
noticeable signature imprinted by reionization heating on the UV luminosity
function, although we note that the resolution of our simulations is insufficient to
model star-forming minihaloes that would be more strongly affected by photoheating. 
\par
Despite the relatively small impact on the cosmic SFR, photoheating is an
important feedback process. First, photoheating amplifies the ability of SNe
to suppress star formation. This amplification is nonlinear and mutual,
demonstrating the need to simultaneously account for both feedback
processes. Second, photoheating smooths out gas density fluctuations in the
IGM and thereby strongly reduces the IGM recombination rate. This makes it
easier to keep the gas ionized, which facilitates reionization. In contrast,
the net impact of SNe on reionization is small. SNe strongly suppress SFRs and
slightly increase the IGM recombination rate as gas is moved from the galaxies
to the IGM. However, this leads only to a small delay in the timing of
reionization, possibly because SNe create additional low-density channels in
the ISM through which ionizing photons can escape, which increases the 
escape fraction of ionizing radiation, or because reionization is driven by the lowest mass
galaxies in which SN feedback is inefficient in our simulations.
\par
\par
In summary, our work demonstrates that both photoheating by the stellar
radiation that reionizes the Universe and the explosion of massive stars as
SNe may have had a strong impact on structure formation and reionization in
the first billion years.
\par

\section*{Acknowledgments}
We are grateful to Volker Springel for letting us use GADGET and the halo
finder Subfind.  We thank Ali Rahmati, Milan Raicevic, Myoungwon Jeon, Craig
Booth, and Volker Bromm for useful discussions, and we thank Benedetta Ciardi
for a careful reading of an early draft. We further thank Ali Rahmati for 
help with executing some of the simulations used in Figure 9. We thank Rychard Bouwens for
providing us with his upwards corrections of UV luminosity densities used in
Figure~\ref{fig2}, and we thank Kenneth Duncan for his tables of observed SFRs
and UV luminosity functions provided to us in electronic form. 
Computer resources for this project have been provided by
the Gauss Centre for Supercomputing/Leibniz Supercomputing Centre under
grant:pr83le. We further acknowledge PRACE for awarding us access to resource
Supermuc based in Germany at LRZ Garching (proposal number 2013091919). Some
of the simulations presented here were run on Odin at the Rechenzentrum
Garching (RZG) and the Max-Planck-Institute for Astrophysics (MPA) and on
Hydra at the RZG. This work was sponsored
with financial support from the Netherlands Organization for Scientific
Research (NWO), also through a VIDI grant and an NWO open competition
grant. We also benefited from funding from NOVA, from the European Research
Council under the European Unions Seventh Framework Programme (FP7/2007-2013)
/ ERC Grant agreement 278594-GasAroundGalaxies and from the Marie Curie
Training Network CosmoComp (PITN-GA-2009-238356). AHP received funding from
the European Union's Seventh Framework Programme (FP7/2007-2013) under grant
agreement number 301096-proFeSsoR. CDV benefitted from Marie Curie
Reintegration Grant PERG06-GA-2009-256573.

\appendix

\section{Dependence on resolution}
\label{App1}
\begin{figure*}
  \begin{center}
     \includegraphics[width=0.44\textwidth,clip=true, trim=0 20 30 0,
      keepaspectratio=true]{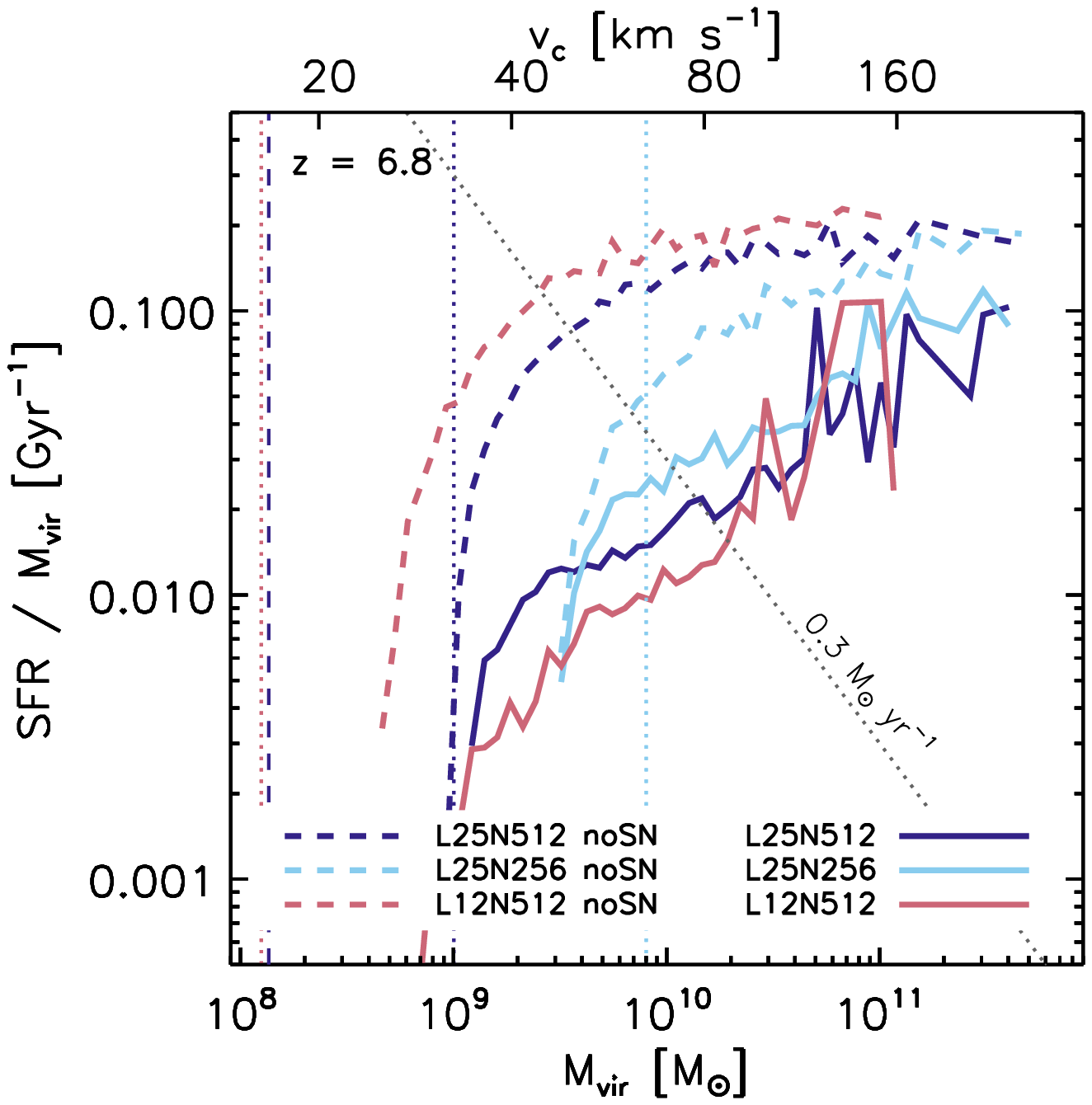}
    \includegraphics[width=0.44\textwidth,clip=true, trim=0 20 30 0,
      keepaspectratio=true]{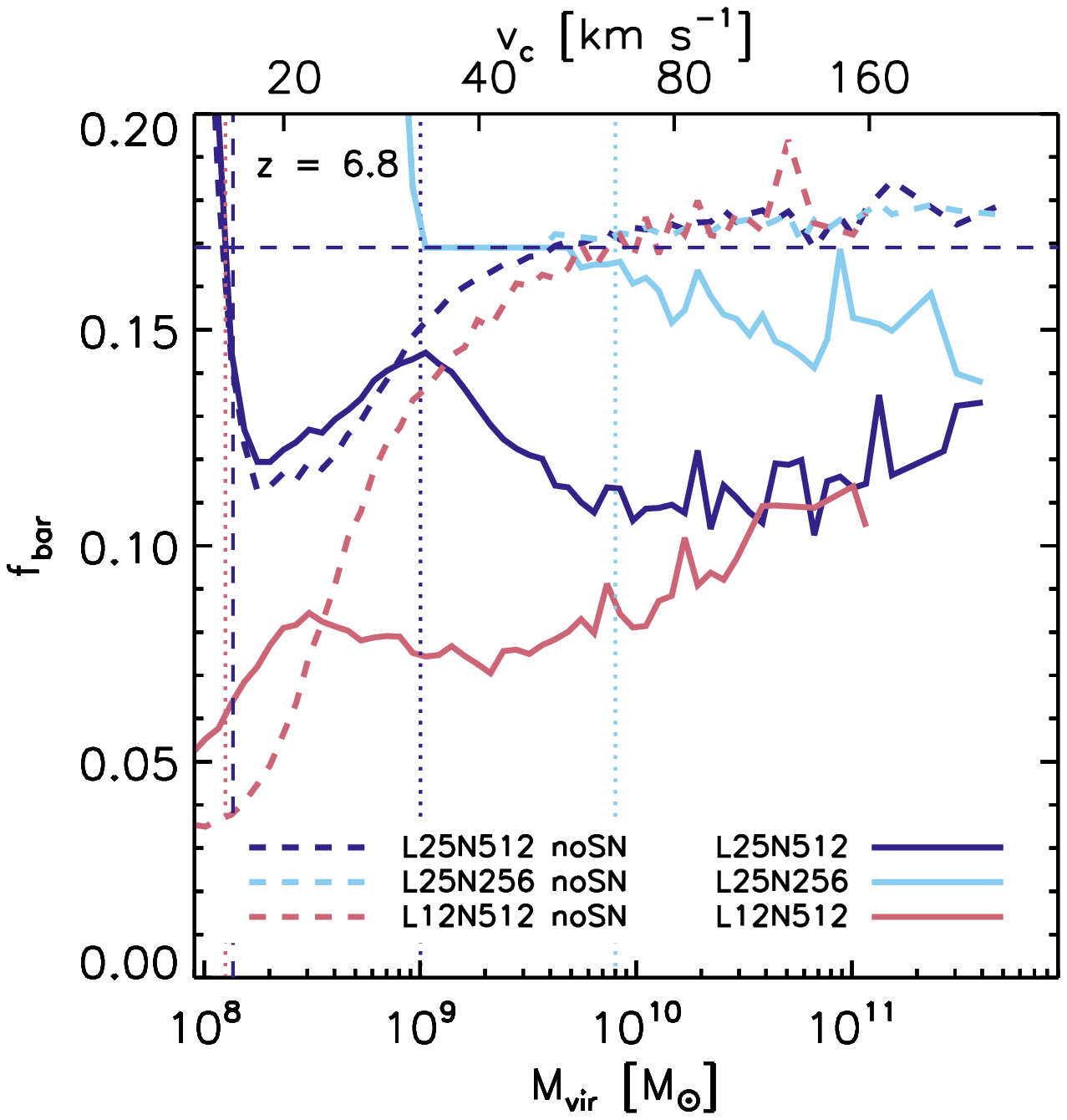}
 \end{center}
  \caption{Impact of resolution on the median SFRs (left) and the median baryonic mass fractions $f_{\rm bar}$
    (right) of galaxies of virial masses $M_{\rm vir}$ at $z \approx 7$. The meaning of the
    dashed and dotted lines is as in Figure~\ref{fig6}. }
  \label{figa1}
\end{figure*}

Here we extend the discussion of the galaxy properties in our reference
simulation in Section~\ref{sec:galprop} (Figure~\ref{fig6}) with a brief
investigation of the dependence on resolution. Figure~\ref{figa1} shows that
in our high-res simulation, in the absence of SN feedback, reionization
suppresses star formation only at masses below $\sim 5 \times 10^8
\Msun$. This is significantly less than at the reference resolution, even
though the IGM in the two simulations is reionized at similar times.  The
decrease in the suppression scale is caused by the increase in the SFRs that,
in the absence of feedback, accompanies an increase in resolution, and for
which photoheating does not entirely compensate in our simulations. However,
the high-res and reference simulations agree closely on the characteristic
scale $\sim 10^9 \Msun$ below which the median SFR is strongly suppressed by
the combined action of radiative and SN feedback. In the low-res simulation,
star formation is suppressed in haloes as massive as $5 \times 10^9 \Msun$,
primarily by the limited resolution.

\label{lastpage}


\begin{thebibliography}{99}

\bibitem[\protect\citeauthoryear{Abel, Norman, 
\& Madau}{1999}]{Abel1999a} Abel T., Norman M.~L., Madau P., 1999, ApJ, 523, 66 

\bibitem[\protect\citeauthoryear{Abel 
\& Haehnelt}{1999b}]{Abel1999b} Abel T., Haehnelt M.~G., 1999, ApJ, 520, L13 

\bibitem[\protect\citeauthoryear{Abel, Wise, 
\& Bryan}{2007}]{Abel2007} Abel T., Wise J.~H., Bryan G.~L., 2007, ApJ, 659, L87

\bibitem[\protect\citeauthoryear{Ahn et al.}{2009}]{Ahn2009} 
Ahn K., Shapiro P.~R., Iliev I.~T., Mellema G., Pen U.-L., 2009, ApJ, 695, 
1430

\bibitem[\protect\citeauthoryear{Ahn et al.}{2012}]{Ahn2012} 
Ahn K., Iliev I.~T., Shapiro P.~R., Mellema G., Koda J., Mao Y., 2012, ApJ, 
756, L16

\bibitem[\protect\citeauthoryear{Altay et al.}{2013}]{Altay2013} 
Altay G., Theuns T., Schaye J., Booth C.~M., Dalla Vecchia C., 2013, MNRAS, 
436, 2689

\bibitem[\protect\citeauthoryear{Alvarez et 
al.}{2006}]{Alvarez2006} Alvarez M.~A., Shapiro P.~R., Ahn K., 
Iliev I.~T., 2006, ApJ, 644, L101 

\bibitem[\protect\citeauthoryear{Alvarez, Wise, 
\& Abel}{2009}]{Alvarez2009} Alvarez M.~A., Wise J.~H., Abel T., 2009, ApJ, 701, L133 

\bibitem[\protect\citeauthoryear{Alvarez, Finlator, 
\& Trenti}{2012}]{Alvarez2012} Alvarez M.~A., Finlator K., Trenti M., 2012, ApJ, 759, L38 

\bibitem[\protect\citeauthoryear{Aubert \& Teyssier}{2010}]{Aubert2008} 
Aubert D., Teyssier R., 2008, MNRAS, 387, 295 

\bibitem[\protect\citeauthoryear{Aubert 
\& Teyssier}{2010}]{Aubert2010} Aubert D., Teyssier R., 2010, ApJ, 724, 244 

\bibitem[\protect\citeauthoryear{Baek, Ferrara, 
\& Semelin}{2012}]{Baek2012} Baek S., Ferrara A., Semelin B., 2012, MNRAS, 423, 774


\bibitem[\protect\citeauthoryear{Barkana 
\& Loeb}{1999}]{Barkana1999} Barkana R., Loeb A., 1999, ApJ, 523, 54 

\bibitem[\protect\citeauthoryear{Barkana 
\& Loeb}{2000}]{Barkana2000} Barkana R., Loeb A., 2000, ApJ, 539, 20 

\bibitem[\protect\citeauthoryear{Barkana 
\& Loeb}{2001}]{Barkana2001} Barkana R., Loeb A., 2001, PhR, 349, 125 

\bibitem[\protect\citeauthoryear{Barkana 
\& Loeb}{2004}]{Barkana2004} Barkana R., Loeb A., 2004, ApJ, 609, 474

\bibitem[\protect\citeauthoryear{Barkana 
\& Loeb}{2006}]{Barkana2006} Barkana R., Loeb A., 2006, MNRAS, 371, 395 

\bibitem[\protect\citeauthoryear{Bate \& Burkert}{1997}]{Bate1997} Bate M.~R., Burkert A., 1997, MNRAS, 288, 1060 

\bibitem[\protect\citeauthoryear{Becker et al.}{2014}]{Becker2014} 
Becker G.~D., Bolton J.~S., Madau P., Pettini M., Ryan-Weber E.~V., 
Venemans B.~P., 2014, arXiv, arXiv:1407.4850 

\bibitem[\protect\citeauthoryear{Benson, Venkatesan, 
\& Shull}{2013}]{Benson2013} Benson A., Venkatesan A., Shull J.~M., 2013, ApJ, 770, 76 

\bibitem[\protect\citeauthoryear{Bolton 
\& Haehnelt}{2007}]{Bolton2007} Bolton J.~S., Haehnelt M.~G., 2007, MNRAS, 382, 325 

\bibitem[\protect\citeauthoryear{Bolton 
\& Becker}{2009}]{Bolton2009} Bolton J.~S., Becker G.~D., 2009, MNRAS, 398, L26 

\bibitem[\protect\citeauthoryear{Bolton 
\& Haehnelt}{2013}]{Bolton2013} Bolton J.~S., Haehnelt M.~G., 2013, MNRAS, 429, 1695 

\bibitem[\protect\citeauthoryear{Booth 
\& Schaye}{2009}]{Booth2009} Booth C.~M., Schaye J., 2009, MNRAS, 398, 53 

\bibitem[\protect\citeauthoryear{Bouwens et al.}{2011}]{Bouwens2011} Bouwens R.~J., et al., 2011, ApJ, 737, 90 

\bibitem[\protect\citeauthoryear{Bouwens et al.}{2014a}]{Bouwens2014} Bouwens R.~J., et al., 2014, arXiv, 
arXiv:1403.4295 

\bibitem[\protect\citeauthoryear{Bouwens et 
al.}{2014b}]{Bouwens2014b} Bouwens R.~J., et al., 2014, ApJ, 795, 126 

\bibitem[\protect\citeauthoryear{Boylan-Kolchin, Bullock, 
\& Garrison-Kimmel}{2014}]{Boylan2014} Boylan-Kolchin M., Bullock J.~S., Garrison-Kimmel S., 2014, MNRAS, 443, L44 

\bibitem[\protect\citeauthoryear{Bromm 
\& Yoshida}{2011}]{Bromm2011} Bromm V., Yoshida N., 2011, ARA\&A, 49, 373 


\bibitem[\protect\citeauthoryear{Bruzual \& Charlot}{2003}]{Bruzual2003} Bruzual G., Charlot S., 2003, MNRAS, 344, 1000 

\bibitem[\protect\citeauthoryear{Cai et al.}{2014}]{Cai2014} 
Cai Z.-Y., Lapi A., Bressan A., De Zotti G., Negrello M., Danese L., 2014, 
ApJ, 785, 65

\bibitem[\protect\citeauthoryear{Cen \& Kimm}{2014}]{Cen2014} Cen R., Kimm T., 2014, ApJ, 782, 32 

\bibitem[\protect\citeauthoryear{Choudhury \& Ferrara}{2007}]{Choudhury2007} 
Choudhury T.~R., Ferrara A., 2007, MNRAS, 380, L6

\bibitem[\protect\citeauthoryear{Ciardi, Stoehr, 
\& White}{2003}]{Ciardi2003} Ciardi B., Stoehr F., White S.~D.~M., 2003, MNRAS, 343, 1101 

\bibitem[\protect\citeauthoryear{Ciardi 
\& Ferrara}{2005}]{Ciardi2005} Ciardi B., Ferrara A., 2005, SSRv, 116, 625


\bibitem[\protect\citeauthoryear{Ciardi et al.}{2006}]{Ciardi2006} 
Ciardi B., Scannapieco E., Stoehr F., Ferrara A., Iliev I.~T., Shapiro 
P.~R., 2006, MNRAS, 366, 689

\bibitem[\protect\citeauthoryear{Ciardi et al.}{2012}]{Ciardi2012} 
Ciardi B., Bolton J.~S., Maselli A., Graziani L., 2012, MNRAS, 423, 558


\bibitem[\protect\citeauthoryear{Coe et al.}{2013}]{Coe2013} 
Coe D., et al., 2013, ApJ, 762, 32

\bibitem[\protect\citeauthoryear{Dalla Vecchia 
\& Schaye}{2012}]{DallaVecchia2012} Dalla Vecchia C., Schaye J., 2012, MNRAS, 426, 140 

\bibitem[\protect\citeauthoryear{Dalla Vecchia 
\& Schaye}{2008}]{DallaVecchia2008} Dalla Vecchia C., Schaye J., 2008, MNRAS, 387, 1431 

\bibitem[\protect\citeauthoryear{Dekel 
\& Silk}{1986}]{Dekel1986} Dekel A., Silk J., 1986, ApJ, 303, 39 

\bibitem[\protect\citeauthoryear{Dijkstra et 
al.}{2004}]{Dijkstra2004} Dijkstra M., Haiman Z., Rees M.~J., 
Weinberg D.~H., 2004, ApJ, 601, 666

\bibitem[\protect\citeauthoryear{Dijkstra, Mesinger, 
\& Wyithe}{2011}]{Dijkstra2011} Dijkstra M., Mesinger A., Wyithe J.~S.~B., 2011, MNRAS, 414, 2139 

\bibitem[\protect\citeauthoryear{Dijkstra}{2014}]{Dijkstra2014} 
Dijkstra M., 2014, arXiv, arXiv:1406.7292 

\bibitem[\protect\citeauthoryear{Di Matteo, Springel, 
\& Hernquist}{2005}]{diMatteo2005} Di Matteo T., Springel V., Hernquist L., 2005, Natur, 433, 604 

\bibitem[\protect\citeauthoryear{Dolag et al.}{2009}]{Dolag2009} 
Dolag K., Borgani S., Murante G., Springel V., 2009, MNRAS, 399, 497 

\bibitem[\protect\citeauthoryear{Dove, Shull, 
\& Ferrara}{2000}]{Dove2000} Dove J.~B., Shull J.~M., Ferrara A.,
  2000, ApJ, 531, 846 


\bibitem[\protect\citeauthoryear{Duffy et al.}{2014}]{Duffy2014} 
Duffy A.~R., Wyithe J.~S.~B., Mutch S.~J., Poole G.~B., 2014, arXiv, 
arXiv:1405.7459 

\bibitem[\protect\citeauthoryear{Duncan et al.}{2014}]{Duncan2014} 
Duncan K., et al., 2014, MNRAS, 444, 2960

\bibitem[\protect\citeauthoryear{Ellis et al.}{2013}]{Ellis2013} 
Ellis R.~S., et al., 2013, ApJ, 763, L7

\bibitem[\protect\citeauthoryear{Emberson, Thomas, 
\& Alvarez}{2013}]{Emberson2013} Emberson J.~D., Thomas R.~M., Alvarez M.~A., 2013, ApJ, 763, 146

\bibitem[\protect\citeauthoryear{Fan, Carilli, 
\& Keating}{2006}]{Fanreview2006} Fan X., Carilli C.~L., Keating B., 2006, ARA\&A, 44, 415

\bibitem[\protect\citeauthoryear{Fan et al.}{2006}]{Fan2006} 
Fan X., et al., 2006, AJ, 132, 117

\bibitem[\protect\citeauthoryear{Ferrara 
\& Loeb}{2013}]{Ferrara2013} Ferrara A., Loeb A., 2013, MNRAS, 431, 2826 


\bibitem[\protect\citeauthoryear{Fontanot, Cristiani, 
\& Vanzella}{2012}]{Fontanot2012} Fontanot F., Cristiani S., Vanzella E., 2012, MNRAS, 425, 1413

\bibitem[\protect\citeauthoryear{Finkelstein et 
al.}{2012}]{Finkelstein2012} Finkelstein S.~L., et al., 2012, ApJ, 758, 
93

\bibitem[\protect\citeauthoryear{Finkelstein et 
al.}{2014}]{Finkelstein2014} Finkelstein S.~L., et al., 2014, arXiv, 
arXiv:1410.5439 

\bibitem[\protect\citeauthoryear{Finlator et 
al.}{2009}]{Finlator2009} Finlator K., {\"O}zel F., Dav{\'e} R., 
Oppenheimer B.~D., 2009, MNRAS, 400, 1049

\bibitem[\protect\citeauthoryear{Finlator, Dav{\'e}, {\"O}zel}{2011}]{Finlator2011} 
Finlator K., Dav{\'e} R., {\"O}zel F., 2011, ApJ, 743, 169 

\bibitem[\protect\citeauthoryear{Finlator}{2012}]{Finlator2012} 
Finlator K., 2012, arXiv, arXiv:1203.4862

\bibitem[\protect\citeauthoryear{Finlator et 
al.}{2012}]{Finlatorclump2012} Finlator K., Oh S.~P., {\"O}zel F., 
Dav{\'e} R., 2012, MNRAS, 427, 2464

\bibitem[\protect\citeauthoryear{Fujita et al.}{2003}]{Fujita2003} 
Fujita A., Martin C.~L., Mac Low M.-M., Abel T., 2003, ApJ, 599, 50

\bibitem[\protect\citeauthoryear{Furlanetto 
\& Oh}{2005}]{Furlanetto2005} Furlanetto S.~R., Oh S.~P., 2005, MNRAS, 363, 1031 

\bibitem[\protect\citeauthoryear{Furlanetto, McQuinn, 
\& Hernquist}{2006}]{Furlanetto2006} Furlanetto S.~R., McQuinn M., Hernquist L., 2006, MNRAS, 365, 115

\bibitem[\protect\citeauthoryear{Furlanetto, Oh, 
\& Briggs}{2006}]{FOB2006} Furlanetto S.~R., Oh S.~P., Briggs F.~H., 2006, PhR, 433, 181

\bibitem[\protect\citeauthoryear{Geen, Slyz, 
\& Devriendt}{2013}]{Geen2013} Geen S., Slyz A., Devriendt J., 2013, MNRAS, 429, 633 

\bibitem[\protect\citeauthoryear{Gnedin 
\& Ostriker}{1997}]{Gnedin1997} Gnedin N.~Y., Ostriker J.~P., 1997, ApJ, 486, 581

\bibitem[\protect\citeauthoryear{Gnedin}{2000}]{Gnedinreion2000} Gnedin 
N.~Y., 2000, ApJ, 535, 530 

\bibitem[\protect\citeauthoryear{Gnedin}{2000b}]{Gnedin2000} Gnedin 
N.~Y., 2000b, ApJ, 542, 535

\bibitem[\protect\citeauthoryear{Gnedin 
\& Abel}{2001}]{Gnedin2001} Gnedin N.~Y., Abel T., 2001, NewA, 6, 437

\bibitem[\protect\citeauthoryear{Gnedin 
\& Fan}{2006}]{Gnedin2006} Gnedin N.~Y., Fan X., 2006, ApJ, 648, 1

\bibitem[\protect\citeauthoryear{Gnedin, Kravtsov, 
\& Chen}{2008}]{Gnedin2008} Gnedin N.~Y., Kravtsov A.~V., Chen H.-W., 2008, ApJ, 672, 765 


\bibitem[\protect\citeauthoryear{Gnedin}{2010}]{Gnedin2010} Gnedin 
N.~Y., 2010, ApJ, 721, L79 

\bibitem[\protect\citeauthoryear{Gnedin}{2014}]{Gnedin2014} Gnedin 
N.~Y., 2014, ApJ, 793, 29 

\bibitem[\protect\citeauthoryear{Gnedin 
\& Kaurov}{2014}]{Gnedin2014b} Gnedin N.~Y., Kaurov A.~A., 2014, ApJ, 793, 30 

\bibitem[\protect\citeauthoryear{Greif \& Bromm}{2006}]{Greif2006} 
Greif T.~H., Bromm V., 2006, MNRAS, 373, 128 

\bibitem[\protect\citeauthoryear{Greif et al.}{2008}]{Greif2008} 
Greif T.~H., Johnson J.~L., Klessen R.~S., Bromm V., 2008, MNRAS, 387, 1021

\bibitem[\protect\citeauthoryear{Haiman, Rees, 
\& Loeb}{1997}]{Haiman1997} Haiman Z., Rees M.~J., Loeb A., 1997, ApJ, 476, 458


\bibitem[\protect\citeauthoryear{Haiman}{2011}]{Haiman2011} Haiman 
Z., 2011, Natur, 472, 47 

\bibitem[\protect\citeauthoryear{Hambrick et 
al.}{2011}]{Hambrick2011} Hambrick D.~C., Ostriker J.~P., Johansson 
P.~H., Naab T., 2011, MNRAS, 413, 2421

\bibitem[\protect\citeauthoryear{Hasegawa 
\& Umemura}{2010}]{Hasegawa2010} Hasegawa K., Umemura M., 2010, MNRAS, 407, 2632 

\bibitem[\protect\citeauthoryear{Hasegawa 
\& Semelin}{2013}]{Hasegawa2013} Hasegawa K., Semelin B., 2013, MNRAS, 428, 154 

\bibitem[\protect\citeauthoryear{Hopkins et 
al.}{2013}]{Hopkins2013} Hopkins P.~F., Keres D., Onorbe J., 
Faucher-Giguere C.-A., Quataert E., Murray N., Bullock J.~S., 2013, arXiv, 
arXiv:1311.2073 

\bibitem[\protect\citeauthoryear{Iliev, Scannapieco, 
\& Shapiro}{2005}]{Iliev2005} Iliev I.~T., Scannapieco E., Shapiro P.~R., 2005, ApJ, 624, 491 

\bibitem[\protect\citeauthoryear{Iliev et al.}{2006}]{Iliev2006} 
Iliev I.~T., et al., 2006, MNRAS, 371, 1057

\bibitem[\protect\citeauthoryear{Iliev et al.}{2006}]{Ilievreion2006} 
Iliev I.~T., Mellema G., Pen U.-L., Merz H., Shapiro P.~R., Alvarez M.~A., 
2006, MNRAS, 369, 1625 

\bibitem[\protect\citeauthoryear{Iliev et al.}{2007}]{Iliev2007} 
Iliev I.~T., Mellema G., Shapiro P.~R., Pen U.-L., 2007, MNRAS, 376, 534 

\bibitem[\protect\citeauthoryear{Iliev et al.}{2009}]{Iliev2009} 
Iliev I.~T., et al., 2009, MNRAS, 400, 1283 

\bibitem[\protect\citeauthoryear{Iliev et al.}{2014}]{Iliev2014} 
Iliev I.~T., Mellema G., Ahn K., Shapiro P.~R., Mao Y., Pen U.-L., 2014, 
MNRAS, 439, 725 

\bibitem[\protect\citeauthoryear{Jaacks et al.}{2012}]{Jaacks2012} 
Jaacks J., Choi J.-H., Nagamine K., Thompson R., Varghese S., 2012, MNRAS, 
420, 1606

\bibitem[\protect\citeauthoryear{Jappsen et 
al.}{2009}]{Jappsen2009} Jappsen A.-K., Klessen R.~S., Glover 
S.~C.~O., Mac Low M.-M., 2009, ApJ, 696, 1065 

\bibitem[\protect\citeauthoryear{Jensen et al.}{2013}]{Jensen2013} 
Jensen H., Laursen P., Mellema G., Iliev I.~T., Sommer-Larsen J., Shapiro 
P.~R., 2013, MNRAS, 428, 1366 


\bibitem[\protect\citeauthoryear{Jeon et al.}{2014a}]{Jeon2014} 
Jeon M., Pawlik A.~H., Bromm V., Milosavljevi{\'c} M., 2014, MNRAS, 440, 
3778 

\bibitem[\protect\citeauthoryear{Jeon et al.}{2014b}]{Jeon2014b} 
Jeon M., Pawlik A.~H., Bromm V., Milosavljevic M., 2014, 2014, MNRAS, 444, 

\bibitem[\protect\citeauthoryear{Jeon et al.}{2015}]{Jeon2015} 
Jeon M., Bromm V., Pawlik A.~H., Milosavljevi{\'c} M., 2015, submitted (arXiv:1501.01002)

\bibitem[\protect\citeauthoryear{Johnson et 
al.}{2009}]{Johnson2009} Johnson J.~L., Greif T.~H., Bromm V., 
Klessen R.~S., Ippolito J., 2009, MNRAS, 399, 37

\bibitem[\protect\citeauthoryear{Johnson 
\& Khochfar}{2011}]{Johnson2011} Johnson J.~L., Khochfar S., 2011, ApJ, 743, 126 

\bibitem[\protect\citeauthoryear{Kaurov 
\& Gnedin}{2014}]{Kaurov2014} Kaurov A.~A., Gnedin N.~Y., 2014, ApJ,
  787, 146 

\bibitem[\protect\citeauthoryear{Kennicutt}{1998}]{Kennicutt1998} 
Kennicutt R.~C., Jr., 1998, ApJ, 498, 541 

\bibitem[\protect\citeauthoryear{Kimm 
\& Cen}{2014}]{Kimm2014} Kimm T., Cen R., 2014, ApJ, 788, 121 

\bibitem[\protect\citeauthoryear{Knevitt et 
al.}{2014}]{Knevitt2014} Knevitt G., Wynn G.~A., Power C., Bolton 
J.~S., 2014, MNRAS, 445, 2034 

\bibitem[\protect\citeauthoryear{Kohler, Gnedin, 
\& Hamilton}{2007}]{Kohler2007} Kohler K., Gnedin N.~Y., Hamilton A.~J.~S., 2007, ApJ, 657, 15 

\bibitem[\protect\citeauthoryear{Komatsu et al.}{2011}]{Komatsu2011} Komatsu E., et al., 2011, ApJS, 192, 18 

\bibitem[\protect\citeauthoryear{Kuhlen 
\& Faucher-Gigu{\`e}re}{2012}]{Kuhlen2012} Kuhlen M., Faucher-Gigu{\`e}re C.-A., 2012, MNRAS, 423, 862 

\bibitem[\protect\citeauthoryear{Kuhlen, Madau, 
\& Krumholz}{2013}]{Kuhlen2013} Kuhlen M., Madau P., Krumholz M.~R., 2013, ApJ, 776, 34 

\bibitem[\protect\citeauthoryear{Lewis \& Bridle}{2002}]{Lewis2002} Lewis A., Bridle S., 2002, PhRvD, 66, 103511 

\bibitem[\protect\citeauthoryear{Lidz et al.}{2008}]{Lidz2008} 
Lidz A., Zahn O., McQuinn M., Zaldarriaga M., Hernquist L., 2008, ApJ, 680, 
962

\bibitem[\protect\citeauthoryear{Loeb}{2009}]{Loeb2009} Loeb A., 
2009, JCAP, 3, 22

\bibitem[\protect\citeauthoryear{Loeb}{2010}]{Loeb2010} Loeb A., 2010, hdfs.book,  

\bibitem[\protect\citeauthoryear{Madau, Haardt, 
\& Rees}{1999}]{MHR1999} Madau P., Haardt F., Rees M.~J., 1999, ApJ, 514, 648

\bibitem[\protect\citeauthoryear{Madau et al.}{2004}]{Madau2004} 
Madau P., Rees M.~J., Volonteri M., Haardt F., Oh S.~P., 2004, ApJ, 604, 484

\bibitem[\protect\citeauthoryear{Maio et al.}{2010}]{Maio2010} 
Maio U., Ciardi B., Dolag K., Tornatore L., Khochfar S., 2010, MNRAS, 407, 
1003 

\bibitem[\protect\citeauthoryear{Maselli, Ferrara, \&
    Ciardi}{2003}]{Maselli2003} Maselli A., Ferrara A., Ciardi B., 2003, MNRAS, 345, 379 

\bibitem[\protect\citeauthoryear{Maselli, Ciardi, 
\& Kanekar}{2009}]{Maselli2009} Maselli A., Ciardi B., Kanekar A., 2009, MNRAS, 393, 171 

\bibitem[\protect\citeauthoryear{Mashian 
\& Loeb}{2013}]{Mashian2013} Mashian N., Loeb A., 2013, JCAP, 12, 17 

\bibitem[\protect\citeauthoryear{McGreer, Mesinger, 
\& Fan}{2011}]{McGreer2011} McGreer I.~D., Mesinger A., Fan X., 2011, MNRAS, 415, 3237 

\bibitem[\protect\citeauthoryear{McLeod et al.}{2014}]{McLeod2014} 
McLeod D.~J., McLure R.~J., Dunlop J.~S., Robertson B.~E., Ellis R.~S., 
Targett T.~T., 2014, arXiv, arXiv:1412.1472 

\bibitem[\protect\citeauthoryear{McQuinn et 
al.}{2007}]{McQuinn2007} McQuinn M., Lidz A., Zahn O., Dutta S., 
Hernquist L., Zaldarriaga M., 2007, MNRAS, 377, 1043 

\bibitem[\protect\citeauthoryear{Mellema et 
al.}{2006}]{Mellema2006} Mellema G., Iliev I.~T., Alvarez M.~A., 
Shapiro P.~R., 2006, NewA, 11, 374 

\bibitem[\protect\citeauthoryear{Mellema et 
al.}{2013}]{Mellema2013} Mellema G., et al., 2013, ExA, 36, 235

\bibitem[\protect\citeauthoryear{Mesinger, Furlanetto, 
\& Cen}{2011}]{Mesinger2011} Mesinger A., Furlanetto S., Cen R., 2011, MNRAS, 411, 955 

\bibitem[\protect\citeauthoryear{Mesinger et 
al.}{2014}]{Mesinger2014} Mesinger A., Aykutalp A., Vanzella E., 
Pentericci L., Ferrara A., Dijkstra M., 2014, arXiv, arXiv:1406.6373 

\bibitem[\protect\citeauthoryear{Miralda-Escud{\'e}, Haehnelt, 
\& Rees}{2000}]{Miralda2000} Miralda-Escud{\'e} J., Haehnelt M., Rees M.~J., 2000, ApJ, 530, 1

\bibitem[\protect\citeauthoryear{Mitra, Ferrara, 
\& Choudhury}{2013}]{Mitra2013} Mitra S., Ferrara A., Choudhury T.~R., 2013, MNRAS, 428, L1

\bibitem[\protect\citeauthoryear{Morales 
\& Wyithe}{2010}]{Morales2010} Morales M.~F., Wyithe J.~S.~B., 2010, ARA\&A, 48, 127 

\bibitem[\protect\citeauthoryear{Mu{\~n}oz 
\& Loeb}{2011}]{Munoz2011} Mu{\~n}oz J.~A., Loeb A., 2011, ApJ, 729, 99

\bibitem[\protect\citeauthoryear{Muratov et 
al.}{2013}]{Muratov2013} Muratov A.~L., Gnedin O.~Y., Gnedin N.~Y., 
Zemp M., 2013, ApJ, 773, 19 

\bibitem[\protect\citeauthoryear{Nagamine, Choi, 
\& Yajima}{2010}]{Nagamine2010} Nagamine K., Choi J.-H., Yajima H., 2010, ApJ, 725, L219

\bibitem[\protect\citeauthoryear{Nakamoto, Umemura, 
\& Susa}{2001}]{Nakamoto2001} Nakamoto T., Umemura M., Susa H., 2001, MNRAS, 321, 593 

\bibitem[\protect\citeauthoryear{Natarajan 
\& Yoshida}{2014}]{Natarajan2014} Natarajan A., Yoshida N., 2014, PTEP, 2014, 6112 

\bibitem[\protect\citeauthoryear{Noh 
\& McQuinn}{2014}]{Noh2014} Noh Y., McQuinn M., 2014, arXiv, arXiv:1401.0737 

\bibitem[\protect\citeauthoryear{Norman et al.}{2013}]{Norman2013} 
Norman M.~L., Reynolds D.~R., So G.~C., Harkness R.~P., 2013, arXiv, 
arXiv:1306.0645

\bibitem[\protect\citeauthoryear{Oesch et al.}{2013}]{Oesch2013} 
Oesch P.~A., et al., 2013, ApJ, 773, 75 

\bibitem[\protect\citeauthoryear{Oesch et al.}{2014}]{Oesch2014} 
Oesch P.~A., et al., 2014, ApJ, 786, 108

\bibitem[\protect\citeauthoryear{Okamoto, Gao, 
\& Theuns}{2008}]{Okamoto2008} Okamoto T., Gao L., Theuns T., 2008, MNRAS, 390, 920 

\bibitem[\protect\citeauthoryear{Okamoto, Yoshikawa, 
\& Umemura}{2012}]{Okamoto2012} Okamoto T., Yoshikawa K., Umemura M., 2012, MNRAS, 419, 2855 

\bibitem[\protect\citeauthoryear{Osterbrock 
\& Ferland}{2006}]{Osterbrock2006} Osterbrock D.~E., Ferland G.~J., 2006, agna.book,  

\bibitem[\protect\citeauthoryear{Ota et al.}{2008}]{Ota2008} 
Ota K., et al., 2008, ApJ, 677, 12 


\bibitem[\protect\citeauthoryear{Ouchi et al.}{2010}]{Ouchi2010} 
Ouchi M., et al., 2010, ApJ, 723, 869

\bibitem[\protect\citeauthoryear{Paardekooper et 
al.}{2011}]{Paardekooper2011} Paardekooper J.-P., Pelupessy F.~I., Altay G., Kruip C.~J.~H., 2011, A\&A, 530, A87 

\bibitem[\protect\citeauthoryear{Paardekooper, Khochfar, 
\& Dalla Vecchia}{2013}]{Paardekooper2013} Paardekooper J.-P., Khochfar S., Dalla Vecchia C., 2013, MNRAS, 429, L94 

\bibitem[\protect\citeauthoryear{Pawlik \& Schaye}{2008}]{Pawlik2008} Pawlik A.~H., Schaye J., 2008, MNRAS, 389, 651 

\bibitem[\protect\citeauthoryear{Pawlik 
\& Schaye}{2009}]{Pawlik2009} Pawlik A.~H., Schaye J., 2009, MNRAS, 396, L46

\bibitem[\protect\citeauthoryear{Pawlik, Schaye, 
\& van Scherpenzeel}{2009}]{Pawlikclump2009} Pawlik A.~H., Schaye J., van Scherpenzeel E., 2009, MNRAS, 394, 1812 


\bibitem[\protect\citeauthoryear{Pawlik \& Schaye}{2011}]{Pawlik2011} Pawlik A.~H., Schaye J., 2011, MNRAS, 412, 1943 

\bibitem[\protect\citeauthoryear{Pawlik, Milosavljevi{\'c}, 
\& Bromm}{2011}]{PawlikBromm2011} Pawlik A.~H., Milosavljevi{\'c} M., Bromm V., 2011, ApJ, 731, 54 

\bibitem[\protect\citeauthoryear{Pawlik, Milosavljevi{\'c}, 
\& Bromm}{2013}]{Pawlik2013} Pawlik A.~H., Milosavljevi{\'c} M., Bromm V., 2013, ApJ, 767, 59 

\bibitem[\protect\citeauthoryear{Petkova 
\& Springel}{2011}]{Petkova2011} Petkova M., Springel V., 2011, MNRAS, 412, 935 

\bibitem[\protect\citeauthoryear{Planck Collaboration}{2014a}]{Planck2014} Planck Collaboration, 2014, A\&A, 571, AA16 

\bibitem[\protect\citeauthoryear{Planck Collaboration et al.}{2015}]{Planck2015} Planck Collaboration, 2015, arXiv, arXiv:1502.01589 

\bibitem[\protect\citeauthoryear{Pritchard 
\& Loeb}{2012}]{Pritchard2012} Pritchard J.~R., Loeb A., 2012, RPPh, 75, 086901 

\bibitem[\protect\citeauthoryear{Rahmati et 
al.}{2013a}]{Rahmati12013} Rahmati A., Pawlik A.~H., Rai{\v c}evi{\'c} M., Schaye J., 2013, MNRAS, 430, 2427

\bibitem[\protect\citeauthoryear{Rahmati et 
al.}{2013b}]{Rahmati2013} Rahmati A., Schaye J., Pawlik A.~H., 
Rai{\v c}evi{\'c} M., 2013, MNRAS, 431, 2261 

\bibitem[\protect\citeauthoryear{Rai{\v c}evi{\'c}, Theuns, 
\& Lacey}{2011}]{Raicevic2011} Rai{\v c}evi{\'c} M., Theuns T., Lacey
  C., 2011, MNRAS, 410, 775 

\bibitem[\protect\citeauthoryear{Rai{\v c}evi{\'c} et 
al.}{2014}]{Raicevic2014} Rai{\v c}evi{\'c} M., Pawlik A.~H., Schaye 
J., Rahmati A., 2014, MNRAS, 437, 2816 


\bibitem[\protect\citeauthoryear{Razoumov et 
al.}{2002}]{Razoumov2002} Razoumov A.~O., Norman M.~L., Abel T., 
Scott D., 2002, ApJ, 572, 695

\bibitem[\protect\citeauthoryear{Razoumov 
\& Sommer-Larsen}{2006}]{Razoumov2006} Razoumov A.~O., Sommer-Larsen J., 2006, ApJ, 651, L89

\bibitem[\protect\citeauthoryear{Rees}{1986}]{Rees1986} Rees 
M.~J., 1986, MNRAS, 218, 25P 

\bibitem[\protect\citeauthoryear{Ricotti, Gnedin, 
\& Shull}{2001}]{Ricotti2001} Ricotti M., Gnedin N.~Y., Shull J.~M., 2001, ApJ, 560, 580

\bibitem[\protect\citeauthoryear{Ricotti 
\& Ostriker}{2004b}]{Ricotti2004b} Ricotti M., Ostriker J.~P., 2004b, MNRAS, 350, 539 

\bibitem[\protect\citeauthoryear{Ricotti 
\& Ostriker}{2004}]{Ricotti2004} Ricotti M., Ostriker J.~P., 2004, MNRAS, 352, 547 

\bibitem[\protect\citeauthoryear{Ricotti, Ostriker, 
\& Gnedin}{2005}]{Ricotti2005} Ricotti M., Ostriker J.~P., Gnedin N.~Y., 2005, MNRAS, 357, 207

\bibitem[\protect\citeauthoryear{Ricotti 
\& Gnedin}{2005}]{Ricotti2005a} Ricotti M., Gnedin N.~Y., 2005, ApJ, 629, 259 

\bibitem[\protect\citeauthoryear{Ricotti, Gnedin, 
\& Shull}{2008}]{Ricotti2008} Ricotti M., Gnedin N.~Y., Shull J.~M., 2008, ApJ, 685, 21


\bibitem[\protect\citeauthoryear{Ricotti}{2010}]{Ricotti2010} 
Ricotti M., 2010, AdAst, 2010

\bibitem[\protect\citeauthoryear{Robertson et 
al.}{2010}]{Robertson2010} Robertson B.~E., Ellis R.~S., Dunlop 
J.~S., McLure R.~J., Stark D.~P., 2010, Natur, 468, 49 

\bibitem[\protect\citeauthoryear{Robertson et al.}{2013}]{Robertson2013} Robertson B.~E., et al., 2013, ApJ, 768, 71 

\bibitem[\protect\citeauthoryear{Rosdahl et 
al.}{2013}]{Rosdahl2013} Rosdahl J., Blaizot J., Aubert D., Stranex 
T., Teyssier R., 2013, MNRAS, 436, 2188 

\bibitem[\protect\citeauthoryear{Salpeter}{1955}]{Salpeter1955} 
Salpeter E.~E., 1955, ApJ, 121, 161 

\bibitem[\protect\citeauthoryear{Sawala et al.}{2014}]{Sawala2014} 
Sawala T., et al., 2014, arXiv, arXiv:1404.3724 

\bibitem[\protect\citeauthoryear{Scannapieco et 
al.}{2012}]{Scannapieco2012} Scannapieco C., et al., 2012, MNRAS, 423, 
1726 

\bibitem[\protect\citeauthoryear{Schaye}{2001}]{Schaye2001} Schaye 
J., 2001, ApJ, 559, 507 

\bibitem[\protect\citeauthoryear{Schaerer}{2003}]{Schaerer2003} Schaerer D., 2003, A\&A, 397, 527 

\bibitem[\protect\citeauthoryear{Schaye}{2004}]{Schaye2004} Schaye J., 2004, ApJ, 609, 667 

\bibitem[\protect\citeauthoryear{Schaye \& Dalla Vecchia}{2008}]{Schaye2008} Schaye J., Dalla Vecchia C., 2008, MNRAS, 383, 1210 

\bibitem[\protect\citeauthoryear{Schaye et al.}{2010}]{Schaye2010} Schaye J., et al., 2010, MNRAS, 402, 1536 

\bibitem[\protect\citeauthoryear{Schaye et al.}{2015}]{Schaye2014}  
Schaye J., et al., 2015, MNRAS, 446, 521 

\bibitem[\protect\citeauthoryear{Schroeder, Mesinger, 
\& Haiman}{2013}]{Schroeder2013} Schroeder J., Mesinger A., Haiman Z., 2013, MNRAS, 428, 3058

\bibitem[\protect\citeauthoryear{Shapiro, Giroux, 
\& Babul}{1994}]{Shapiro1994} Shapiro P.~R., Giroux M.~L., Babul A., 1994, ApJ, 427, 25 

\bibitem[\protect\citeauthoryear{Shull 
\& Venkatesan}{2008}]{Shull2008} Shull J.~M., Venkatesan A., 2008, ApJ, 685, 1 

\bibitem[\protect\citeauthoryear{Shull et al.}{2012}]{Shull2012} 
Shull J.~M., Harness A., Trenti M., Smith B.~D., 2012, ApJ, 747, 100

\bibitem[\protect\citeauthoryear{Smit et al.}{2012}]{Smit2012} Smit R., Bouwens R.~J., Franx M., Illingworth G.~D., Labb{\'e} I., Oesch P.~A., van Dokkum P.~G., 2012, ApJ, 756, 14 

\bibitem[\protect\citeauthoryear{So et al.}{2014}]{So2014} So 
G.~C., Norman M.~L., Reynolds D.~R., Wise J.~H., 2014, ApJ, 789, 149

\bibitem[\protect\citeauthoryear{Sobacchi 
\& Mesinger}{2014}]{Sobacchi2014} Sobacchi E., Mesinger A., 2014, MNRAS, 440, 1662

\bibitem[\protect\citeauthoryear{Springel et 
al.}{2001}]{Springel2001} Springel V., White S.~D.~M., Tormen G., 
Kauffmann G., 2001, MNRAS, 328, 726

\bibitem[\protect\citeauthoryear{Springel \& Hernquist}{2002}]{Springel2002} Springel V., Hernquist L., 2002, MNRAS, 333, 649 

\bibitem[\protect\citeauthoryear{Springel}{2005}]{Springel2005} Springel V., 2005, MNRAS, 364, 1105 

\bibitem[\protect\citeauthoryear{Stiavelli, Fall, 
\& Panagia}{2004}]{Stiavelli2004} Stiavelli M., Fall S.~M., Panagia N., 2004, ApJ, 610, L1 

\bibitem[\protect\citeauthoryear{Tanaka et al.}{2014}]{Tanaka2014} 
Tanaka S., Yoshikawa K., Okamoto T., Hasegawa K., 2014, arXiv, 
arXiv:1410.0763 

\bibitem[\protect\citeauthoryear{Taylor 
\& Lidz}{2014}]{Taylor2014} Taylor J., Lidz A., 2014, MNRAS, 437, 2542 

\bibitem[\protect\citeauthoryear{Thoul 
\& Weinberg}{1996}]{Thoul1996} Thoul A.~A., Weinberg D.~H., 1996, ApJ, 465, 608 

\bibitem[\protect\citeauthoryear{Trac 
\& Cen}{2007}]{Trac2007} Trac H., Cen R., 2007, ApJ, 671, 1 

\bibitem[\protect\citeauthoryear{Trac 
\& Gnedin}{2011}]{Trac2011} Trac H.~Y., Gnedin N.~Y., 2011, ASL, 4, 228 

\bibitem[\protect\citeauthoryear{Trenti et al.}{2010}]{Trenti2010} 
Trenti M., Stiavelli M., Bouwens R.~J., Oesch P., Shull J.~M., Illingworth 
G.~D., Bradley L.~D., Carollo C.~M., 2010, ApJ, 714, L202 

\bibitem[\protect\citeauthoryear{Verner et al.}{1996}]{Verner1996} 
Verner D.~A., Ferland G.~J., Korista K.~T., Yakovlev D.~G., 1996, ApJ, 465, 
487 

\bibitem[\protect\citeauthoryear{Vogelsberger et 
al.}{2013}]{Vogelsberger2013} Vogelsberger M., Genel S., Sijacki D., 
Torrey P., Springel V., Hernquist L., 2013, MNRAS, 436, 3031

\bibitem[\protect\citeauthoryear{Volonteri 
\& Gnedin}{2009}]{Volenteri2009} Volonteri M., Gnedin N.~Y., 2009, ApJ, 703, 2113 

\bibitem[\protect\citeauthoryear{White}{1996}]{White1996} 
White S. D. M., 1996, in Schaeffer R., Silk J., Spiro M., \& Zinn-
Justin J. ed., Cosmology and Large Scale Structure Formation
and Evolution of Galaxies, p. 349

\bibitem[\protect\citeauthoryear{Wiersma, Schaye, 
\& Smith}{2009}]{Wiersma2009} Wiersma R.~P.~C., Schaye J., Smith B.~D., 2009, MNRAS, 393, 99 

\bibitem[\protect\citeauthoryear{Wise 
\& Abel}{2005}]{Wise2005} Wise J.~H., Abel T., 2005, ApJ, 629, 615

\bibitem[\protect\citeauthoryear{Wise 
\& Abel}{2008}]{Wise2008} Wise J.~H., Abel T., 2008, ApJ, 685, 40

\bibitem[\protect\citeauthoryear{Wise, Turk, 
\& Abel}{2008}]{Wise2008b} Wise J.~H., Turk M.~J., Abel T., 2008, ApJ, 682, 745 

\bibitem[\protect\citeauthoryear{Wise 
\& Cen}{2009}]{Wise2009} Wise J.~H., Cen R., 2009, ApJ, 693, 984

\bibitem[\protect\citeauthoryear{Wise et al.}{2014}]{Wise2014} 
Wise J.~H., Demchenko V.~G., Halicek M.~T., Norman M.~L., Turk M.~J., Abel 
T., Smith B.~D., 2014, arXiv, arXiv:1403.6123 

\bibitem[\protect\citeauthoryear{Wolcott-Green, Haiman, \&
    Bryan}{2011}]{Wolcott2011} Wolcott-Green J., Haiman Z., Bryan
  G.~L., 2011, MNRAS, 418, 838 

\bibitem[\protect\citeauthoryear{Wyithe \& Bolton}{2011}]{Wyithe2011} Wyithe J.~S.~B., Bolton J.~S., 2011, MNRAS, 412, 1926 

\bibitem[\protect\citeauthoryear{Wyithe 
\& Loeb}{2013}]{Wyithe2013} Wyithe J.~S.~B., Loeb A., 2013, MNRAS, 428, 2741 

\bibitem[\protect\citeauthoryear{Xu et al.}{2014}]{Xu2014} Xu 
H., Ahn K., Wise J.~H., Norman M.~L., O'Shea B.~W., 2014, ApJ, 791, 110 

\bibitem[\protect\citeauthoryear{Yajima et al.}{2009}]{Yajima2009} 
Yajima H., Umemura M., Mori M., Nakamoto T., 2009, MNRAS, 398, 715 

\bibitem[\protect\citeauthoryear{Yajima, Choi, 
\& Nagamine}{2011}]{Yajima2011} Yajima H., Choi J.-H., Nagamine K., 2011, MNRAS, 412, 411

\bibitem[\protect\citeauthoryear{Yajima, Choi, 
\& Nagamine}{2012}]{Yajima2012} Yajima H., Choi J.-H., Nagamine K., 2012, MNRAS, 427, 2889

\bibitem[\protect\citeauthoryear{Zackrisson et 
al.}{2011}]{Zackrisson2011} Zackrisson E., Rydberg C.-E., Schaerer D., 
{\"O}stlin G., Tuli M., 2011, ApJ, 740, 13 

\bibitem[\protect\citeauthoryear{Zaroubi et 
al.}{2012}]{Zaroubi2012} Zaroubi S., et al., 2012, MNRAS, 425, 2964

\bibitem[\protect\citeauthoryear{Zaroubi}{2013}]{Zaroubi2013} 
Zaroubi S., 2013, ASSL, 396, 45 

\end{thebibliography}
\end{document}